\documentclass[twocolumn,aps,prd,10pt,superscriptaddress,longbibliography]{revtex4-1}

\usepackage{amssymb,amsmath}
\usepackage{dcolumn}
\usepackage{glossaries}
\usepackage{graphicx}
\usepackage[version=4]{mhchem}
\usepackage[usenames,dvipsnames,svgnames]{xcolor}
\usepackage{siunitx}
\usepackage{hyperref}
\hypersetup{
pdfnewwindow=true,      
colorlinks=true,        
linkcolor=Blue,         
citecolor=Blue,         
filecolor=Blue,         
urlcolor=Blue           
}

\newacronym{aimd}{AIMD}{\emph{ab initio} molecular dynamics}
\newacronym{bcc}{BCC}{body-centered cubic}
\newacronym{dft}{DFT}{density functional theory}
\newacronym{fcc}{FCC}{face-centered cubic}
\newacronym{fwhm}{FWHM}{full width at half maximum}
\newacronym{gpu}{GPU}{graphics processing units}
\newacronym{hcp}{HCP}{hexagonal close packed}
\newacronym{lda}{LDA}{local density approximation}
\newacronym{md}{MD}{molecular dynamics}
\newacronym{mlp}{MLP}{machine-learned potential}
\newacronym{mof}{MOF}{metal-organic framework}
\newacronym{nep}{NEP}{neuroevolution potential}
\newacronym{nqe}{NQE}{nuclear quantum effect}
\newacronym{pimd}{PIMD}{path integral molecular dynamics}
\newacronym{pbe}{PBE}{Perdew-Burke-Ernzerhof}
\newacronym{qha}{QHA}{quasi-harmonic approximation}
\newacronym{rdf}{RDF}{radial distribution function}
\newacronym{rmse}{RMSE}{root mean square error}
\newacronym{rpmd}{RPMD}{ring-polymer molecular dynamics}
\newacronym{sed}{SED}{spectral energy density}
\newacronym{sm}{SM}{Supplemental Material}
\newacronym{tec}{TEC}{thermal expansion coefficient}
\newacronym{trpmd}{TRPMD}{thermostatted ring-polymer molecular dynamics}

\DeclareSIUnit\angstrom{\text{Å}}
\DeclareSIUnit{\atom}{atom}
\DeclareSIUnit{\step}{step}
\DeclareSIUnit{\atomstepsecond}{\atom\step\per\second}
\renewcommand{\vec}[1]{\ensuremath\boldsymbol{#1}}

\begin{document}

\title{Highly efficient path-integral molecular dynamics simulations with GPUMD using neuroevolution potentials: Case studies on thermal properties of materials}

\author{Penghua Ying}
\thanks{These authors contributed equally to this work.}
\affiliation{Department of Physical Chemistry, School of Chemistry, Tel Aviv University, Tel Aviv, 6997801, Israel}

\author{Wenjiang Zhou}
\thanks{These authors contributed equally to this work.}
\affiliation{Department of Energy and Resources Engineering, Peking University, Beijing 100871, China}
\affiliation{School of Advanced Engineering, Great Bay University, Dongguan 523000, China}

\author{Lucas Svensson}
\thanks{These authors contributed equally to this work.}
\affiliation{
  Department of Physics,
  Chalmers University of Technology,
  41926 Gothenburg, Sweden
}
\affiliation{
    Wallenberg Initiative Materials Science for Sustainability, Chalmers University of Technology, 41926 Gothenburg, Sweden
}

\author{Esmée Berger}
\affiliation{
  Department of Physics,
  Chalmers University of Technology,
  41926 Gothenburg, Sweden
}

\author{Erik Fransson}
\affiliation{
  Department of Physics,
  Chalmers University of Technology,
  41926 Gothenburg, Sweden
}

\author{Fredrik Eriksson}
\affiliation{
  Department of Physics,
  Chalmers University of Technology,
  41926 Gothenburg, Sweden
}

\author{Ke Xu}
\affiliation{Department of Electronic Engineering and Materials Science and Technology Research Center, The Chinese University of Hong Kong, Shatin, N.T., Hong Kong SAR, 999077, China}

 \author{Ting Liang}
\affiliation{Department of Electronic Engineering and Materials Science and Technology Research Center, The Chinese University of Hong Kong, Shatin, N.T., Hong Kong SAR, 999077, China}

\author{Jianbin Xu}
\affiliation{Department of Electronic Engineering and Materials Science and Technology Research Center, The Chinese University of Hong Kong, Shatin, N.T., Hong Kong SAR, 999077, China}

\author{Bai Song}
\email {songbai@pku.edu.cn}
\affiliation{Department of Energy and Resources Engineering, Peking University, Beijing 100871, China}
\affiliation{Department of Advanced Manufacturing and Robotics, Peking University, Beijing 100871, China}
\affiliation{National Key Laboratory of Advanced MicroNanoManufacture Technology, Beijing 100871, China}

\author{Shunda Chen}
\email{phychensd@gmail.com}
\affiliation{Department of Civil and Environmental Engineering, George Washington University,
Washington, DC 20052, USA}

\author{Paul Erhart}
\email{erhart@chalmers.se}
\affiliation{
  Department of Physics,
  Chalmers University of Technology,
  41926 Gothenburg, Sweden
}
\affiliation{
    Wallenberg Initiative Materials Science for Sustainability, Chalmers University of Technology, 41926 Gothenburg, Sweden
}

\author{Zheyong Fan}
 \email{brucenju@gmail.com}
 \affiliation{College of Physical Science and Technology, Bohai University, Jinzhou 121013, China}

\date{\today}

\begin{abstract}
Path-integral molecular dynamics (PIMD) simulations are crucial for accurately capturing nuclear quantum effects in materials.
However, their computational intensity and reliance on multiple software packages often limit their applicability at large scales.
Here, we present an integration of PIMD methods, including thermostatted ring-polymer molecular dynamics (TRPMD), into the open-source GPUMD package, combined with highly accurate and efficient machine-learned neuroevolution potential (NEP) models.
This approach achieves almost the accuracy of first-principles calculations with the computational efficiency of empirical potentials, enabling large-scale atomistic simulations that incorporate nuclear quantum effects.
We demonstrate the efficacy of the combined NEP-PIMD approach by examining various thermal properties of diverse materials, including lithium hydride (LiH), three porous metal-organic frameworks (MOFs), liquid water, and elemental aluminum. 
For LiH, our NEP-PIMD simulations successfully capture the isotope effect, reproducing the experimentally observed dependence of the lattice parameter on the reduced mass. 
For MOFs, our results reveal that achieving good agreement with experimental data requires consideration of both nuclear quantum effects and dispersive interactions. 
For water, our PIMD simulations capture the significant impact of nuclear quantum effects on its microscopic structure.
For aluminum, the TRPMD method effectively captures thermal expansion and phonon properties, aligning well with quantum mechanical predictions.
This efficient NEP-PIMD approach opens new avenues for exploring complex material properties influenced by nuclear quantum effects, with potential applications across a broad range of materials.
\end{abstract}

\maketitle

\section{Introduction\label{intro}}
Since Rahman's pioneering work in 1964 \cite{rahman1964correlations}, \gls{md} simulations have been playing a central role in modeling physical and chemical properties of matter.
Two essential components of \gls{md} simulations are the interatomic potential and the integrator, both of which can be treated classically or quantum-mechanically.
When it comes to describing the interatomic interactions, classical empirical models usually lack the required accuracy, while first-principles methods such as quantum-mechanical \gls{dft} calculations have been the standard when the accuracy in force calculations is crucial, despite being computationally very demanding. 
In recent years, the situation has improved with the advent of \glspl{mlp} or force fields \cite{Unke2021cr} that can achieve nearly quantum-mechanical accuracy with orders of magnitude enhancement on the computational efficiency over \gls{dft} calculations.
With respect to integrators, there are classical ones based on classical statistical mechanics, leading to classical \gls{md}, as well as quantum-mechanical ones based on path-integral statistical mechanics \cite{feynman1965quantum}, leading to \gls{pimd} \cite{Parrinello1984jcp}.
\Gls{pimd} can account for \glspl{nqe} \cite{markland2018nuclear} by employing multiple replicas for each atom, which substantially increases computational cost.

Due to the improved computational efficiency of \glspl{mlp} over \gls{dft} calculations, \gls{mlp}-\gls{pimd} simulations have gained popularity for studying materials with notable \glspl{nqe}, particularly in water  \cite{ko2019mp, cheng2019ab, xu2020prb, reinhardt2021quantum, kapil2022first, bore2023realistic, Bocus2023NC, chen2024jced, berrens2024nuclear}.
These previous studies often used separate packages for force calculation and integration, typically combining the \textsc{lammps} package \cite{thompson2022lammps} with a \gls{mlp} model for forces and the \textsc{i-pi} package  \cite{ceriotti2014pi, kapil2019pi, litman2024ipi} for \gls{pimd} integration and sampling.
Although this modular approach is flexible, it can lead to suboptimal computational performance due to the overhead of interfacing different software packages. 

In this study, we implement and benchmark an integrated \gls{mlp}-\gls{pimd} approach within the \textsc{gpumd} package \cite{fan2017cpc}, leveraging the highly efficient \gls{nep} approach \cite{fan2021neuroevolution, fan2022jpcm, fan2022gpumd, song2023generalpurpose}.
\Gls{nep} models have demonstrated extremely high computational efficiency comparable to typical empirical force fields \cite{xu2023accurate,ying2023sub,Liu2023prb} and have been employed to study physical properties that require extensive spatiotemporal sampling, such as fracture  \cite{ying2023atomistic, yu2024fracture}, thermal transport \cite{wang2023quantum, liang2023mechanisms, xu2023accurate, ying2023sub, eriksson2023tuning, fan2024combining, dong2024molecular}, and phase transitions \cite{FraWikErh23, fransson2023revealing, FraRosErh24, FraWikErh24} as well as nucleation processes \cite{ahlawat2024size}, among others \cite{Liu2023prb, chen2024jced, chen2024intricate, song2024solute}.
By integrating \gls{pimd} methods directly into the \textsc{gpumd} package, we introduce the \gls{nep}-\gls{pimd} approach, a highly efficient computational tool capable of capturing both accurate interatomic forces and \glspl{nqe}.   

To demonstrate the effectiveness of the \gls{nep}-\gls{pimd} approach, we investigate thermal expansion and related properties in materials with strong \glspl{nqe}, including crystalline lithium hydride (LiH), three different \glspl{mof}, liquid water, and elemental aluminum. LiH is ideal for exploring the isotope effect \cite{lindsay2016isotope, zhou2024isotope}, with experimental data available for isotope-dependent lattice parameters \cite{vidal1986accurate,anderson1970isotopic}.
For \glspl{mof}, experimental studies \cite{rowsell2005gas, zhou2008origin, lock2010elucidating, lock2013scrutinizing, wu2008negative, peterson2010local, schneider2019tuning, sapnik2018compositional, burtch2019negative} show complex thermal expansion behavior, with both positive and negative coefficients. 
To our knowledge, theoretical investigations of \glspl{mof} thermal expansion using \gls{pimd} are limited, with only one study \cite{lamaire2019importance} based on an empirical force field.
Water is a prototypical material for studying \glspl{nqe} on the microscopic structure of a condensed system in the liquid phase. 
Aluminum is particularly well-suited for investigating quantum dynamics in the form of phonon properties, as it can be accurately described using low-order perturbation theory at low temperatures, providing a suitable reference for results obtained through \gls{trpmd}.
Leveraging our highly efficient \gls{nep}-\gls{pimd} approach, we systematically investigate the convergence of results with respect to the number of replicas in large-scale \gls{pimd} simulations, which effectively capture \glspl{nqe}.
Our approach paves the way for accurate and efficient large-scale modeling of materials with \glspl{nqe} using \gls{pimd}.

\section{Methods and models\label{meth}}

\subsection{Path-integral molecular dynamics implemented in GPUMD}

\subsubsection{PIMD integration algorithms}

\Gls{pimd} is an \gls{md} method based on the path-integral formulation of quantum mechanics \cite{feynman1965quantum}, as first proposed by Parrinello and Rahman \cite{Parrinello1984jcp}.
The crucial observation for deriving \gls{pimd} is that the quantum partition function for $N$ particles can be approximately cast to a classical partition function of $NP$ ($P \to \infty$) particles with the following Hamiltonian:
\begin{equation*}
    H_P = H_P^0 + U_P,
\end{equation*}
where
\begin{equation}
\label{equation:HP0}
H_P^0 = \sum_{i=1}^N\sum_{j=1}^P \left[\frac{m_i\omega_P^2\left(\mathbf{r}^{(i)}_j-\mathbf{r}^{(i)}_{j+1}\right)^2}{2} 
+ \frac{\left[\mathbf{p}^{(i)}_j\right]^2}{2m_i'} \right]
\end{equation}
and 
\begin{equation}
\label{equation:UP}
U_P = \sum_{j=1}^P U\left(\mathbf{r}^{(1)}_j,\mathbf{r}^{(2)}_j,\cdots,\mathbf{r}^{(N)}_j\right).
\end{equation}
That is, each quantum (physical) particle (indexed by $i$) is represented as a collection of $P$ classical particles (replicas, indexed by $j$). For each $i$ we have $\mathbf{r}^{(i)}_1 = \mathbf{r}^{(i)}_{P+1}$, and the relevant interaction terms in Eq.~\eqref{equation:HP0} represent a ring of replicas connected by springs with frequency $\omega_P=k_{\rm B} T P/\hbar$.
Therefore, each physical particle is approximated by a ``ring polymer'' with $P$ ``beads''.
Each of the $P$ replicas of the $N$ physical particles is still governed by the potential function $U$ of the system, as can be seen from Eq.~\eqref{equation:UP}.
While the mass $m_i'$ in the kinetic energy terms is not necessarily the physical mass $m_i$, we follow the convention \cite{Parrinello1984jcp} of taking them to be equal.

From the Hamiltonian, one can derive equations of motion and derive integration algorithms.
For simplicity, we follow the work of Cerrioti \textit{et al.} \cite{ceriotti2010jcp} to present the algorithms in terms of a single one-dimensional physical particle, with mass $m$, position $q$, and momentum $p$. First, we consider the free ring polymer consisting of $P$ beads with the Hamiltonian: 
\begin{equation*}
    H_P^0 = \sum_{j=1}^P \left(\frac{p_j^2}{2m} + \frac{1}{2} m \omega_P^2 (q_j - q_{j+1})^2 \right).
\end{equation*}
Next, we define the following transform ($k=0$ to $P-1$): \cite{ceriotti2010jcp}
\begin{align}
    \label{equation:transform_to_normal_1}
    \tilde{q}_k = \sum_{j=1}^P q_j C_{jk}, \\
    \label{equation:transform_to_normal_2}
    \tilde{p}_k = \sum_{j=1}^P p_j C_{jk},
\end{align}
where
\begin{equation*}
\label{equation:c10-Cjk}
   C_{jk}
   = \begin{cases}
   \frac{1}{\sqrt{P}} & (k=0)\\
   \sqrt{2/P} \cos(2\pi jk/P) & (1 \leq k \leq P/2 - 1)\\
   \frac{1}{\sqrt{P}}(-1)^j & (k = P/2)\\
   \sqrt{2/P} \sin(2\pi jk/P) & (P/2+1 \leq k \leq P - 1).
   \end{cases}
\end{equation*}
This leads to a representation of the free ring-polymer Hamiltonian in terms of uncoupled harmonic oscillators:
\begin{equation*}
    H_P^0 = \sum_{k=0}^{P-1} \left(\frac{\tilde{p}_k^2}{2m} + \frac{1}{2} m \omega_k^2 \tilde{q}_k^2 \right),
\end{equation*}
where
\begin{equation*}
    \omega_k = 2\omega_P \sin (k\pi/P).
\end{equation*}
This set of equations of motion can be integrated for one time step $\Delta t$ as follows \cite{ceriotti2010jcp}:
\begin{align*}
    \tilde{p}_k &\leftarrow \cos (\omega_k \Delta t) \tilde{p}_k - m \omega_k \sin (\omega_k \Delta t) \tilde{q}_k,
    \\
    \tilde{q}_k &\leftarrow \frac{1}{m \omega_k} \sin (\omega_k \Delta t) \tilde{p}_k + \cos (\omega_k \Delta t) \tilde{q}_k.
\end{align*}
Korol \textit{et al.} \cite{Korol2019jcp} applied the Cayley transform to construct a more robust algorithm: 
\begin{align}
    \tilde{p}_k &\leftarrow \frac{1 - (\omega_k \Delta t/2)^2}{1 + (\omega_k \Delta t/2)^2} \tilde{p}_k - m \omega_k \frac{ \omega_k \Delta t}{1 + (\omega_k \Delta t/2)^2} \tilde{q}_k,
    \label{equation:Cayley1}
    \\
    \tilde{q}_k &\leftarrow \frac{1}{m \omega_k} \frac{ \omega_k \Delta t}{1 + (\omega_k \Delta t/2)^2} \tilde{p}_k + \frac{1 - (\omega_k \Delta t/2)^2}{1 + (\omega_k \Delta t/2)^2} \tilde{q}_k.
    \label{equation:Cayley2}
    \end{align}
After integration, we can change the normal mode variables back to the original ones:
\begin{align}
    q_j &= \sum_{k=0}^{P-1}  \tilde{q}_k C_{kj},
    \label{equation:transform_from_normal_1}
    \\
    p_j &= \sum_{k=0}^{P-1}  \tilde{p}_k C_{kj}.
    \label{equation:transform_from_normal_2}
\end{align}

The above integration algorithm is for free ring polymers. To enable extension to interacting systems, the bead momenta need to be updated before and after the above set of operations, as follows:
\begin{equation}
    p_j \leftarrow p_j - \frac{\partial U}{\partial q_j} \frac{\Delta t}{2}.
    \label{equation:velocity-update}
\end{equation}
To thermostat the beads, both the Langevin-type thermostat  \cite{ceriotti2010jcp} and the massive Nos\'{e}-Hoover chain thermostat \cite{tuckerman_book} have been used. When using the Langevin-type thermostat, the integration over the bead momenta should be applied before and after the non-thermostatted operations as follows:
\begin{align}
    \label{equation:c10:langevin-1}
    \tilde{p}_k &\leftarrow \sum_{j=1}^P p_j C_{jk}, 
    \\
    \label{equation:c10:langevin-2}
    \tilde{p}_k &\leftarrow c_{1k} \tilde{p}_k +\sqrt{\frac{m}{\beta_P}} c_{2k} \xi_k,
    \\
    \label{equation:c10:langevin-3}
    p_j &\leftarrow \sum_{k=0}^{P-1} C_{jk} \tilde{p}_k. 
    \end{align}
Here $\xi_k$ are normally distributed random numbers with zero mean and unit variance,
$c_{1k} = e^{-(\Delta t/2)\gamma_k}$,
$c_{2k} = \sqrt{1-(c_{1k})^2}$, $\gamma_0=1/\tau_T$, and $\gamma_k = \omega_k$ ($k>0$). 

The time parameter $\tau_T$ is an input chosen for the centroid ($k=0$) mode. When all the bead modes are thermostatted, the algorithm is known as \gls{pimd}. Without thermostatting, the algorithm is known as \gls{rpmd}, which has been proposed as a method for approximating time-correlation functions \cite{Craig2004JCP,Habershon2014ARPC}. If only the internal bead modes ($k>0$) are thermostatted, the algorithm is known as \gls{trpmd} \cite{rossi_how_2014}.

The above algorithms can be easily extended to multi-particle systems in three dimensions. 
The potential energy of the system can be evaluated by averaging over the replicas, while the kinetic energy can be more conveniently calculated using the virial estimate \cite{tuckerman_book}: 
\begin{equation*}
    K = \frac{3}{2} Nk_{\rm B} T - \frac{1}{2} \sum_{i=1}^N \frac{1}{P} \sum_{j=1}^P \left(\mathbf{r}_j^{(i)} - \mathbf{r}_{\rm c}^{(i)}\right) \cdot \mathbf{F}_j^{(i)}.
\end{equation*}
Here, 
\begin{equation*}
    \mathbf{r}_{\rm c}^{(i)} = \frac{1}{P} \sum_{j=1}^{P} \mathbf{r}_j^{(i)}
\end{equation*}
is the centroid position of atom $i$, for which periodic boundary conditions should be taken into account. The force $\mathbf{F}_j^{(i)}$ is defined as 
\begin{equation*}
    \mathbf{F}_j^{(i)} = -\frac{\partial}{\partial \mathbf{r}_j^{(i)}} U\left(\mathbf{r}^{(1)}_j,\mathbf{r}^{(2)}_j,\cdots,\mathbf{r}^{(N)}_j\right).
\end{equation*}
The rank-2 virial tensor is calculated in a similar manner:
\begin{equation*}
    \mathbf{W} = - \sum_{i=1}^N \frac{1}{P} \sum_{j=1}^P \left(\mathbf{r}_j^{(i)} - \mathbf{r}_{\rm c}^{(i)}\right) \otimes \mathbf{F}_j^{(i)} + \sum_{i=1}^N \frac{1}{P} \sum_{j=1}^P \mathbf{W}_k^{(i)}.
\end{equation*}
Here, $\mathbf{W}_k^{(i)}$ is the per-atom virial tensor for the $k$-th replica of the $i$-th atom as calculated from the potential, which has been derived explicitly for \gls{nep} \cite{fan2021neuroevolution,fan2022gpumd}.
The pressure tensor is then calculated as 
\begin{equation}
\label{equation:pressure}
    \mathbf{P} = \frac{N k_{\rm B} T  + \mathbf{W}}{V},
\end{equation}
where $V$ is the volume of the system.

The algorithms presented above have been implemented into the open-source \textsc{gpumd} package \cite{fan2017cpc} during the course of this study.
The implementation is fully on \gls{gpu} using CUDA programming, with minimal data transfer between \gls{gpu} and the host.
The overall computational workflow of our implementation is as follows. 
\begin{enumerate}
\item Preparation.
    \begin{enumerate}
        \item Allocate \gls{gpu} memory for the bead variables, such as position, velocity, potential energy, force, and virial. There are $P$ sets of these variables.
        \item Pre-compute the transformation matrix $C_{jk}$ and store the data in \gls{gpu} memory.
        \item Initialize the pseudo random number generator and the seeds based on the cuRAND library in CUDA.
    \end{enumerate} 
\item Iterate the integration loop.
    \begin{enumerate}
        \item In the case of \gls{pimd} and \gls{trpmd}, apply the Langevin thermostat according to Eqs.~\eqref{equation:c10:langevin-1} to \eqref{equation:c10:langevin-3}.
        In \gls{trpmd}, only the non-centroid modes (internal bead modes) are thermostatted.
        \item Update velocities according to Eq.~\eqref{equation:velocity-update}, for half time step $\Delta t/2$.
        \item Update positions and velocities for the free ring-polymer system.
        \begin{enumerate}
            \item Transform to normal modes according to Eqs.~\eqref{equation:transform_to_normal_1} and \eqref{equation:transform_to_normal_2}.
            \item Perform the time stepping for the normal mode variables according to Eqs.~\eqref{equation:Cayley1} and \eqref{equation:Cayley2}.
            \item Transform back from normal modes according to Eqs.~\eqref{equation:transform_from_normal_1} and \eqref{equation:transform_from_normal_2}.
        \end{enumerate}
        \item Calculate forces according to the updated bead positions. 
        \item Update velocities according to Eq.~\eqref{equation:velocity-update}, for the half time step $\Delta t/2$.
        \item When required, control the pressure using a Berendsen-like algorithm \cite{berendsen1984molecular}, with the instant pressure calculated according to Eq.~\eqref{equation:pressure}.
        Although this algorithm does not lead to a true isothermal–isobaric ensemble, it is sufficient for accurately determining the average simulation cell dimensions.
    \end{enumerate}
\end{enumerate}

\subsection{Phonon properties from TRPMD}
To showcase the application of the implemented \gls{trpmd} method, we examine the variations in phonon frequencies and phonon damping parameters in aluminum as a function of temperature. Phonon properties are derived from atomic velocities sampled during \gls{trpmd} simulations. Specifically, the velocities are used to compute the phonon \gls{sed} with the Python package \textsc{dynasor} \cite{fransson_dynasortool_2021}. The \gls{sed} is a measure of how the kinetic energy of the system is distributed over the different phonon modes making it closely related to the phonon dispersion. It can be obtained by defining $q_j^{\mu \alpha}(\boldsymbol{q}, t)$ as the
contribution to the $j$-th normal coordinate coming from the displacement $u_{n \mu}^\alpha$ of the $\mu$-th atom in Cartesian direction $\alpha$, i.e.,

\begin{equation*}
    q_j^{\mu \alpha}(\boldsymbol{q}, t)=\frac{1}{\sqrt{N}} \sum_n \sqrt{m_\mu} u_{n \mu}^\alpha(\boldsymbol{q}, t) A_{\mu j}^{\alpha *}(\boldsymbol{q}) e^{i \boldsymbol{q} \cdot \boldsymbol{R}_n^0}.
\end{equation*}
Here, $n$ ranges from 1 to $N$, where $N$ is the total number of unit cells in the crystal, $m_\mu$ represents the mass of the $\mu$-th atom, $A_{\mu j}^{\alpha *}(\boldsymbol{q})$ is the complex conjugate of the vibrational eigenvector, and $\boldsymbol{R}_n^0$ denotes the position of the $n$-th unit cell relative to the origin.

The average kinetic energy of phonon mode $j$ can be expressed as a time average in terms of $q_j^{\mu \alpha}(\boldsymbol{q}, t)$ \cite{thomas_predicting_2010}
\begin{align*}
    \begin{split}
        \overline{\mathcal{T}}_j &= \sum_{\mu \alpha} \lim _{\tau_0 \rightarrow \infty} \frac{1}{\tau_0} \int_0^{\tau_0} \frac{1}{2} \dot{q}_j^{\mu \alpha *}(\boldsymbol{q}, t) \, \dot{q}_j^{\mu \alpha}(\boldsymbol{q}, t) \, d t \\
        &= \lim _{\tau_0 \rightarrow \infty} \frac{1}{2 \tau_0} \int_0^{\infty} \sum_{\mu \alpha}\left|\dot{q}_j^{\mu \alpha}(\boldsymbol{q}, t)\right|^2 d t .
    \end{split}
\end{align*}
Through the use of Parseval’s theorem the average kinetic energy may be expressed in the frequency domain as
\begin{equation*}
    \overline{\mathcal{T}}_j = \int_0^{\infty} \sum_{\mu \alpha} \lim _{\tau_0 \rightarrow \infty} \frac{1}{2 \tau_0}\left|\dot{q}_j^{\mu \alpha}(\boldsymbol{q}, \omega)\right|^2 d \omega.
\end{equation*}
The \gls{sed} is simply defined as the integrand
\begin{equation*}
    \Phi(\boldsymbol{q}, \omega)=\sum_{\mu \alpha} \lim _{\tau_0 \rightarrow \infty} \frac{1}{2 \tau_0}\left|\dot{q}_j^{\mu \alpha}(\boldsymbol{q}, \omega)\right|^2.
\end{equation*}
By applying the convolution theorem and the fact that the conjugate of a Fourier transformed function $x(t)$ is equal to the Fourier transform of $x^*(-t)$, $\left|\dot{q}_j^{\mu \alpha}(\boldsymbol{q}, \omega)\right|^2$ can be simplified and the \gls{sed} becomes 
\begin{align*}
    \begin{split}
        \Phi(\boldsymbol{q}, \omega) = \frac{1}{4 \pi} \sum_{\mu \alpha} \int_0^\infty C_{\dot{q} \dot{q}}(\tau)  e^{-i \omega \tau} d \tau.
    \end{split}
\end{align*}
Here $C_{\dot{q} \dot{q}}(\tau)$ is the auto correlation function of the time derivative of $q_j^{\mu \alpha}(\boldsymbol{q}, t)$ and is defined as
\begin{equation*}
    C_{\dot{q} \dot{q}}(\tau) \equiv \lim_{\tau_0 \rightarrow \infty} \frac{1}{\tau_0}
        \int_0^{\tau_0}  \dot{q}_j^{\mu \alpha *}(\boldsymbol{q}, t - \tau)  \, \dot{q}_j^{\mu \alpha}(\boldsymbol{q}, t) \, dt.
\end{equation*}

Phonon frequencies and damping parameters are determined by fitting the velocity correlation function for a damped harmonic oscillator, expressed in the frequency domain, to the peaks of the calculated \gls{sed}.
The equation of the velocity correlation function for a damped harmonic oscillator is given by
\begin{equation*}
    b(\omega) = B \frac{2 \Gamma \omega^2}{(\omega^2 - \omega_0^2)^2 + (\Gamma \omega)^2},
    \label{eq:velocity_DHO_freq}
\end{equation*}
where $\omega_0$ is the phonon frequency of the mode, $\Gamma$ is the damping parameter, related to the phonon lifetime as $\Gamma = 2 / \tau_{ph}$, and $B$ is the amplitude. The phonon frequencies correspond to the centroid frequencies of the peaks, while the damping parameters represent the \gls{fwhm} of the peaks.

\subsection{NEP machine-learned potential models}

\subsubsection{NEP formalism}

To construct accurate potential models driving the \gls{pimd} simulations, we use the \gls{nep} approach \cite{fan2021neuroevolution,fan2022jpcm,fan2022gpumd}. This is a neural-network-based \gls{mlp} trained using the separable natural evolution strategy (SNES) \cite{Schaul2011High} algorithm. Similar to many other \glspl{mlp}, such as the Behler-Parrinello neural network potential \cite{Behler2007prl}, the total potential energy $U$ of a system of $N$ atoms is expressed as the sum of $N$ site energies, $U=\sum_i U_i$, and each site energy $U_i$ is modeled as a function of set of descriptors, $U_i=U_i(\{ q^i_{\nu}\})$. Each descriptor $q^i_{\nu}$ depends on the atomic positions and is invariant under translation, rotation and permutation of atoms of the same species. The explicit forms of the descriptor in \gls{nep} are detailed in Ref.~\onlinecite{fan2022gpumd}. As for the machine-learning model representing the function $U_i=U_i(\{ q^i_{\nu}\})$, a feedforward neural network with a single hidden layer is used. For multi-species systems, we utilize the NEP4 version, which was recently introduced for many-species metals and their alloys \cite{song2023generalpurpose}.

\subsubsection{NEP for LiH}

For LiH, we use an iterative method to train the \gls{nep} model to accurately fit the potential energy surface \cite{dong2024molecular}. A total of 603 structures were sampled, including the MD simulations at various temperatures (\SI{100}{\kelvin} to \SI{1000}{\kelvin}), and with random strain perturbations (in the range of $\pm1.5\%$ and $\pm3\%$). Each structure contains 250 atoms. The Vienna Ab initio Simulation Package  (VASP) with the projector-augmented wave method \cite{PAW1994, planewavebasis1996} is used to obtain the energy, forces, and virial of each structure. In the \gls{dft} calculations,  the Perdew-Zunger functional \cite{PerdewZungerfunctional1981} with the local density approximation (LDA) is used to describe the exchange-correlation of electrons.
The cutoff energy and the energy convergence threshold were chosen as  \qty{600}{\electronvolt} and \qty{1e-8 }{\electronvolt}, respectively. The $\vec{k}$-point mesh was  set to \numproduct{2x2x2}. Then, \num{447} and \num{156} structures were randomly selected to form the training and testing datasets, respectively. 

During the training processes, the cutoff radii for the radial and angular descriptor components were both set to \qty{5}{\angstrom}. The parity plots and accuracy metrics are shown in  Fig.~S1 of the \gls{sm}. The \gls{rmse} values of the total energy, atomic forces, and virial for the training dataset are \qty{0.1}{\meV\per\atom}, \qty{9.5}{\meV\per\angstrom}, and \qty{0.8}{\meV\per\atom}, respectively. In the testing dataset, the corresponding \gls{rmse} values are \qty{0.1}{\meV\per\atom}, \qty{9.4}{\meV\per\angstrom}, and \qty{0.8}{\meV\per\atom}. To further validate the \gls{mlp} accuracy, we computed the phonon dispersion using the finite-difference method. As shown in  Fig.~S2 of the \gls{sm}, our trained \gls{mlp} accurately describes the lattice dynamics and can be reliably used to simulate the thermal expansion of LiH.

\subsubsection{NEP for MOFs}

Using three \glspl{mof} ---MOF-5, HKUST-1, and ZIF-8--- as examples of soft porous crystals, we explore the \glspl{nqe} on their thermal expansion behavior through \gls{pimd} simulations powered by \glspl{mlp}.
For this purpose, we utilized three machine-learned \gls{nep} models, one for each \gls{mof}, previously developed and validated against \gls{dft} calculations at the \gls{pbe} level \cite{ying2023sub}.
Before investigating the thermal expansion behavior of these three \glspl{mof}, we assessed the reliability of the \gls{nep} models in \gls{pimd} simulations, given that spring interactions between beads might generate configurations not encompassed by the dataset of the previously developed \gls{nep} models \cite{ying2023sub}.
While the original \gls{nep} models \cite{ying2023sub} were trained using \gls{pbe} reference calculations that did not include long-range dispersion interactions, incorporating these interactions is crucial for accurately modeling the thermodynamic behavior of \gls{mof} crystals \cite{walker2010flexibility, wieme2018tuning, ying2023combining}.
Here, we therefore used a newly developed \gls{nep}-D3 approach \cite{ying2023combining} to perform \gls{pimd} simulations, integrating the original \gls{nep} models with dispersion interactions using the \gls{dft}-D3 method with the Becke-Johnson damping function \cite{grimme2011effect}. 
Specifically, the cutoff radius for the D3 potential and the calculation of coordination numbers are set to \qty{12}{\angstrom} and \qty{6}{\angstrom}, respectively, to balance accuracy and efficiency \cite{ying2023combining}.

To validate this approach, \gls{pimd} simulations driven by the \gls{nep}-D3 models were conducted on the primitive cells of MOF-5, HKUST-1, and ZIF-8 using a ring polymer comprising \num{64} beads, gradually heating from \qty{100}{\kelvin} to \qty{500}{\kelvin} over \qty{1}{\nano\second}. During these simulations, all six cell components were independently adjusted to maintain zero pressure. For each \gls{mof}, \num{50} snapshots were uniformly selected from the trajectory, and static \gls{dft} calculations were performed on these snapshots. The static \gls{dft} calculation setup was identical to that used for the previous reference dataset \cite{ying2023sub}. As shown in Fig.~S3 of the \gls{sm}, the \gls{nep}-D3 and \gls{pbe}-D3 approaches yield consistent results for total energy, forces, and virials, with energy, force, and virial \gls{rmse} values of \qty{0.4}{\meV\per\atom}, \qty{53.1}{\meV\per\angstrom}, and \qty{7.7}{\meV\per\atom}, respectively, demonstrating the reliability of the \gls{nep}-D3 model for modeling thermal expansion behavior in \gls{pimd} simulations.

\subsubsection{NEP for liquid water}
For liquid water, the \gls{nep} model from Ref.~\onlinecite{XuRosSch24} is employed, which was trained on a data set containing \num{1888} structures \cite{zhang2018prl} calculated based on the strongly constrained and appropriately normed (SCAN) functional \cite{HuiCha16}. For more details on the training of this model, see Ref.~\onlinecite{XuRosSch24}.

\subsubsection{NEP for elemental aluminum}

For elemental aluminum, bootstrapping and active learning were utilized for constructing the training dataset, up to three iterations.
The training data included: (1) rattled structures based on the \gls{fcc} and \gls{hcp} phases, (2) structures from energy-volume curves for the \gls{fcc}, \gls{hcp}, diamond, and \gls{bcc} phases; (3) structures generated via simulations of heating under pressures ranging from \num{-5} to \qty{10}{\giga\pascal}, including molten configurations; (4) configurations of (111), (110), and (100) surfaces, as well as vacancy configurations (rattled).
In total, the reference dataset contains \num{1050} configurations and \num{52187} atoms.
Reference data were generated using \gls{dft} calculations as implemented in VASP and the van-der-Waals density functional method with consistent exchange \cite{DioRydSch04, BerHyl2014}.

In the \gls{nep} model, the cutoff radii for radial and angular descriptor parts were set to \qty{6}{\angstrom} and \qty{4}{\angstrom}, respectively.
For three-body terms, $l_\text{max}$, was set to 4. The neural network has 40 neurons in the hidden layer.
The ensemble model contains 5 submodels, created through bagging.
The \gls{nep} model for aluminum was validated against various physical properties, including bulk phases, surface properties, melting behavior, and phonon spectra.
The \gls{rmse} values obtained by averaging over the ensemble models are \qty{1.2}{\meV\per\atom}, \qty{30}{\meV\per\angstrom}, \qty{14}{\meV\per\atom} for energies, forces, and virials, respectively.

\subsection{NEP-PIMD simulation details}

In this work, we study LiH, water, aluminum, and three typical \glspl{mof}, including MOF-5 \cite{li1999design}, HKUST-1 \cite{chui1999chemically}, and ZIF-8 \cite{huang2006ligand}.
For LiH, we used a \numproduct{10x10x10} cubic supercell with \num{8000} atoms.
For the \glspl{mof}, we employed orthogonal \numproduct{4x4x4} supercells, containing \num{27136} atoms for MOF-5, \num{39936} atoms for HKUST-1, and \num{17664} atoms for ZIF-8, respectively.
In Fig.~S4 of the \gls{sm}, we illustrate the supercells of LiH and the three \glspl{mof} used in the \gls{md} simulations.
The liquid water simulations were conducted using a cubic cell with \num{41472} atoms. 
For aluminum, we used a \numproduct{12x12x12} cubic supercell with \num{6912} atoms for the bead convergence test and a \numproduct{24x24x24} triclinic supercell with \num{13824} atoms for the \gls{trpmd} simulations.

We used a time step of \qty{1}{\femto\second} for LiH and aluminum, and  \qty{0.5}{\femto\second} for \glspl{mof} and water.
It might be safe to use a time step of \qty{1}{\femto\second} for \glspl{mof} as well, but we decided to follow the previous work \cite{ying2023sub}, which used a time step of \qty{0.5}{\femto\second} in classical \gls{md} simulations of heat transport.
To study thermal expansion, we control the isotropic pressure with a target value of 0 GPa in the \gls{pimd} simulations.
For LiH, the total simulation time for each temperature was \qty{50}{\pico\second}, and the last \qty{25}{\pico\second} were used to calculate the cell dimensions.
For the \glspl{mof}, the total simulation time for each temperature was \qty{30}{\pico\second}, and the last \qty{10}{\pico\second} were used to calculate the cell dimensions.
The total simulation time for water was \qty{30}{\pico\second} for each temperature, of which the last \qty{20}{\pico\second} were used to compute the \glspl{rdf}.
For aluminum, the total simulation time for each temperature was \qty{510}{\pico\second}, and the last \qty{500}{\pico\second} were used to calculate the cell dimensions and evaluate the phonon properties.

From the above simulation details, we see that our implementation of the \gls{nep}-\gls{pimd} approach in the \textsc{gpumd} package achieves state-of-the-art efficiency, allowing for extensive simulations on nanosecond timescales for systems containing tens of thousand atoms using a large number of beads, sufficient to converge the ring polymer, i.e., approach the $P\rightarrow\infty$ limit in Eq.~\eqref{equation:HP0}.

\section{Results and Discussion}

\subsection{LiH}
Here, we first consider the natural isotopic abundance with 7.6\% $\mathrm{^7Li}$ and 92.4\% $\mathrm{^6Li}$, while H is practically pure $\mathrm{^1H}$ (99.99\%).
There is a balance between computational accuracy and efficiency with respect to the number of beads.
The \glspl{nqe} are expected to become more dominant at lower temperatures.

\begin{figure}
\centering
\includegraphics[width=1\columnwidth]{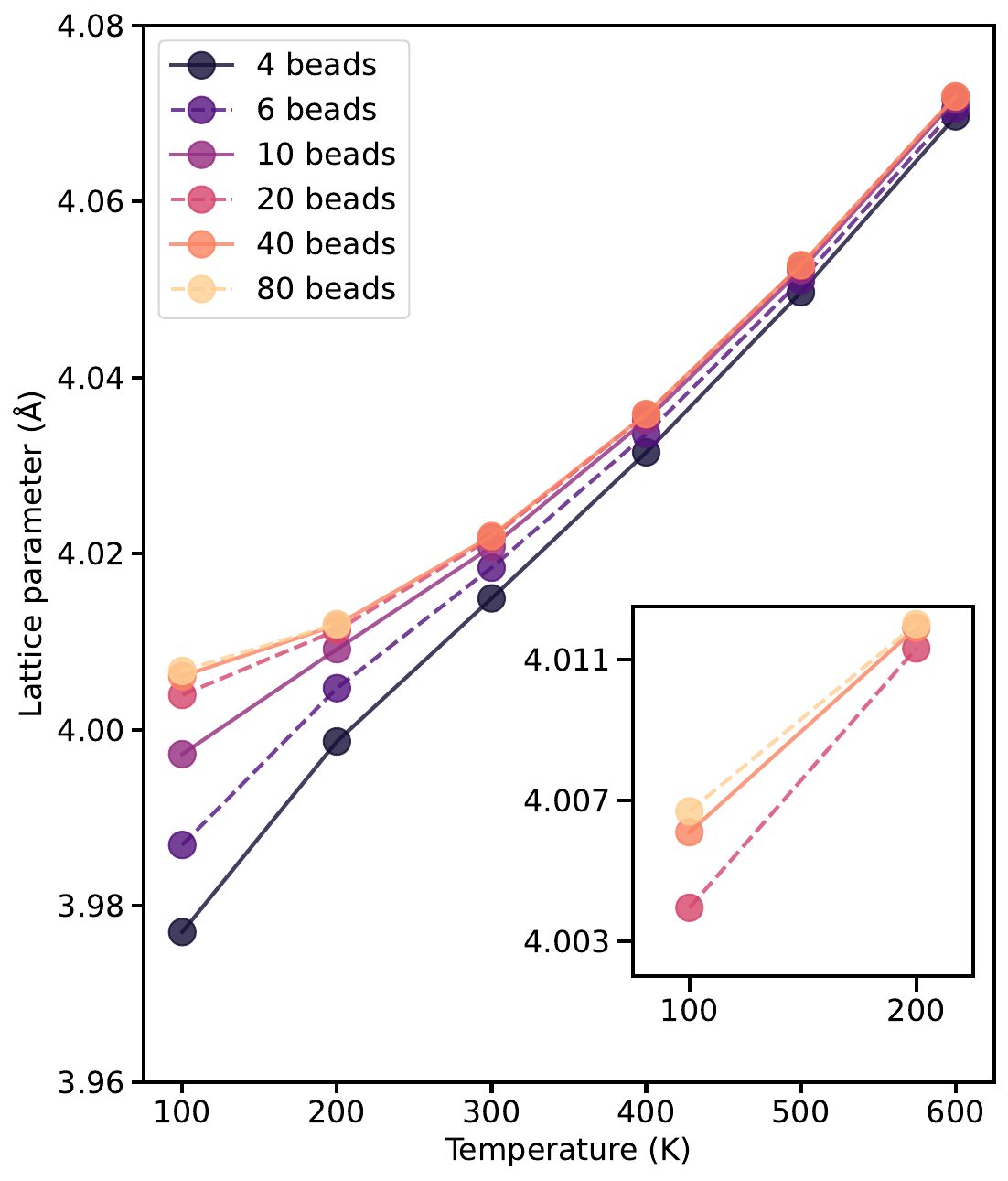}
\caption{
    Convergence of \gls{nep}-\gls{pimd} simulations with respect to the number of beads for the lattice parameter of LiH at various temperatures.
    The inset shows a zoomed-in view of the \gls{pimd} results at \qty{100}{\kelvin} and \qty{200}{\kelvin}.
}
\label{fig:bead}
\end{figure}

\begin{figure*}
\centering
\includegraphics[width=2\columnwidth]{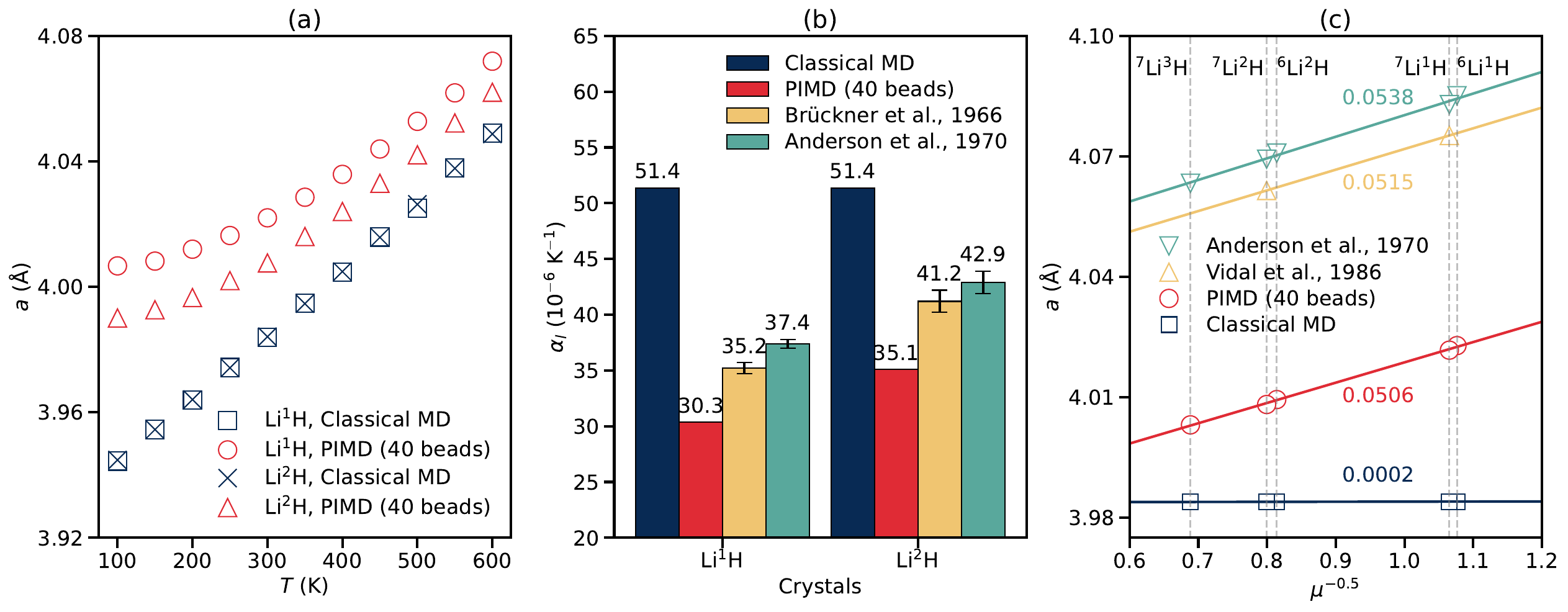}
\caption{
    Isotope effects on the thermal expansion of LiH. (a) Lattice parameters of $\mathrm{Li^1H}$ and $\mathrm{Li^2H}$ as a function of temperature.
    (b) Calculated room-temperature linear thermal expansion coefficients of $\mathrm{Li^1H}$ and $\mathrm{Li^2H}$. Experimental data is from Refs.~\citenum{bruckner1966untersuchungen} and \citenum{anderson1970isotopic}.
    (c) Room-temperature lattice parameters as a function of the reduced mass $\mu$.
    For direct comparison, the simulation temperature for panel (c) is chosen as \qty{298.15}{\kelvin}, matching the temperature at which data was acquired in Ref.~\citenum{anderson1970isotopic}.
    The data in Ref.~\citenum{vidal1986accurate} was obtained at \qty{300}{\kelvin}.
}
\label{fig:isotope}
\end{figure*}

Indeed, as shown in \autoref{fig:bead}, one can see that we need at least \num{40} beads to achieve full convergence for temperatures of \qty{200}{\kelvin} and below. 
Considering the heavier atomic mass and thus weaker \glspl{nqe} in other isotopic systems of LiH, 40 beads should be sufficient and are employed in all simulations of LiH.

Subsequently, we investigate the thermal expansion and the isotope-dependent lattice parameters of LiH.
We first present the temperature-dependent lattice parameters ($a$) of $\rm{Li^1H}$ and $\rm{Li^2H}$ in 
\autoref{fig:isotope}a. 
In our calculations, the lattice parameters of $\mathrm{Li^1H}$ and $\mathrm{Li^2H}$ from classical \gls{md} simulations are nearly identical, and both show a linear temperature dependence. 
This is in contrast to previous experimental observations \cite{vidal1986accurate, anderson1970isotopic}.
Upon inclusion of the \glspl{nqe}, the correct trend and the isotope-induced differences are captured, particularly at low temperatures. 
We define the linear thermal expansion coefficient as:
\begin{equation*}
    \alpha_{l} = \frac{\partial{\ln(a)}}{\partial{T}}. 
\end{equation*}
The discrete temperature-dependent $a$ values are utilized to estimate $\alpha_{l}$ and the temperature step is set to \qty{50}{\kelvin}.
\autoref{fig:isotope}b shows the room-temperature linear thermal expansion coefficients of $\rm{Li^1H}$ and $\rm{Li^2H}$ calculated using both \gls{pimd} and classical \gls{md} simulations. 
Our \gls{pimd} results for both $\rm{Li^1H}$ and $\rm{Li^2H}$ are in good agreement with earlier experimental measurements. 

We further vary the atomic masses of Li and H to investigate whether the isotope effect on the lattice parameters are captured in \gls{nep}-\gls{pimd} simulations. 
Considering the experimental values measured by Anderson \textit{et al.} \cite{anderson1970isotopic}, the simulations are performed at \qty{298.15}{\kelvin} and five isotopic compositions are studied, including $\rm{^6Li^1H}$, $\rm{^7Li^1H}$, $\rm{^6Li^2H}$, $\rm{^7Li^2H}$, and $\rm{^7Li^3H}$. 
The experimental studies revealed that the lattice parameters of isotopically engineered LiH can be evaluated using the formula $a = A \mu^{-1/2}+B$ \cite{anderson1970isotopic}, where $\mu$ is the reduced mass defined as $\mathrm{1/(m_{Li}^{-1}+m_{H}^{-1})}$, and $A$ and $B$ are two coefficients. 
In \autoref{fig:isotope}c, we plot the room-temperature lattice parameters as a function of the reduced mass $\mu$.
The calculated $a$ values are about $1.5\%$ smaller than the measured data. 
This is acceptable since \gls{dft} calculations based on the \gls{lda} tend to underestimate the lattice parameter. 
Considering the isotope mass effect, the fitting coefficient $A$ is \qty{0.051}{\angstrom\sqrt{\gram}\per\sqrt{\mol}}, which is in excellent agreement with 
the experimental value of about \qty{0.054}{\angstrom\sqrt{\gram}\per\sqrt{\mol}} \cite{anderson1970isotopic}. 
In contrast, the classical \gls{md} simulations predict a substantially weaker dependence. 
These findings demonstrate that our \gls{nep}-\gls{pimd} simulations yield a high accuracy in modeling the thermal expansion of LiH and its isotope effects.

\subsection{MOFs}

\begin{figure*}[htb]
\centering
\includegraphics[width=2\columnwidth]{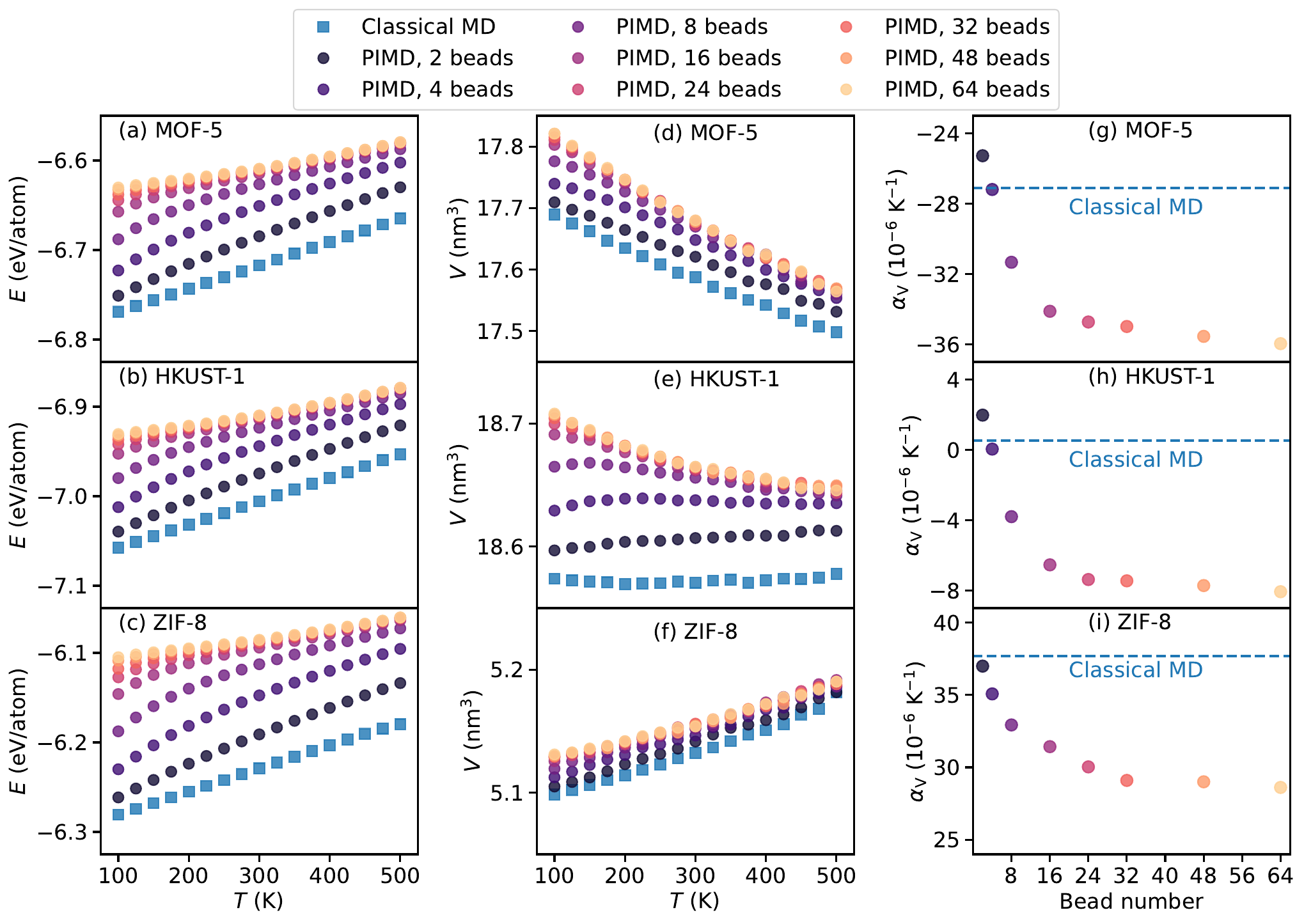}
\caption{
    The evolution of (a--c) energy ($E$) and (d--f) volume ($V$)  as a function of temperature $T$ using different bead numbers in \gls{pimd} simulations for MOF-5 (top), HKUST-1 (middle), and ZIF-8 (bottom).
    In panels (g--i), the volumetric \gls{tec} ($\alpha_{V}$) are obtained by fitting the $V(T)$ results using Eq.~\eqref{equation:tec}.
    The classical \gls{md} results (blue squares or dashed lines) are also provided for comparison.
}
\label{fig:mof}
\end{figure*}

\autoref{fig:mof}a--c shows the energy profile as a function of temperature using different numbers of beads.
In contrast to classical \gls{md} simulations, \gls{pimd} simulations predict a higher energy for all three \glspl{mof} due to zero-point energy contributions.
Additionally, the energies tend to converge at bead numbers larger than 32 for all temperatures, leading to a significantly lower heat capacity when compared to classical \gls{md}.
For the temperature-dependent volume as shown in \autoref{fig:mof}d--f, the \gls{pimd} simulations predict larger volumes than classical \gls{md}, particularly at lower temperatures where \glspl{nqe} are more pronounced.
The number of beads required to reach convergence is consistent with those observed for the energy.   

Based on the $V(T)$ results, the temperature-independent volumetric \gls{tec} at zero pressure can be estimated as
\begin{equation}
    \label{equation:tec}
    \alpha_{V} = \frac{\partial{\ln(V)}}{\partial{T}}. 
\end{equation}
\autoref{fig:mof}g--i shows the estimated $\alpha_{V}$, fitted using the above equation from \gls{pimd} simulation, as a function of the number of beads.
For all \glspl{mof}, $\alpha_{V}$ decreases as the number of beads increases and reaches convergence at around 48 beads. The relative difference in the predicted $\alpha_{V}$ between 48 and 64 beads is less than 1.5\% for MOF-5 and ZIF-8, and 4.5\% for HKUST-1.
Therefore, in the subsequent discussion, all \gls{pimd} results are based on simulations using 64 beads.

\begin{figure}[!]
\centering
\includegraphics[width=\columnwidth]{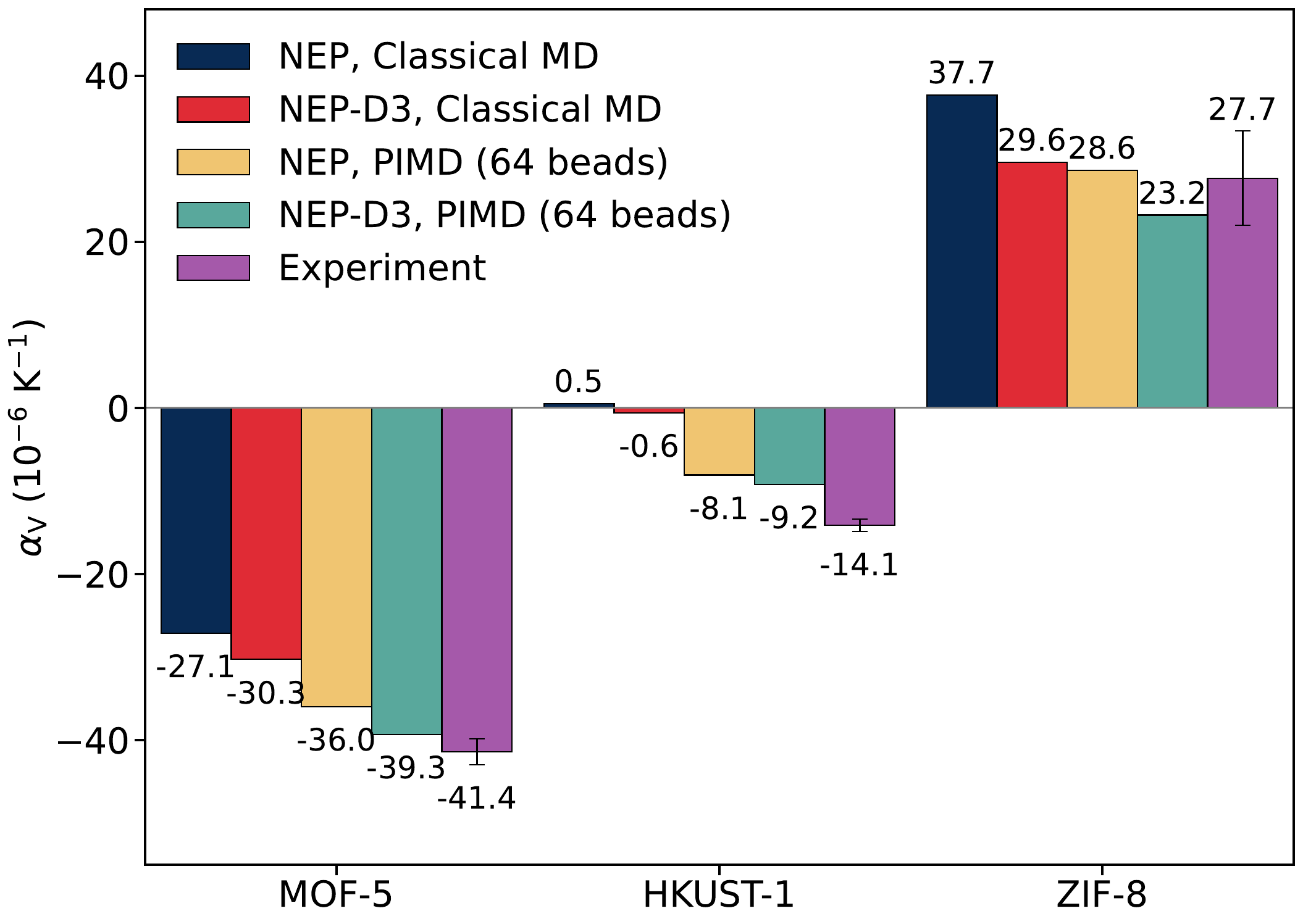}
\caption{\label{fig:tec} Comparison of volumetric \gls{tec} of three \glspl{mof} predicted by classical \gls{md} and \gls{pimd} (64 beads) simulations with experimental data. For each \gls{md} approach, both \gls{nep} and \gls{nep}-D3 results are provided to examine the long-range dispersion effects. The presented experimental results for each \gls{mof} are obtained by averaging several previous studies \cite{rowsell2005gas, zhou2008origin, lock2010elucidating, lock2013scrutinizing, wu2008negative, peterson2010local, schneider2019tuning, sapnik2018compositional, burtch2019negative}, with the corresponding error bar denoting the standard error of the mean as detailed in Table~S1 of the \gls{sm}.}
\end{figure}

To quantify the impact of \glspl{nqe} and long-range dispersion interactions on the \gls{tec} of \glspl{mof}, we estimated the $V(T)$ (see Fig.~S5 of the \gls{sm}) and corresponding $\alpha_{V}$ (see \autoref{fig:tec}) from four different sets of \gls{md} simulations.
These sets include classical \gls{md} and \gls{pimd}, each driven by either \gls{nep} or \gls{nep}-D3 models.
It is evident that both dispersion interactions and \glspl{nqe} are crucial for accurately estimating $\alpha_{V}$, as the \gls{nep}-D3 with \gls{pimd} approach aligns most closely with previous experimental measurements (see Table~S1 of the \gls{sm}) \cite{rowsell2005gas, zhou2008origin, lock2010elucidating, lock2013scrutinizing, wu2008negative, peterson2010local, schneider2019tuning, sapnik2018compositional, burtch2019negative}.
For HKUST-1, excluding \glspl{nqe}, the classical \gls{md} simulations predict a near-zero (NEP-D3) or even positive (NEP) $\alpha_{V}$, resulting in a qualitatively incorrect prediction.

\subsection{Water}

\begin{figure}
    \centering
    \includegraphics[width=\columnwidth]{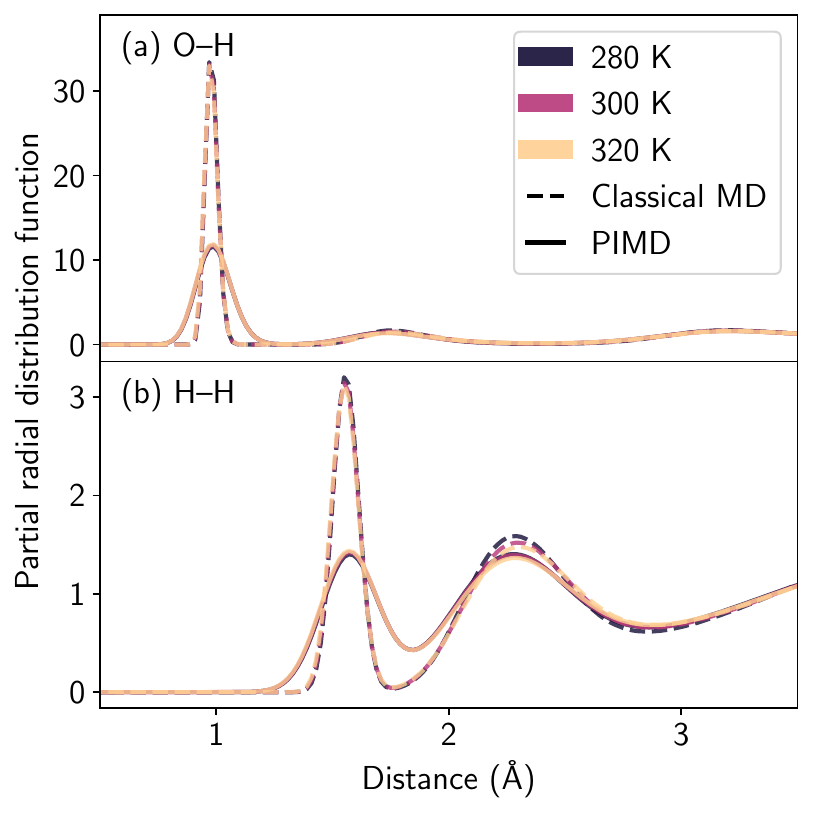}
    \caption{
        Partial oxygen--hydrogen (a) and hydrogen-hydrogen (b) radial distribution functions from classical \gls{md} (dashed line) and 64-bead \gls{pimd} (solid line) simulations, for three different temperatures \qty{280}{\kelvin} (dark color), \qty{300}{\kelvin} (intermediate color), and \qty{320}{\kelvin} (light color).
    }
    \label{fig:water-rdfs}
\end{figure}

\begin{figure}
    \centering
    \includegraphics[width=\columnwidth]{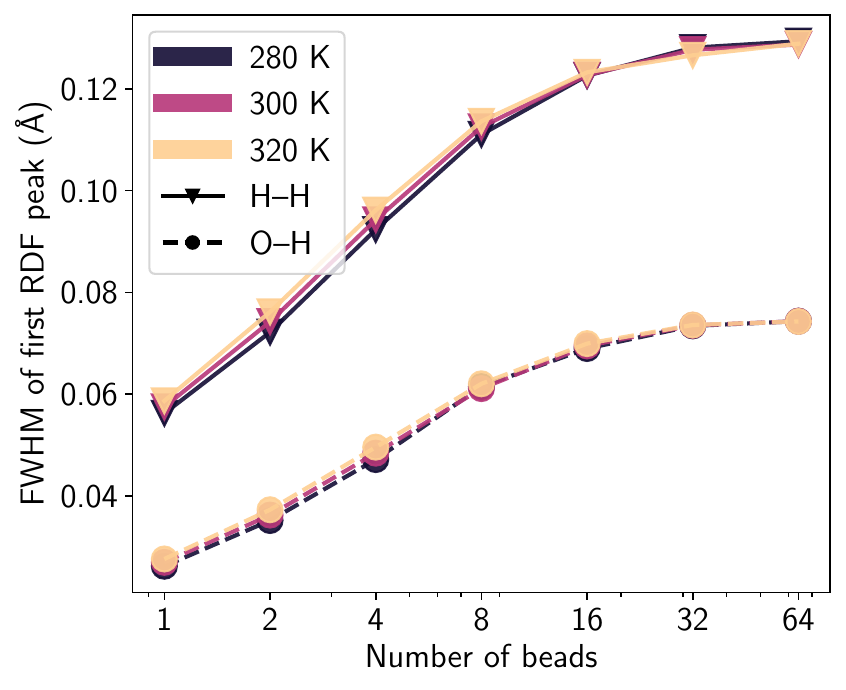}
    \caption{
        \Gls{fwhm} of the first nearest-neighbor peak in the hydrogen--hydrogen (solid triangles) and oxygen--hydrogen (dashed circles) \glspl{rdf} as a function of the number of \gls{pimd} beads for three different temperatures, revealing similar convergence behavior of the \gls{fwhm} with number of beads, irrespective of temperature.
    }
    \label{fig:water-fwhm}
\end{figure}

\begin{figure}
    \centering
    \includegraphics[width=\columnwidth]{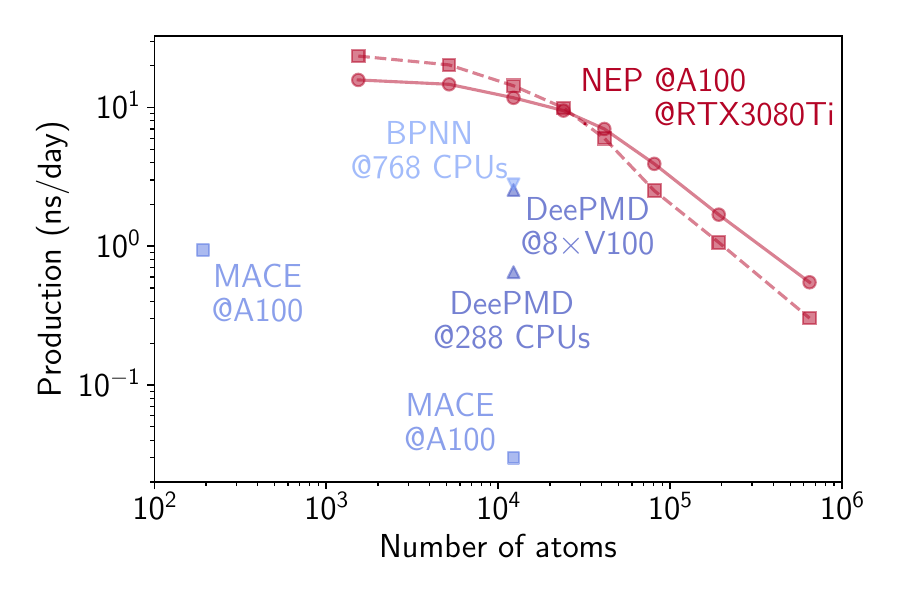}
    \caption{
        Computational efficiency of different \glspl{mlp} for water measured in terms of the \unit{\nano\second} that can be simulated per day of wall-clock time.
        The timing is given assuming a classical simulation (equivalent to one bead) in line with Table~III in Ref.~\cite{litman2024ipi}, which is also the source of the timings shown for BPNN \cite{RavAdvSch24}, DeePMD \cite{HanZhaCar18}, and MACE \cite{BatBenChi24} models (all using the same time step of \qty{0.5}{\femto\second}).
        For \gls{nep} timing is shown for both A100 (solid) and RTX3080Ti (dashed) cards.
    }
    \label{fig:water-timing}
\end{figure}

While we in the previous sections studied the effect of \glspl{nqe} on the \emph{macroscopic} structure, we now examine its impact on the \emph{microscopic} structure.
Here, this is exemplified by the partial \glspl{rdf} of water (\autoref{fig:water-rdfs}), which are significantly affected by \glspl{nqe} because of the high hydrogen content of water.

The \glspl{nqe} are particularly apparent in the first nearest-neighbor peak in both the oxygen--hydrogen (\autoref{fig:water-rdfs}a) and hydrogen--hydrogen (\autoref{fig:water-rdfs}b) \glspl{rdf}, which are significantly broadened compared to the classical limit. To converge these calculations one requires at least \num{32} beads in the temperature range of \num{280} to \qty{320}{\kelvin} considered here, as illustrated by the \gls{fwhm} of the first nearest-neighbor coordination peaks (\autoref{fig:water-fwhm}).

We once again observe that the effect of \glspl{nqe} is more substantial at lower temperatures.
This can be seen, e.g., by noting that the peak height difference between the second peak of the hydrogen--hydrogen \gls{rdf} from classical \gls{md} and \gls{pimd} simulations is larger at \qty{280}{\kelvin} than at \qty{320}{\kelvin} (\autoref{fig:water-rdfs}b; see also Figure~S6k--o, and Figure S7). 
The same effect is also observed in the first peak of the oxygen--oxygen \gls{rdf} (Figure~S6a--e; also see Figure S7).
Overall the impact of \glspl{nqe} on the O--O \gls{rdf} is, however, much smaller than for the H--H and O--H \glspl{rdf} due to the larger mass of O.
All of these results agree well with experiment and previous \gls{pimd} simulations \cite{ko2019mp, cheng2019ab, CheAmbMic16, xu2020prb, MorCar08, CerFanKus16}.

The case of water also allows us to compare the computational efficiency of \gls{nep} with other \glspl{mlp} (\autoref{fig:water-timing}).
For the latter we resort to data from Ref.~\citenum{litman2024ipi} for the BPNN \cite{RavAdvSch24}, DeePMD \cite{HanZhaCar18}, and MACE \cite{BatBenChi24} models.
We consider the timing for a classical simulation (equivalent to one bead) for consistency with Ref.~\citenum{litman2024ipi}.

The results demonstrate that for the system sizes considered here the \gls{nep} model is at least about one order of magnitude faster than the next efficient models (BPNN and DeePMD).
This applies even though the latter were run on many hundred CPU cores (BPNN) and 8 GPUs (DeePMD), respectively, while the \gls{nep} data were obtained using a single A100 GPU.
We also note that the performance of the latter diminishes only slightly when running on consumer GPUs as illustrated here by a RTX3080Ti card.

It is also noteworthy that the overhead associated with the driver-force evaluator approach is almost negligible for BPNN, DeePMD, and MACE thanks to a very efficient implementation of the interface in \textsc{i-PI} \cite{litman2024ipi}.
For those \glspl{mlp} the cost of a force evaluation in their present implementations is at least on the order of \qty{10}{\milli\second}, which is larger than the typical overhead per step in \textsc{i-PI} \cite{litman2024ipi}.
In contrast, for \gls{nep} models, a typical force evaluation is approximately on the order of \qty{1}{\milli\second} (\autoref{fig:water-timing}; see also Figure S8 for a comparison of other \gls{nep} models used in this work).
The cost per step would therefore be significantly affected by the communication overhead.
The direct combination of \gls{pimd} with \gls{nep} in \textsc{gpumd} avoids this extra cost, providing a much more efficient approach.

\subsection{Aluminum}

\begin{figure}
\centering
\includegraphics[width=\columnwidth]{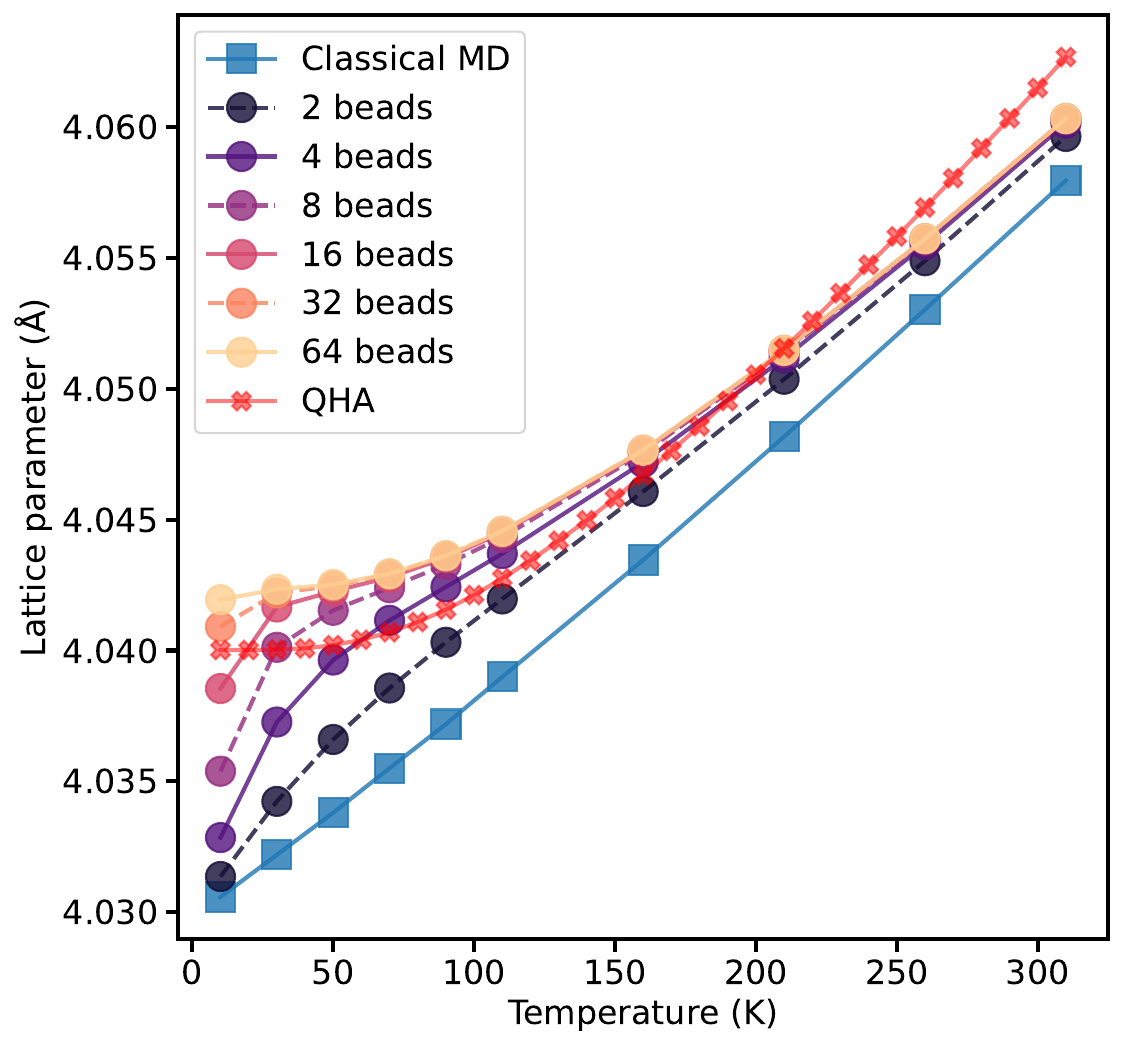}
\caption{
    Temperature dependence of the lattice parameter of aluminum obtained from \gls{qha}, classical \gls{md}, and \gls{pimd} simulations with different numbers of beads.
}
\label{fig:Al}
\end{figure}

\begin{figure*}
\centering
\includegraphics[width=\linewidth]{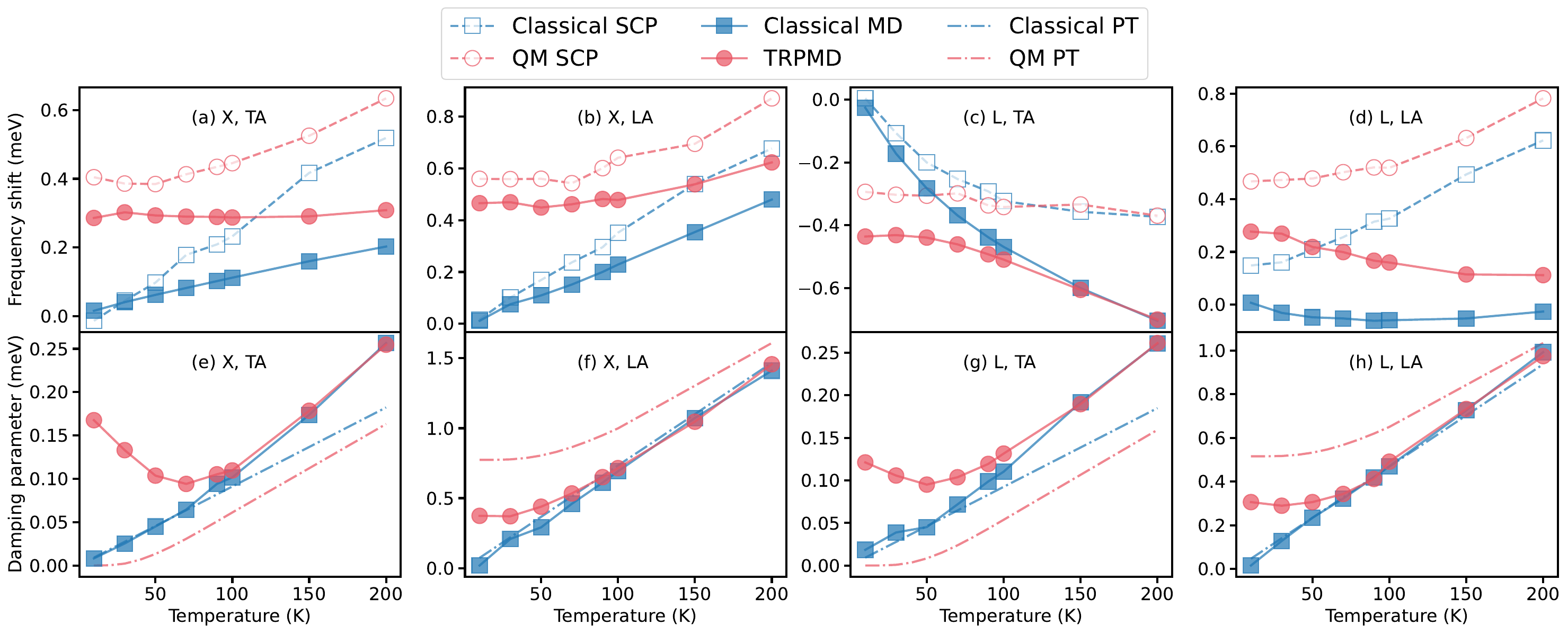}
\caption{
    Phonon frequency shift (a--d) and phonon damping parameters (e--h) of the transversal acoustic (TA) and longitudinal acoustic (LA) $X$ and $L$-modes in elemental aluminum as a function of temperature.
    For comparison, the frequency shift and damping parameters have been calculated using self-consistent phonons (SCP) and perturbation theory (PT), respectively.
    These calculations have been performed using both classical and quantum mechanical (QM) approaches.
}
\label{fig:phonon}
\end{figure*}

\autoref{fig:Al} shows the convergence test of the lattice parameter of aluminum at various temperatures, obtained from \gls{pimd} simulations. Similar to the convergence tests for LiH, the \glspl{nqe} dominate at lower temperatures. The results indicate that 64 beads are sufficient to achieve convergence even for the lowest temperatures considered here. Additionally, the lattice parameter has been determined using the \gls{qha} method as implemented in \textsc{phonopy} \cite{phonopy-phono3py-JPCM, phonopy-phono3py-JPSJ}. At low temperatures, the \gls{pimd} and \gls{qha} lattice parameters exhibit similar temperature dependence, as expected. At higher temperatures, the \gls{qha} results diverge from those obtained through \gls{pimd} simulations due to the incomplete treatment of anharmonicity in the former. The simulations of Al described from here on were performed using 16 beads, except for temperatures under \qty{50}{\kelvin}, where 64 beads were employed to ensure convergence.

Next, we analyze the phonon properties of aluminum using \gls{trpmd} to demonstrate that the implemented method can yield quantum dynamical properties. 
For this analysis, the lattice parameter is kept fixed at \qty{4.05}{\angstrom} to make sure the \glspl{nqe} we observe actually originate from the dynamics rather than thermal expansion. 
\autoref{fig:phonon}a--d shows the phonon frequency shift as a function of temperature for the transversal and longitudinal $X$ and $L$-modes, where $X$ and $L$ are the reciprocal points $(1/2, \, 0, \, 1/2)$ and $(1/2,\,  1/2,\,  1/2)$, respectively. 
The frequency shift is determined by subtracting the zero-temperature frequency, obtained with \textsc{phonopy} from the frequency observed at each respective temperature. 
Additionally, for comparison, the phonon frequency shift is determined through self-consistent phonons, both classically and quantum mechanically. 
The \textsc{hiphive} package \cite{eriksson_hiphive_2019} is used for this analysis. 
Examining the frequency shift obtained from classical \gls{md} and classical self-consistent phonons, we observe that it approaches zero as the temperature approaches absolute zero for all four modes. 
This behavior is expected since the zero-temperature frequency is calculated with small atomic displacements. 
In the classical scenario, there is no zero-point energy, and atomic displacements diminish and approach zero as the temperature decreases to absolute zero. 
However, in the quantum mechanical case, the frequency shift is not expected to reach zero due to quantum fluctuations. 
Observing the frequency shifts obtained from \gls{trpmd} and quantum mechanical self-consistent phonons, we see that this holds true, as these shifts are finite even at low temperatures.
This demonstrates that \gls{trpmd} simulations account for \glspl{nqe}, while classical \gls{md} simulations do not.
The discrepancy between the self-consistent phonons results and the \gls{trpmd} simulations can be attributed to the incomplete inclusion of anharmonicity in the former.
We note that a similar overestimation has also been observed for other materials \cite{FraRosEri23}.

Along with the phonon frequency shift, we also highlight the critical role of \glspl{nqe} in calculating phonon lifetimes.
Our study demonstrates how \gls{trpmd} enables the incorporation of these effects into simulations.
\autoref{fig:phonon}e--h shows the temperature dependence of the damping parameter $\Gamma$, which is inversely proportional to the phonon lifetime $\tau_{\text{ph}}$
\begin{equation*}
    \Gamma = \frac{2}{\tau_{\text{ph}}}.
\end{equation*}
For comparison the damping has also been obtained with perturbation calculations conducted with \textsc{kaldo} \cite{kaldo}. 

It is evident that the damping obtained through classical \gls{md} approaches zero, meaning phonon lifetimes tend toward infinity, as the temperature goes to zero.
This characteristic makes damping a more convenient quantity for analysis compared to the lifetimes.
By contrast, due to the inclusion of \glspl{nqe}, the \gls{trpmd} damping remains finite as the temperature approaches zero.
When comparing the damping obtained through \gls{trpmd} simulations to that calculated using quantum mechanical perturbation theory, a similar temperature dependence is observed, reinforcing the conclusion that \gls{trpmd} effectively captures the quantum dynamical behavior of the damping.
For the two transversal modes, the damping obtained through \gls{trpmd} increases when the temperature falls below \qty{70}{\kelvin} and \qty{50}{\kelvin}, respectively.
This increase is also observed in the longitudinal $L$-mode, though it is less pronounced.
This behavior is likely due to the coupling between the dynamics of the beads and the ring-polymer centroid dynamics \cite{rossi_how_2014}.
While \gls{trpmd} dampens this coupling, it does not entirely eliminate it.
By contrast, the damping of the transversal modes predicted by quantum mechanical perturbation theory tends to zero as the temperature nears absolute zero.
This is likely due to the absence of higher-order anharmonic terms, as \textsc{kaldo} only incorporates force constants up to the third order.

\section{Conclusions}

In summary, we have integrated the \gls{pimd} method with machine-learned \gls{nep} models into the \textsc{gpumd} package, enabling efficient and accurate \gls{md} simulations that account for \glspl{nqe}.
The effectiveness of the \gls{nep}-\gls{pimd} approach is demonstrated by studying thermal properties of four different types of materials: ionic LiH, three porous \glspl{mof}, liquid water, and elemental aluminum.

Our results show that including the \glspl{nqe} is crucial for accurately modeling the thermal expansion of LiH, MOFs and aluminum, achievable with \gls{nep}-\gls{pimd} simulations. 
Specifically, the isotope effect on the lattice parameter of LiH predicted by \gls{nep}-\gls{pimd} simulations exhibits a dependence on the reduced mass, in good agreement with the experimental observations, whereas classical \gls{md} simulations predict a negligible isotope dependence. 
For the porous \glspl{mof}, our results indicate that incorporating dispersive interactions into the \gls{nep} models, along with \glspl{nqe} via \gls{pimd}, brings the simulated values closer to experimental data. 
Furthermore, accounting for \glspl{nqe} in liquid water significantly affects the microscopic structure, which is crucial for obtaining simulated structural properties that align more closely with experimental results.
In the case of elemental aluminum, incorporating \glspl{nqe} is essential for accurately capturing thermal expansion and phonon properties. 
While classical \gls{md} simulations show both the phonon frequency shift and damping parameter approaching zero as the temperature decreases, \gls{trpmd} simulations, which include \glspl{nqe}, predict finite values that align well with quantum mechanical perturbation theory.

The integration of the \gls{nep}-\gls{pimd} into the \textsc{gpumd} package opens new pathways for investigating the properties of a wide range of materials affected by \glspl{nqe}, enabling large-scale \gls{pimd} simulations with high fidelity and efficiency.
Notably, during the preparation of our manuscript, this implementation of the \gls{nep}-\gls{pimd} approach into the open-source \textsc{gpumd} package has already been utilized to study some static and dynamic properties in liquid water \cite{wangzhou2024jctc}, crystalline silicon \cite{folkner2024elastic}, and \ce{Cs3Bi2I6Cl3} \cite{zeng2024lattice}.

\begin{acknowledgments}
This work has been supported by the National Natural Science Foundation of China  (No. 11974059), the Swedish Research Council (Nos. 2020-04935 and 2021-05072), and the Chalmers Initiative for Advancement of Neutron and Synchrotron Techniques.
P.~Y. was supported by the Israel Academy of Sciences and Humanities \& Council for Higher Education Excellence Fellowship Program for International Postdoctoral Researchers.
W.~Z. and B.~S. acknowledge support from National Natural Science Foundation of China (No. 52076002), the New Cornerstone Science Foundation through the XPLORER PRIZE, and the high-performance computing platform of Peking University.
K.~X, T.~L. and J.~X. acknowledge support from the National Key R\&D Project from Ministry of Science and Technology of China (No. 2022YFA1203100), and RGC GRF (No. 14220022).
Some of the computations were enabled by resources provided by the National Academic Infrastructure for Supercomputing in Sweden (NAISS) at PDC, C3SE, and NSC, partially funded by the Swedish Research Council through grant agreement no. 2022-06725.
\end{acknowledgments}

\vspace{0.5cm} 

\noindent{\textbf{Conflict of Interest}}

The authors have no conflicts to disclose.

\noindent{\textbf{Data availability}}

The source code and documentation for \textsc{gpumd} are available
at \url{https://github.com/brucefan1983/GPUMD} and \url{https://gpumd.org}, respectively.
Representative input and output files for thermal expansion calculations of \glspl{mof} and LiH are freely available at \url{https://github.com/hityingph/supporting-info}.
The training datasets and the trained \gls{nep} models for \glspl{mof} and LiH are freely available at \url{https://gitlab.com/brucefan1983/nep-data}, the \gls{nep} model and data for water are available at \url{https://doi.org/10.5281/zenodo.10257363}, and the \gls{nep} model for aluminum along with the \gls{dft} reference data used for its construction are available at \url{https://doi.org/10.5281/zenodo.13712924}.

\bibliography{refs}

\begin{thebibliography}{95}%
\makeatletter
\providecommand \@ifxundefined [1]{%
 \@ifx{#1\undefined}
}%
\providecommand \@ifnum [1]{%
 \ifnum #1\expandafter \@firstoftwo
 \else \expandafter \@secondoftwo
 \fi
}%
\providecommand \@ifx [1]{%
 \ifx #1\expandafter \@firstoftwo
 \else \expandafter \@secondoftwo
 \fi
}%
\providecommand \natexlab [1]{#1}%
\providecommand \enquote  [1]{``#1''}%
\providecommand \bibnamefont  [1]{#1}%
\providecommand \bibfnamefont [1]{#1}%
\providecommand \citenamefont [1]{#1}%
\providecommand \href@noop [0]{\@secondoftwo}%
\providecommand \href [0]{\begingroup \@sanitize@url \@href}%
\providecommand \@href[1]{\@@startlink{#1}\@@href}%
\providecommand \@@href[1]{\endgroup#1\@@endlink}%
\providecommand \@sanitize@url [0]{\catcode `\\12\catcode `\$12\catcode `\&12\catcode `\#12\catcode `\^12\catcode `\_12\catcode `\%12\relax}%
\providecommand \@@startlink[1]{}%
\providecommand \@@endlink[0]{}%
\providecommand \url  [0]{\begingroup\@sanitize@url \@url }%
\providecommand \@url [1]{\endgroup\@href {#1}{\urlprefix }}%
\providecommand \urlprefix  [0]{URL }%
\providecommand \Eprint [0]{\href }%
\providecommand \doibase [0]{http://dx.doi.org/}%
\providecommand \selectlanguage [0]{\@gobble}%
\providecommand \bibinfo  [0]{\@secondoftwo}%
\providecommand \bibfield  [0]{\@secondoftwo}%
\providecommand \translation [1]{[#1]}%
\providecommand \BibitemOpen [0]{}%
\providecommand \bibitemStop [0]{}%
\providecommand \bibitemNoStop [0]{.\EOS\space}%
\providecommand \EOS [0]{\spacefactor3000\relax}%
\providecommand \BibitemShut  [1]{\csname bibitem#1\endcsname}%
\let\auto@bib@innerbib\@empty
\bibitem [{\citenamefont {Rahman}(1964)}]{rahman1964correlations}%
  \BibitemOpen
  \bibfield  {author} {\bibinfo {author} {\bibfnamefont {Aneesur}\ \bibnamefont {Rahman}},\ }\bibfield  {title} {\enquote {\bibinfo {title} {Correlations in the motion of atoms in liquid argon},}\ }\href {\doibase 10.1103/PhysRev.136.A405} {\bibfield  {journal} {\bibinfo  {journal} {Physical Review}\ }\textbf {\bibinfo {volume} {136}},\ \bibinfo {pages} {A405} (\bibinfo {year} {1964})}\BibitemShut {NoStop}%
\bibitem [{\citenamefont {Unke}\ \emph {et~al.}(2021)\citenamefont {Unke}, \citenamefont {Chmiela}, \citenamefont {Sauceda}, \citenamefont {Gastegger}, \citenamefont {Poltavsky}, \citenamefont {Schütt}, \citenamefont {Tkatchenko},\ and\ \citenamefont {Müller}}]{Unke2021cr}%
  \BibitemOpen
  \bibfield  {author} {\bibinfo {author} {\bibfnamefont {Oliver~T.}\ \bibnamefont {Unke}}, \bibinfo {author} {\bibfnamefont {Stefan}\ \bibnamefont {Chmiela}}, \bibinfo {author} {\bibfnamefont {Huziel~E.}\ \bibnamefont {Sauceda}}, \bibinfo {author} {\bibfnamefont {Michael}\ \bibnamefont {Gastegger}}, \bibinfo {author} {\bibfnamefont {Igor}\ \bibnamefont {Poltavsky}}, \bibinfo {author} {\bibfnamefont {Kristof~T.}\ \bibnamefont {Schütt}}, \bibinfo {author} {\bibfnamefont {Alexandre}\ \bibnamefont {Tkatchenko}}, \ and\ \bibinfo {author} {\bibfnamefont {Klaus-Robert}\ \bibnamefont {Müller}},\ }\bibfield  {title} {\enquote {\bibinfo {title} {{Machine learning force fields}},}\ }\href {\doibase 10.1021/acs.chemrev.0c01111} {\bibfield  {journal} {\bibinfo  {journal} {Chemical Reviews}\ }\textbf {\bibinfo {volume} {121}},\ \bibinfo {pages} {10142--10186} (\bibinfo {year} {2021})}\BibitemShut {NoStop}%
\bibitem [{\citenamefont {Feynman}\ and\ \citenamefont {Hibbs}(1965)}]{feynman1965quantum}%
  \BibitemOpen
  \bibfield  {author} {\bibinfo {author} {\bibfnamefont {Richard~P}\ \bibnamefont {Feynman}}\ and\ \bibinfo {author} {\bibfnamefont {Albert~R}\ \bibnamefont {Hibbs}},\ }\href@noop {} {\emph {\bibinfo {title} {{Quantum Mechanics and Path Integrals}}}}\ (\bibinfo  {publisher} {McGraw-Hill},\ \bibinfo {address} {New York},\ \bibinfo {year} {1965})\BibitemShut {NoStop}%
\bibitem [{\citenamefont {Parrinello}\ and\ \citenamefont {Rahman}(1984)}]{Parrinello1984jcp}%
  \BibitemOpen
  \bibfield  {author} {\bibinfo {author} {\bibfnamefont {M.}~\bibnamefont {Parrinello}}\ and\ \bibinfo {author} {\bibfnamefont {A.}~\bibnamefont {Rahman}},\ }\bibfield  {title} {\enquote {\bibinfo {title} {{Study of an F center in molten KCl}},}\ }\href {\doibase 10.1063/1.446740} {\bibfield  {journal} {\bibinfo  {journal} {The Journal of Chemical Physics}\ }\textbf {\bibinfo {volume} {80}},\ \bibinfo {pages} {860--867} (\bibinfo {year} {1984})}\BibitemShut {NoStop}%
\bibitem [{\citenamefont {Markland}\ and\ \citenamefont {Ceriotti}(2018)}]{markland2018nuclear}%
  \BibitemOpen
  \bibfield  {author} {\bibinfo {author} {\bibfnamefont {Thomas~E}\ \bibnamefont {Markland}}\ and\ \bibinfo {author} {\bibfnamefont {Michele}\ \bibnamefont {Ceriotti}},\ }\bibfield  {title} {\enquote {\bibinfo {title} {Nuclear quantum effects enter the mainstream},}\ }\href {\doibase 10.1038/s41570-017-0109} {\bibfield  {journal} {\bibinfo  {journal} {Nature Reviews Chemistry}\ }\textbf {\bibinfo {volume} {2}},\ \bibinfo {pages} {0109} (\bibinfo {year} {2018})}\BibitemShut {NoStop}%
\bibitem [{\citenamefont {Ko}\ \emph {et~al.}(2019)\citenamefont {Ko}, \citenamefont {Zhang}, \citenamefont {Santra}, \citenamefont {Wang}, \citenamefont {E}, \citenamefont {Jr},\ and\ \citenamefont {Car}}]{ko2019mp}%
  \BibitemOpen
  \bibfield  {author} {\bibinfo {author} {\bibfnamefont {Hsin-Yu}\ \bibnamefont {Ko}}, \bibinfo {author} {\bibfnamefont {Linfeng}\ \bibnamefont {Zhang}}, \bibinfo {author} {\bibfnamefont {Biswajit}\ \bibnamefont {Santra}}, \bibinfo {author} {\bibfnamefont {Han}\ \bibnamefont {Wang}}, \bibinfo {author} {\bibfnamefont {Weinan}\ \bibnamefont {E}}, \bibinfo {author} {\bibfnamefont {Robert A.~DiStasio}\ \bibnamefont {Jr}}, \ and\ \bibinfo {author} {\bibfnamefont {Roberto}\ \bibnamefont {Car}},\ }\bibfield  {title} {\enquote {\bibinfo {title} {Isotope effects in liquid water via deep potential molecular dynamics},}\ }\href {\doibase 10.1080/00268976.2019.1652366} {\bibfield  {journal} {\bibinfo  {journal} {Molecular Physics}\ }\textbf {\bibinfo {volume} {117}},\ \bibinfo {pages} {3269--3281} (\bibinfo {year} {2019})}\BibitemShut {NoStop}%
\bibitem [{\citenamefont {Cheng}\ \emph {et~al.}(2019)\citenamefont {Cheng}, \citenamefont {Engel}, \citenamefont {Behler}, \citenamefont {Dellago},\ and\ \citenamefont {Ceriotti}}]{cheng2019ab}%
  \BibitemOpen
  \bibfield  {author} {\bibinfo {author} {\bibfnamefont {Bingqing}\ \bibnamefont {Cheng}}, \bibinfo {author} {\bibfnamefont {Edgar~A}\ \bibnamefont {Engel}}, \bibinfo {author} {\bibfnamefont {J{\"o}rg}\ \bibnamefont {Behler}}, \bibinfo {author} {\bibfnamefont {Christoph}\ \bibnamefont {Dellago}}, \ and\ \bibinfo {author} {\bibfnamefont {Michele}\ \bibnamefont {Ceriotti}},\ }\bibfield  {title} {\enquote {\bibinfo {title} {Ab initio thermodynamics of liquid and solid water},}\ }\href {\doibase 10.1073/pnas.1815117116} {\bibfield  {journal} {\bibinfo  {journal} {Proceedings of the National Academy of Sciences}\ }\textbf {\bibinfo {volume} {116}},\ \bibinfo {pages} {1110--1115} (\bibinfo {year} {2019})}\BibitemShut {NoStop}%
\bibitem [{\citenamefont {Xu}\ \emph {et~al.}(2020)\citenamefont {Xu}, \citenamefont {Zhang}, \citenamefont {Zhang}, \citenamefont {Chen}, \citenamefont {Santra},\ and\ \citenamefont {Wu}}]{xu2020prb}%
  \BibitemOpen
  \bibfield  {author} {\bibinfo {author} {\bibfnamefont {Jianhang}\ \bibnamefont {Xu}}, \bibinfo {author} {\bibfnamefont {Chunyi}\ \bibnamefont {Zhang}}, \bibinfo {author} {\bibfnamefont {Linfeng}\ \bibnamefont {Zhang}}, \bibinfo {author} {\bibfnamefont {Mohan}\ \bibnamefont {Chen}}, \bibinfo {author} {\bibfnamefont {Biswajit}\ \bibnamefont {Santra}}, \ and\ \bibinfo {author} {\bibfnamefont {Xifan}\ \bibnamefont {Wu}},\ }\bibfield  {title} {\enquote {\bibinfo {title} {{Isotope effects in molecular structures and electronic properties of liquid water via deep potential molecular dynamics based on the SCAN functional}},}\ }\href {\doibase 10.1103/PhysRevB.102.214113} {\bibfield  {journal} {\bibinfo  {journal} {Physical Review B}\ }\textbf {\bibinfo {volume} {102}},\ \bibinfo {pages} {214113} (\bibinfo {year} {2020})}\BibitemShut {NoStop}%
\bibitem [{\citenamefont {Reinhardt}\ and\ \citenamefont {Cheng}(2021)}]{reinhardt2021quantum}%
  \BibitemOpen
  \bibfield  {author} {\bibinfo {author} {\bibfnamefont {Aleks}\ \bibnamefont {Reinhardt}}\ and\ \bibinfo {author} {\bibfnamefont {Bingqing}\ \bibnamefont {Cheng}},\ }\bibfield  {title} {\enquote {\bibinfo {title} {Quantum-mechanical exploration of the phase diagram of water},}\ }\href {\doibase 10.1038/s41467-020-20821-w} {\bibfield  {journal} {\bibinfo  {journal} {Nature communications}\ }\textbf {\bibinfo {volume} {12}},\ \bibinfo {pages} {588} (\bibinfo {year} {2021})}\BibitemShut {NoStop}%
\bibitem [{\citenamefont {Kapil}\ \emph {et~al.}(2022)\citenamefont {Kapil}, \citenamefont {Schran}, \citenamefont {Zen}, \citenamefont {Chen}, \citenamefont {Pickard},\ and\ \citenamefont {Michaelides}}]{kapil2022first}%
  \BibitemOpen
  \bibfield  {author} {\bibinfo {author} {\bibfnamefont {Venkat}\ \bibnamefont {Kapil}}, \bibinfo {author} {\bibfnamefont {Christoph}\ \bibnamefont {Schran}}, \bibinfo {author} {\bibfnamefont {Andrea}\ \bibnamefont {Zen}}, \bibinfo {author} {\bibfnamefont {Ji}~\bibnamefont {Chen}}, \bibinfo {author} {\bibfnamefont {Chris~J}\ \bibnamefont {Pickard}}, \ and\ \bibinfo {author} {\bibfnamefont {Angelos}\ \bibnamefont {Michaelides}},\ }\bibfield  {title} {\enquote {\bibinfo {title} {The first-principles phase diagram of monolayer nanoconfined water},}\ }\href {\doibase 10.1038/s41586-022-05036-x} {\bibfield  {journal} {\bibinfo  {journal} {Nature}\ }\textbf {\bibinfo {volume} {609}},\ \bibinfo {pages} {512--516} (\bibinfo {year} {2022})}\BibitemShut {NoStop}%
\bibitem [{\citenamefont {adn F.~Paesani}(2023)}]{bore2023realistic}%
  \BibitemOpen
  \bibfield  {author} {\bibinfo {author} {\bibfnamefont {S.~L.~Bore}\ \bibnamefont {adn F.~Paesani}},\ }\bibfield  {title} {\enquote {\bibinfo {title} {Realistic phase diagram of water from “first principles” data-driven quantum simulations},}\ }\href {\doibase 10.1038/s41467-023-38855-1} {\bibfield  {journal} {\bibinfo  {journal} {Nature communications}\ }\textbf {\bibinfo {volume} {14}},\ \bibinfo {pages} {3349} (\bibinfo {year} {2023})}\BibitemShut {NoStop}%
\bibitem [{\citenamefont {Bocus}\ \emph {et~al.}(2023)\citenamefont {Bocus}, \citenamefont {Goeminne}, \citenamefont {Lamaire}, \citenamefont {Cools-Ceuppens}, \citenamefont {Verstraelen},\ and\ \citenamefont {Van~Speybroeck}}]{Bocus2023NC}%
  \BibitemOpen
  \bibfield  {author} {\bibinfo {author} {\bibfnamefont {Massimo}\ \bibnamefont {Bocus}}, \bibinfo {author} {\bibfnamefont {Ruben}\ \bibnamefont {Goeminne}}, \bibinfo {author} {\bibfnamefont {Aran}\ \bibnamefont {Lamaire}}, \bibinfo {author} {\bibfnamefont {Maarten}\ \bibnamefont {Cools-Ceuppens}}, \bibinfo {author} {\bibfnamefont {Toon}\ \bibnamefont {Verstraelen}}, \ and\ \bibinfo {author} {\bibfnamefont {Veronique}\ \bibnamefont {Van~Speybroeck}},\ }\bibfield  {title} {\enquote {\bibinfo {title} {Nuclear quantum effects on zeolite proton hopping kinetics explored with machine learning potentials and path integral molecular dynamics},}\ }\href {\doibase 10.1038/s41467-023-36666-y} {\bibfield  {journal} {\bibinfo  {journal} {Nature Communications}\ }\textbf {\bibinfo {volume} {14}},\ \bibinfo {pages} {1008} (\bibinfo {year} {2023})}\BibitemShut {NoStop}%
\bibitem [{\citenamefont {Chen}\ \emph {et~al.}(2024{\natexlab{a}})\citenamefont {Chen}, \citenamefont {Berrens}, \citenamefont {Chan}, \citenamefont {Fan},\ and\ \citenamefont {Donadio}}]{chen2024jced}%
  \BibitemOpen
  \bibfield  {author} {\bibinfo {author} {\bibfnamefont {Zekun}\ \bibnamefont {Chen}}, \bibinfo {author} {\bibfnamefont {Margaret~L.}\ \bibnamefont {Berrens}}, \bibinfo {author} {\bibfnamefont {Kam-Tung}\ \bibnamefont {Chan}}, \bibinfo {author} {\bibfnamefont {Zheyong}\ \bibnamefont {Fan}}, \ and\ \bibinfo {author} {\bibfnamefont {Davide}\ \bibnamefont {Donadio}},\ }\bibfield  {title} {\enquote {\bibinfo {title} {{Thermodynamics of water and ice from a fast and scalable first-principles neuroevolution potential}},}\ }\href {\doibase 10.1021/acs.jced.3c00561} {\bibfield  {journal} {\bibinfo  {journal} {Journal of Chemical \& Engineering Data}\ }\textbf {\bibinfo {volume} {69}},\ \bibinfo {pages} {128--140} (\bibinfo {year} {2024}{\natexlab{a}})}\BibitemShut {NoStop}%
\bibitem [{\citenamefont {Berrens}\ \emph {et~al.}(2024)\citenamefont {Berrens}, \citenamefont {Kundu}, \citenamefont {Calegari~Andrade}, \citenamefont {Pham}, \citenamefont {Galli},\ and\ \citenamefont {Donadio}}]{berrens2024nuclear}%
  \BibitemOpen
  \bibfield  {author} {\bibinfo {author} {\bibfnamefont {Margaret~L.}\ \bibnamefont {Berrens}}, \bibinfo {author} {\bibfnamefont {Arpan}\ \bibnamefont {Kundu}}, \bibinfo {author} {\bibfnamefont {Marcos~F.}\ \bibnamefont {Calegari~Andrade}}, \bibinfo {author} {\bibfnamefont {Tuan~Anh}\ \bibnamefont {Pham}}, \bibinfo {author} {\bibfnamefont {Giulia}\ \bibnamefont {Galli}}, \ and\ \bibinfo {author} {\bibfnamefont {Davide}\ \bibnamefont {Donadio}},\ }\bibfield  {title} {\enquote {\bibinfo {title} {Nuclear quantum effects on the electronic structure of water and ice},}\ }\href {\doibase 10.1021/acs.jpclett.4c01315} {\bibfield  {journal} {\bibinfo  {journal} {The Journal of Physical Chemistry Letters}\ }\textbf {\bibinfo {volume} {15}},\ \bibinfo {pages} {6818--6825} (\bibinfo {year} {2024})}\BibitemShut {NoStop}%
\bibitem [{\citenamefont {Thompson}\ \emph {et~al.}(2022)\citenamefont {Thompson}, \citenamefont {Aktulga}, \citenamefont {Berger}, \citenamefont {Bolintineanu}, \citenamefont {Brown}, \citenamefont {Crozier}, \citenamefont {In't~Veld}, \citenamefont {Kohlmeyer}, \citenamefont {Moore}, \citenamefont {Nguyen} \emph {et~al.}}]{thompson2022lammps}%
  \BibitemOpen
  \bibfield  {author} {\bibinfo {author} {\bibfnamefont {Aidan~P}\ \bibnamefont {Thompson}}, \bibinfo {author} {\bibfnamefont {H~Metin}\ \bibnamefont {Aktulga}}, \bibinfo {author} {\bibfnamefont {Richard}\ \bibnamefont {Berger}}, \bibinfo {author} {\bibfnamefont {Dan~S}\ \bibnamefont {Bolintineanu}}, \bibinfo {author} {\bibfnamefont {W~Michael}\ \bibnamefont {Brown}}, \bibinfo {author} {\bibfnamefont {Paul~S}\ \bibnamefont {Crozier}}, \bibinfo {author} {\bibfnamefont {Pieter~J}\ \bibnamefont {In't~Veld}}, \bibinfo {author} {\bibfnamefont {Axel}\ \bibnamefont {Kohlmeyer}}, \bibinfo {author} {\bibfnamefont {Stan~G}\ \bibnamefont {Moore}}, \bibinfo {author} {\bibfnamefont {Trung~Dac}\ \bibnamefont {Nguyen}},  \emph {et~al.},\ }\bibfield  {title} {\enquote {\bibinfo {title} {{LAMMPS}-a flexible simulation tool for particle-based materials modeling at the atomic, meso, and continuum scales},}\ }\href {\doibase 10.1016/j.cpc.2021.108171} {\bibfield  {journal} {\bibinfo  {journal} {Computer Physics Communications}\
  }\textbf {\bibinfo {volume} {271}},\ \bibinfo {pages} {108171} (\bibinfo {year} {2022})}\BibitemShut {NoStop}%
\bibitem [{\citenamefont {Ceriotti}\ \emph {et~al.}(2014)\citenamefont {Ceriotti}, \citenamefont {More},\ and\ \citenamefont {Manolopoulos}}]{ceriotti2014pi}%
  \BibitemOpen
  \bibfield  {author} {\bibinfo {author} {\bibfnamefont {Michele}\ \bibnamefont {Ceriotti}}, \bibinfo {author} {\bibfnamefont {Joshua}\ \bibnamefont {More}}, \ and\ \bibinfo {author} {\bibfnamefont {David~E}\ \bibnamefont {Manolopoulos}},\ }\bibfield  {title} {\enquote {\bibinfo {title} {i-{PI}: A python interface for ab initio path integral molecular dynamics simulations},}\ }\href {\doibase 10.1016/j.cpc.2013.10.027} {\bibfield  {journal} {\bibinfo  {journal} {Computer Physics Communications}\ }\textbf {\bibinfo {volume} {185}},\ \bibinfo {pages} {1019--1026} (\bibinfo {year} {2014})}\BibitemShut {NoStop}%
\bibitem [{\citenamefont {Kapil}\ \emph {et~al.}(2019)\citenamefont {Kapil}, \citenamefont {Rossi}, \citenamefont {Marsalek}, \citenamefont {Petraglia}, \citenamefont {Litman}, \citenamefont {Spura}, \citenamefont {Cheng}, \citenamefont {Cuzzocrea}, \citenamefont {Mei{\ss}ner}, \citenamefont {Wilkins} \emph {et~al.}}]{kapil2019pi}%
  \BibitemOpen
  \bibfield  {author} {\bibinfo {author} {\bibfnamefont {Venkat}\ \bibnamefont {Kapil}}, \bibinfo {author} {\bibfnamefont {Mariana}\ \bibnamefont {Rossi}}, \bibinfo {author} {\bibfnamefont {Ondrej}\ \bibnamefont {Marsalek}}, \bibinfo {author} {\bibfnamefont {Riccardo}\ \bibnamefont {Petraglia}}, \bibinfo {author} {\bibfnamefont {Yair}\ \bibnamefont {Litman}}, \bibinfo {author} {\bibfnamefont {Thomas}\ \bibnamefont {Spura}}, \bibinfo {author} {\bibfnamefont {Bingqing}\ \bibnamefont {Cheng}}, \bibinfo {author} {\bibfnamefont {Alice}\ \bibnamefont {Cuzzocrea}}, \bibinfo {author} {\bibfnamefont {Robert~H}\ \bibnamefont {Mei{\ss}ner}}, \bibinfo {author} {\bibfnamefont {David~M}\ \bibnamefont {Wilkins}},  \emph {et~al.},\ }\bibfield  {title} {\enquote {\bibinfo {title} {i-{PI} 2.0: A universal force engine for advanced molecular simulations},}\ }\href {\doibase 10.1016/j.cpc.2018.09.020} {\bibfield  {journal} {\bibinfo  {journal} {Computer Physics Communications}\ }\textbf {\bibinfo {volume} {236}},\ \bibinfo
  {pages} {214--223} (\bibinfo {year} {2019})}\BibitemShut {NoStop}%
\bibitem [{\citenamefont {Litman}\ \emph {et~al.}(2024)\citenamefont {Litman}, \citenamefont {Kapil}, \citenamefont {Feldman}, \citenamefont {Tisi}, \citenamefont {Begušić}, \citenamefont {Fidanyan}, \citenamefont {Fraux}, \citenamefont {Higer}, \citenamefont {Kellner}, \citenamefont {Li}, \citenamefont {Pós}, \citenamefont {Stocco}, \citenamefont {Trenins}, \citenamefont {Hirshberg}, \citenamefont {Rossi},\ and\ \citenamefont {Ceriotti}}]{litman2024ipi}%
  \BibitemOpen
  \bibfield  {author} {\bibinfo {author} {\bibfnamefont {Yair}\ \bibnamefont {Litman}}, \bibinfo {author} {\bibfnamefont {Venkat}\ \bibnamefont {Kapil}}, \bibinfo {author} {\bibfnamefont {Yotam M.~Y.}\ \bibnamefont {Feldman}}, \bibinfo {author} {\bibfnamefont {Davide}\ \bibnamefont {Tisi}}, \bibinfo {author} {\bibfnamefont {Tomislav}\ \bibnamefont {Begušić}}, \bibinfo {author} {\bibfnamefont {Karen}\ \bibnamefont {Fidanyan}}, \bibinfo {author} {\bibfnamefont {Guillaume}\ \bibnamefont {Fraux}}, \bibinfo {author} {\bibfnamefont {Jacob}\ \bibnamefont {Higer}}, \bibinfo {author} {\bibfnamefont {Matthias}\ \bibnamefont {Kellner}}, \bibinfo {author} {\bibfnamefont {Tao~E.}\ \bibnamefont {Li}}, \bibinfo {author} {\bibfnamefont {Eszter~S.}\ \bibnamefont {Pós}}, \bibinfo {author} {\bibfnamefont {Elia}\ \bibnamefont {Stocco}}, \bibinfo {author} {\bibfnamefont {George}\ \bibnamefont {Trenins}}, \bibinfo {author} {\bibfnamefont {Barak}\ \bibnamefont {Hirshberg}}, \bibinfo {author} {\bibfnamefont {Mariana}\
  \bibnamefont {Rossi}}, \ and\ \bibinfo {author} {\bibfnamefont {Michele}\ \bibnamefont {Ceriotti}},\ }\bibfield  {title} {\enquote {\bibinfo {title} {{i-{PI} 3.0: A flexible and efficient framework for advanced atomistic simulations}},}\ }\href {\doibase 10.1063/5.0215869} {\bibfield  {journal} {\bibinfo  {journal} {The Journal of Chemical Physics}\ }\textbf {\bibinfo {volume} {161}},\ \bibinfo {pages} {062504} (\bibinfo {year} {2024})}\BibitemShut {NoStop}%
\bibitem [{\citenamefont {Fan}\ \emph {et~al.}(2017)\citenamefont {Fan}, \citenamefont {Chen}, \citenamefont {Vierimaa},\ and\ \citenamefont {Harju}}]{fan2017cpc}%
  \BibitemOpen
  \bibfield  {author} {\bibinfo {author} {\bibfnamefont {Zheyong}\ \bibnamefont {Fan}}, \bibinfo {author} {\bibfnamefont {Wei}\ \bibnamefont {Chen}}, \bibinfo {author} {\bibfnamefont {Ville}\ \bibnamefont {Vierimaa}}, \ and\ \bibinfo {author} {\bibfnamefont {Ari}\ \bibnamefont {Harju}},\ }\bibfield  {title} {\enquote {\bibinfo {title} {Efficient molecular dynamics simulations with many-body potentials on graphics processing units},}\ }\href {\doibase https://doi.org/10.1016/j.cpc.2017.05.003} {\bibfield  {journal} {\bibinfo  {journal} {Computer Physics Communications}\ }\textbf {\bibinfo {volume} {218}},\ \bibinfo {pages} {10--16} (\bibinfo {year} {2017})}\BibitemShut {NoStop}%
\bibitem [{\citenamefont {Fan}\ \emph {et~al.}(2021)\citenamefont {Fan}, \citenamefont {Zeng}, \citenamefont {Zhang}, \citenamefont {Wang}, \citenamefont {Song}, \citenamefont {Dong}, \citenamefont {Chen},\ and\ \citenamefont {Ala-Nissila}}]{fan2021neuroevolution}%
  \BibitemOpen
  \bibfield  {author} {\bibinfo {author} {\bibfnamefont {Zheyong}\ \bibnamefont {Fan}}, \bibinfo {author} {\bibfnamefont {Zezhu}\ \bibnamefont {Zeng}}, \bibinfo {author} {\bibfnamefont {Cunzhi}\ \bibnamefont {Zhang}}, \bibinfo {author} {\bibfnamefont {Yanzhou}\ \bibnamefont {Wang}}, \bibinfo {author} {\bibfnamefont {Keke}\ \bibnamefont {Song}}, \bibinfo {author} {\bibfnamefont {Haikuan}\ \bibnamefont {Dong}}, \bibinfo {author} {\bibfnamefont {Yue}\ \bibnamefont {Chen}}, \ and\ \bibinfo {author} {\bibfnamefont {Tapio}\ \bibnamefont {Ala-Nissila}},\ }\bibfield  {title} {\enquote {\bibinfo {title} {Neuroevolution machine learning potentials: Combining high accuracy and low cost in atomistic simulations and application to heat transport},}\ }\href {\doibase 10.1103/PhysRevB.104.104309} {\bibfield  {journal} {\bibinfo  {journal} {Physical Review B}\ }\textbf {\bibinfo {volume} {104}},\ \bibinfo {pages} {104309} (\bibinfo {year} {2021})}\BibitemShut {NoStop}%
\bibitem [{\citenamefont {Fan}(2022)}]{fan2022jpcm}%
  \BibitemOpen
  \bibfield  {author} {\bibinfo {author} {\bibfnamefont {Zheyong}\ \bibnamefont {Fan}},\ }\bibfield  {title} {\enquote {\bibinfo {title} {Improving the accuracy of the neuroevolution machine learning potential for multi-component systems},}\ }\href {\doibase 10.1088/1361-648X/ac462b} {\bibfield  {journal} {\bibinfo  {journal} {Journal of Physics: Condensed Matter}\ }\textbf {\bibinfo {volume} {34}},\ \bibinfo {pages} {125902} (\bibinfo {year} {2022})}\BibitemShut {NoStop}%
\bibitem [{\citenamefont {Fan}\ \emph {et~al.}(2022)\citenamefont {Fan}, \citenamefont {Wang}, \citenamefont {Ying}, \citenamefont {Song}, \citenamefont {Wang}, \citenamefont {Wang}, \citenamefont {Zeng}, \citenamefont {Xu}, \citenamefont {Lindgren}, \citenamefont {Rahm} \emph {et~al.}}]{fan2022gpumd}%
  \BibitemOpen
  \bibfield  {author} {\bibinfo {author} {\bibfnamefont {Zheyong}\ \bibnamefont {Fan}}, \bibinfo {author} {\bibfnamefont {Yanzhou}\ \bibnamefont {Wang}}, \bibinfo {author} {\bibfnamefont {Penghua}\ \bibnamefont {Ying}}, \bibinfo {author} {\bibfnamefont {Keke}\ \bibnamefont {Song}}, \bibinfo {author} {\bibfnamefont {Junjie}\ \bibnamefont {Wang}}, \bibinfo {author} {\bibfnamefont {Yong}\ \bibnamefont {Wang}}, \bibinfo {author} {\bibfnamefont {Zezhu}\ \bibnamefont {Zeng}}, \bibinfo {author} {\bibfnamefont {Ke}~\bibnamefont {Xu}}, \bibinfo {author} {\bibfnamefont {Eric}\ \bibnamefont {Lindgren}}, \bibinfo {author} {\bibfnamefont {J~Magnus}\ \bibnamefont {Rahm}},  \emph {et~al.},\ }\bibfield  {title} {\enquote {\bibinfo {title} {{GPUMD: A package for constructing accurate machine-learned potentials and performing highly efficient atomistic simulations}},}\ }\href {\doibase https://doi.org/10.1063/5.0106617} {\bibfield  {journal} {\bibinfo  {journal} {The Journal of Chemical Physics}\ }\textbf {\bibinfo {volume}
  {157}},\ \bibinfo {pages} {114801} (\bibinfo {year} {2022})}\BibitemShut {NoStop}%
\bibitem [{\citenamefont {Song}\ \emph {et~al.}(2023)\citenamefont {Song}, \citenamefont {Zhao}, \citenamefont {Liu}, \citenamefont {Wang}, \citenamefont {Lindgren}, \citenamefont {Wang}, \citenamefont {Chen}, \citenamefont {Xu}, \citenamefont {Liang}, \citenamefont {Ying}, \citenamefont {Xu}, \citenamefont {Zhao}, \citenamefont {Shi}, \citenamefont {Wang}, \citenamefont {Lyu}, \citenamefont {Zeng}, \citenamefont {Liang}, \citenamefont {Dong}, \citenamefont {Sun}, \citenamefont {Chen}, \citenamefont {Zhang}, \citenamefont {Guo}, \citenamefont {Qian}, \citenamefont {Sun}, \citenamefont {Erhart}, \citenamefont {Ala-Nissila}, \citenamefont {Su},\ and\ \citenamefont {Fan}}]{song2023generalpurpose}%
  \BibitemOpen
  \bibfield  {author} {\bibinfo {author} {\bibfnamefont {Keke}\ \bibnamefont {Song}}, \bibinfo {author} {\bibfnamefont {Rui}\ \bibnamefont {Zhao}}, \bibinfo {author} {\bibfnamefont {Jiahui}\ \bibnamefont {Liu}}, \bibinfo {author} {\bibfnamefont {Yanzhou}\ \bibnamefont {Wang}}, \bibinfo {author} {\bibfnamefont {Eric}\ \bibnamefont {Lindgren}}, \bibinfo {author} {\bibfnamefont {Yong}\ \bibnamefont {Wang}}, \bibinfo {author} {\bibfnamefont {Shunda}\ \bibnamefont {Chen}}, \bibinfo {author} {\bibfnamefont {Ke}~\bibnamefont {Xu}}, \bibinfo {author} {\bibfnamefont {Ting}\ \bibnamefont {Liang}}, \bibinfo {author} {\bibfnamefont {Penghua}\ \bibnamefont {Ying}}, \bibinfo {author} {\bibfnamefont {Nan}\ \bibnamefont {Xu}}, \bibinfo {author} {\bibfnamefont {Zhiqiang}\ \bibnamefont {Zhao}}, \bibinfo {author} {\bibfnamefont {Jiuyang}\ \bibnamefont {Shi}}, \bibinfo {author} {\bibfnamefont {Junjie}\ \bibnamefont {Wang}}, \bibinfo {author} {\bibfnamefont {Shuang}\ \bibnamefont {Lyu}}, \bibinfo {author} {\bibfnamefont {Zezhu}\
  \bibnamefont {Zeng}}, \bibinfo {author} {\bibfnamefont {Shirong}\ \bibnamefont {Liang}}, \bibinfo {author} {\bibfnamefont {Haikuan}\ \bibnamefont {Dong}}, \bibinfo {author} {\bibfnamefont {Ligang}\ \bibnamefont {Sun}}, \bibinfo {author} {\bibfnamefont {Yue}\ \bibnamefont {Chen}}, \bibinfo {author} {\bibfnamefont {Zhuhua}\ \bibnamefont {Zhang}}, \bibinfo {author} {\bibfnamefont {Wanlin}\ \bibnamefont {Guo}}, \bibinfo {author} {\bibfnamefont {Ping}\ \bibnamefont {Qian}}, \bibinfo {author} {\bibfnamefont {Jian}\ \bibnamefont {Sun}}, \bibinfo {author} {\bibfnamefont {Paul}\ \bibnamefont {Erhart}}, \bibinfo {author} {\bibfnamefont {Tapio}\ \bibnamefont {Ala-Nissila}}, \bibinfo {author} {\bibfnamefont {Yanjing}\ \bibnamefont {Su}}, \ and\ \bibinfo {author} {\bibfnamefont {Zheyong}\ \bibnamefont {Fan}},\ }\href {https://arxiv.org/abs/2311.04732} {\enquote {\bibinfo {title} {General-purpose machine-learned potential for 16 elemental metals and their alloys},}\ } (\bibinfo {year} {2023}),\ \Eprint
  {http://arxiv.org/abs/2311.04732} {arXiv:2311.04732 [cond-mat.mtrl-sci]} \BibitemShut {NoStop}%
\bibitem [{\citenamefont {Xu}\ \emph {et~al.}(2023)\citenamefont {Xu}, \citenamefont {Hao}, \citenamefont {Liang}, \citenamefont {Ying}, \citenamefont {Xu}, \citenamefont {Wu},\ and\ \citenamefont {Fan}}]{xu2023accurate}%
  \BibitemOpen
  \bibfield  {author} {\bibinfo {author} {\bibfnamefont {Ke}~\bibnamefont {Xu}}, \bibinfo {author} {\bibfnamefont {Yongchao}\ \bibnamefont {Hao}}, \bibinfo {author} {\bibfnamefont {Ting}\ \bibnamefont {Liang}}, \bibinfo {author} {\bibfnamefont {Penghua}\ \bibnamefont {Ying}}, \bibinfo {author} {\bibfnamefont {Jianbin}\ \bibnamefont {Xu}}, \bibinfo {author} {\bibfnamefont {Jianyang}\ \bibnamefont {Wu}}, \ and\ \bibinfo {author} {\bibfnamefont {Zheyong}\ \bibnamefont {Fan}},\ }\bibfield  {title} {\enquote {\bibinfo {title} {Accurate prediction of heat conductivity of water by a neuroevolution potential},}\ }\href {https://pubs.aip.org/aip/jcp/article/158/20/204114/2892524} {\bibfield  {journal} {\bibinfo  {journal} {The Journal of Chemical Physics}\ }\textbf {\bibinfo {volume} {158}} (\bibinfo {year} {2023})}\BibitemShut {NoStop}%
\bibitem [{\citenamefont {Ying}\ \emph {et~al.}(2023{\natexlab{a}})\citenamefont {Ying}, \citenamefont {Liang}, \citenamefont {Xu}, \citenamefont {Zhang}, \citenamefont {Xu}, \citenamefont {Zhong},\ and\ \citenamefont {Fan}}]{ying2023sub}%
  \BibitemOpen
  \bibfield  {author} {\bibinfo {author} {\bibfnamefont {Penghua}\ \bibnamefont {Ying}}, \bibinfo {author} {\bibfnamefont {Ting}\ \bibnamefont {Liang}}, \bibinfo {author} {\bibfnamefont {Ke}~\bibnamefont {Xu}}, \bibinfo {author} {\bibfnamefont {Jin}\ \bibnamefont {Zhang}}, \bibinfo {author} {\bibfnamefont {Jianbin}\ \bibnamefont {Xu}}, \bibinfo {author} {\bibfnamefont {Zheng}\ \bibnamefont {Zhong}}, \ and\ \bibinfo {author} {\bibfnamefont {Zheyong}\ \bibnamefont {Fan}},\ }\bibfield  {title} {\enquote {\bibinfo {title} {Sub-micrometer phonon mean free paths in metal--organic frameworks revealed by machine learning molecular dynamics simulations},}\ }\href {\doibase 10.1021/acsami.3c07770} {\bibfield  {journal} {\bibinfo  {journal} {ACS Applied Materials \& Interfaces}\ }\textbf {\bibinfo {volume} {15}},\ \bibinfo {pages} {36412--36422} (\bibinfo {year} {2023}{\natexlab{a}})}\BibitemShut {NoStop}%
\bibitem [{\citenamefont {Liu}\ \emph {et~al.}(2023)\citenamefont {Liu}, \citenamefont {Byggm\"astar}, \citenamefont {Fan}, \citenamefont {Qian},\ and\ \citenamefont {Su}}]{Liu2023prb}%
  \BibitemOpen
  \bibfield  {author} {\bibinfo {author} {\bibfnamefont {Jiahui}\ \bibnamefont {Liu}}, \bibinfo {author} {\bibfnamefont {Jesper}\ \bibnamefont {Byggm\"astar}}, \bibinfo {author} {\bibfnamefont {Zheyong}\ \bibnamefont {Fan}}, \bibinfo {author} {\bibfnamefont {Ping}\ \bibnamefont {Qian}}, \ and\ \bibinfo {author} {\bibfnamefont {Yanjing}\ \bibnamefont {Su}},\ }\bibfield  {title} {\enquote {\bibinfo {title} {Large-scale machine-learning molecular dynamics simulation of primary radiation damage in tungsten},}\ }\href {\doibase 10.1103/PhysRevB.108.054312} {\bibfield  {journal} {\bibinfo  {journal} {Physical Review B}\ }\textbf {\bibinfo {volume} {108}},\ \bibinfo {pages} {054312} (\bibinfo {year} {2023})}\BibitemShut {NoStop}%
\bibitem [{\citenamefont {Ying}\ \emph {et~al.}(2023{\natexlab{b}})\citenamefont {Ying}, \citenamefont {Dong}, \citenamefont {Liang}, \citenamefont {Fan}, \citenamefont {Zhong},\ and\ \citenamefont {Zhang}}]{ying2023atomistic}%
  \BibitemOpen
  \bibfield  {author} {\bibinfo {author} {\bibfnamefont {Penghua}\ \bibnamefont {Ying}}, \bibinfo {author} {\bibfnamefont {Haikuan}\ \bibnamefont {Dong}}, \bibinfo {author} {\bibfnamefont {Ting}\ \bibnamefont {Liang}}, \bibinfo {author} {\bibfnamefont {Zheyong}\ \bibnamefont {Fan}}, \bibinfo {author} {\bibfnamefont {Zheng}\ \bibnamefont {Zhong}}, \ and\ \bibinfo {author} {\bibfnamefont {Jin}\ \bibnamefont {Zhang}},\ }\bibfield  {title} {\enquote {\bibinfo {title} {Atomistic insights into the mechanical anisotropy and fragility of monolayer fullerene networks using quantum mechanical calculations and machine-learning molecular dynamics simulations},}\ }\href {\doibase 10.1016/j.eml.2022.101929} {\bibfield  {journal} {\bibinfo  {journal} {Extreme Mechanics Letters}\ }\textbf {\bibinfo {volume} {58}},\ \bibinfo {pages} {101929} (\bibinfo {year} {2023}{\natexlab{b}})}\BibitemShut {NoStop}%
\bibitem [{\citenamefont {Yu}\ \emph {et~al.}(2024)\citenamefont {Yu}, \citenamefont {Zhao}, \citenamefont {Guo},\ and\ \citenamefont {Zhang}}]{yu2024fracture}%
  \BibitemOpen
  \bibfield  {author} {\bibinfo {author} {\bibfnamefont {Maolin}\ \bibnamefont {Yu}}, \bibinfo {author} {\bibfnamefont {Zhiqiang}\ \bibnamefont {Zhao}}, \bibinfo {author} {\bibfnamefont {Wanlin}\ \bibnamefont {Guo}}, \ and\ \bibinfo {author} {\bibfnamefont {Zhuhua}\ \bibnamefont {Zhang}},\ }\bibfield  {title} {\enquote {\bibinfo {title} {Fracture toughness of two-dimensional materials dominated by edge energy anisotropy},}\ }\href {\doibase 10.1016/j.jmps.2024.105579} {\bibfield  {journal} {\bibinfo  {journal} {Journal of the Mechanics and Physics of Solids}\ }\textbf {\bibinfo {volume} {186}},\ \bibinfo {pages} {105579} (\bibinfo {year} {2024})}\BibitemShut {NoStop}%
\bibitem [{\citenamefont {Wang}\ \emph {et~al.}(2023)\citenamefont {Wang}, \citenamefont {Fan}, \citenamefont {Qian}, \citenamefont {Caro},\ and\ \citenamefont {Ala-Nissila}}]{wang2023quantum}%
  \BibitemOpen
  \bibfield  {author} {\bibinfo {author} {\bibfnamefont {Yanzhou}\ \bibnamefont {Wang}}, \bibinfo {author} {\bibfnamefont {Zheyong}\ \bibnamefont {Fan}}, \bibinfo {author} {\bibfnamefont {Ping}\ \bibnamefont {Qian}}, \bibinfo {author} {\bibfnamefont {Miguel~A}\ \bibnamefont {Caro}}, \ and\ \bibinfo {author} {\bibfnamefont {Tapio}\ \bibnamefont {Ala-Nissila}},\ }\bibfield  {title} {\enquote {\bibinfo {title} {Quantum-corrected thickness-dependent thermal conductivity in amorphous silicon predicted by machine learning molecular dynamics simulations},}\ }\href {\doibase 10.1103/PhysRevB.107.054303} {\bibfield  {journal} {\bibinfo  {journal} {Physical Review B}\ }\textbf {\bibinfo {volume} {107}},\ \bibinfo {pages} {054303} (\bibinfo {year} {2023})}\BibitemShut {NoStop}%
\bibitem [{\citenamefont {Liang}\ \emph {et~al.}(2023)\citenamefont {Liang}, \citenamefont {Ying}, \citenamefont {Xu}, \citenamefont {Ye}, \citenamefont {Ling}, \citenamefont {Fan},\ and\ \citenamefont {Xu}}]{liang2023mechanisms}%
  \BibitemOpen
  \bibfield  {author} {\bibinfo {author} {\bibfnamefont {Ting}\ \bibnamefont {Liang}}, \bibinfo {author} {\bibfnamefont {Penghua}\ \bibnamefont {Ying}}, \bibinfo {author} {\bibfnamefont {Ke}~\bibnamefont {Xu}}, \bibinfo {author} {\bibfnamefont {Zhenqiang}\ \bibnamefont {Ye}}, \bibinfo {author} {\bibfnamefont {Chao}\ \bibnamefont {Ling}}, \bibinfo {author} {\bibfnamefont {Zheyong}\ \bibnamefont {Fan}}, \ and\ \bibinfo {author} {\bibfnamefont {Jianbin}\ \bibnamefont {Xu}},\ }\bibfield  {title} {\enquote {\bibinfo {title} {Mechanisms of temperature-dependent thermal transport in amorphous silica from machine-learning molecular dynamics},}\ }\href {\doibase 10.1103/PhysRevB.108.184203} {\bibfield  {journal} {\bibinfo  {journal} {Physical Review B}\ }\textbf {\bibinfo {volume} {108}},\ \bibinfo {pages} {184203} (\bibinfo {year} {2023})}\BibitemShut {NoStop}%
\bibitem [{\citenamefont {Eriksson}\ \emph {et~al.}(2023)\citenamefont {Eriksson}, \citenamefont {Fransson}, \citenamefont {Linderälv}, \citenamefont {Fan},\ and\ \citenamefont {Erhart}}]{eriksson2023tuning}%
  \BibitemOpen
  \bibfield  {author} {\bibinfo {author} {\bibfnamefont {Fredrik}\ \bibnamefont {Eriksson}}, \bibinfo {author} {\bibfnamefont {Erik}\ \bibnamefont {Fransson}}, \bibinfo {author} {\bibfnamefont {Christopher}\ \bibnamefont {Linderälv}}, \bibinfo {author} {\bibfnamefont {Zheyong}\ \bibnamefont {Fan}}, \ and\ \bibinfo {author} {\bibfnamefont {Paul}\ \bibnamefont {Erhart}},\ }\bibfield  {title} {\enquote {\bibinfo {title} {Tuning the through-plane lattice thermal conductivity in van der waals structures through rotational (dis)ordering},}\ }\href {\doibase 10.1021/acsnano.3c09717} {\bibfield  {journal} {\bibinfo  {journal} {ACS Nano}\ }\textbf {\bibinfo {volume} {17}},\ \bibinfo {pages} {25565--25574} (\bibinfo {year} {2023})}\BibitemShut {NoStop}%
\bibitem [{\citenamefont {Fan}\ \emph {et~al.}(2024)\citenamefont {Fan}, \citenamefont {Xiao}, \citenamefont {Wang}, \citenamefont {Ying}, \citenamefont {Chen},\ and\ \citenamefont {Dong}}]{fan2024combining}%
  \BibitemOpen
  \bibfield  {author} {\bibinfo {author} {\bibfnamefont {Zheyong}\ \bibnamefont {Fan}}, \bibinfo {author} {\bibfnamefont {Yang}\ \bibnamefont {Xiao}}, \bibinfo {author} {\bibfnamefont {Yanzhou}\ \bibnamefont {Wang}}, \bibinfo {author} {\bibfnamefont {Penghua}\ \bibnamefont {Ying}}, \bibinfo {author} {\bibfnamefont {Shunda}\ \bibnamefont {Chen}}, \ and\ \bibinfo {author} {\bibfnamefont {Haikuan}\ \bibnamefont {Dong}},\ }\bibfield  {title} {\enquote {\bibinfo {title} {Combining linear-scaling quantum transport and machine-learning molecular dynamics to study thermal and electronic transports in complex materials},}\ }\href {\doibase 10.1088/1361-648X/ad31c2} {\bibfield  {journal} {\bibinfo  {journal} {Journal of Physics: Condensed Matter}\ }\textbf {\bibinfo {volume} {36}},\ \bibinfo {pages} {245901} (\bibinfo {year} {2024})}\BibitemShut {NoStop}%
\bibitem [{\citenamefont {Dong}\ \emph {et~al.}(2024)\citenamefont {Dong}, \citenamefont {Shi}, \citenamefont {Ying}, \citenamefont {Xu}, \citenamefont {Liang}, \citenamefont {Wang}, \citenamefont {Zeng}, \citenamefont {Wu}, \citenamefont {Zhou}, \citenamefont {Xiong}, \citenamefont {Chen},\ and\ \citenamefont {Fan}}]{dong2024molecular}%
  \BibitemOpen
  \bibfield  {author} {\bibinfo {author} {\bibfnamefont {Haikuan}\ \bibnamefont {Dong}}, \bibinfo {author} {\bibfnamefont {Yongbo}\ \bibnamefont {Shi}}, \bibinfo {author} {\bibfnamefont {Penghua}\ \bibnamefont {Ying}}, \bibinfo {author} {\bibfnamefont {Ke}~\bibnamefont {Xu}}, \bibinfo {author} {\bibfnamefont {Ting}\ \bibnamefont {Liang}}, \bibinfo {author} {\bibfnamefont {Yanzhou}\ \bibnamefont {Wang}}, \bibinfo {author} {\bibfnamefont {Zezhu}\ \bibnamefont {Zeng}}, \bibinfo {author} {\bibfnamefont {Xin}\ \bibnamefont {Wu}}, \bibinfo {author} {\bibfnamefont {Wenjiang}\ \bibnamefont {Zhou}}, \bibinfo {author} {\bibfnamefont {Shiyun}\ \bibnamefont {Xiong}}, \bibinfo {author} {\bibfnamefont {Shunda}\ \bibnamefont {Chen}}, \ and\ \bibinfo {author} {\bibfnamefont {Zheyong}\ \bibnamefont {Fan}},\ }\bibfield  {title} {\enquote {\bibinfo {title} {{Molecular dynamics simulations of heat transport using machine-learned potentials: A mini-review and tutorial on GPUMD with neuroevolution potentials}},}\ }\href {\doibase
  10.1063/5.0200833} {\bibfield  {journal} {\bibinfo  {journal} {Journal of Applied Physics}\ }\textbf {\bibinfo {volume} {135}},\ \bibinfo {pages} {161101} (\bibinfo {year} {2024})}\BibitemShut {NoStop}%
\bibitem [{\citenamefont {Fransson}\ \emph {et~al.}(2023{\natexlab{a}})\citenamefont {Fransson}, \citenamefont {Wiktor},\ and\ \citenamefont {Erhart}}]{FraWikErh23}%
  \BibitemOpen
  \bibfield  {author} {\bibinfo {author} {\bibfnamefont {Erik}\ \bibnamefont {Fransson}}, \bibinfo {author} {\bibfnamefont {Julia}\ \bibnamefont {Wiktor}}, \ and\ \bibinfo {author} {\bibfnamefont {Paul}\ \bibnamefont {Erhart}},\ }\bibfield  {title} {\enquote {\bibinfo {title} {Phase transitions in inorganic halide perovskites from machine-learned potentials},}\ }\href {\doibase 10.1021/acs.jpcc.3c01542} {\bibfield  {journal} {\bibinfo  {journal} {Journal of Physical Chemistry C}\ }\textbf {\bibinfo {volume} {127}},\ \bibinfo {pages} {13773} (\bibinfo {year} {2023}{\natexlab{a}})}\BibitemShut {NoStop}%
\bibitem [{\citenamefont {Fransson}\ \emph {et~al.}(2023{\natexlab{b}})\citenamefont {Fransson}, \citenamefont {Rahm}, \citenamefont {Wiktor},\ and\ \citenamefont {Erhart}}]{fransson2023revealing}%
  \BibitemOpen
  \bibfield  {author} {\bibinfo {author} {\bibfnamefont {Erik}\ \bibnamefont {Fransson}}, \bibinfo {author} {\bibfnamefont {J.~Magnus}\ \bibnamefont {Rahm}}, \bibinfo {author} {\bibfnamefont {Julia}\ \bibnamefont {Wiktor}}, \ and\ \bibinfo {author} {\bibfnamefont {Paul}\ \bibnamefont {Erhart}},\ }\bibfield  {title} {\enquote {\bibinfo {title} {Revealing the free energy landscape of halide perovskites: Metastability and transition characters in {CsPbBr3} and {MAPbI3}},}\ }\href {\doibase 10.1021/acs.chemmater.3c01740} {\bibfield  {journal} {\bibinfo  {journal} {Chemistry of Materials}\ }\textbf {\bibinfo {volume} {35}},\ \bibinfo {pages} {8229--8238} (\bibinfo {year} {2023}{\natexlab{b}})}\BibitemShut {NoStop}%
\bibitem [{\citenamefont {Fransson}\ \emph {et~al.}(2024{\natexlab{a}})\citenamefont {Fransson}, \citenamefont {Rosander}, \citenamefont {Erhart},\ and\ \citenamefont {Wahnström}}]{FraRosErh24}%
  \BibitemOpen
  \bibfield  {author} {\bibinfo {author} {\bibfnamefont {Erik}\ \bibnamefont {Fransson}}, \bibinfo {author} {\bibfnamefont {Petter}\ \bibnamefont {Rosander}}, \bibinfo {author} {\bibfnamefont {Paul}\ \bibnamefont {Erhart}}, \ and\ \bibinfo {author} {\bibfnamefont {Göran}\ \bibnamefont {Wahnström}},\ }\bibfield  {title} {\enquote {\bibinfo {title} {Understanding correlations in \ce{BaZrO3}: Structure and dynamics on the nano-scale},}\ }\href {\doibase 10.1021/acs.chemmater.3c02548} {\bibfield  {journal} {\bibinfo  {journal} {Chemistry of Materials}\ }\textbf {\bibinfo {volume} {36}},\ \bibinfo {pages} {514} (\bibinfo {year} {2024}{\natexlab{a}})}\BibitemShut {NoStop}%
\bibitem [{\citenamefont {Fransson}\ \emph {et~al.}(2024{\natexlab{b}})\citenamefont {Fransson}, \citenamefont {Wiktor},\ and\ \citenamefont {Erhart}}]{FraWikErh24}%
  \BibitemOpen
  \bibfield  {author} {\bibinfo {author} {\bibfnamefont {Erik}\ \bibnamefont {Fransson}}, \bibinfo {author} {\bibfnamefont {Julia}\ \bibnamefont {Wiktor}}, \ and\ \bibinfo {author} {\bibfnamefont {Paul}\ \bibnamefont {Erhart}},\ }\bibfield  {title} {\enquote {\bibinfo {title} {Impact of organic spacers and dimensionality on templating of halide perovskites},}\ }\href {\doibase 10.1021/acsenergylett.4c01283} {\bibfield  {journal} {\bibinfo  {journal} {ACS Energy Letters}\ }\textbf {\bibinfo {volume} {9}},\ \bibinfo {pages} {3947} (\bibinfo {year} {2024}{\natexlab{b}})}\BibitemShut {NoStop}%
\bibitem [{\citenamefont {Ahlawat}(2024)}]{ahlawat2024size}%
  \BibitemOpen
  \bibfield  {author} {\bibinfo {author} {\bibfnamefont {Paramvir}\ \bibnamefont {Ahlawat}},\ }\href@noop {} {\enquote {\bibinfo {title} {Size dependent solid-solid crystallization of halide perovskites},}\ } (\bibinfo {year} {2024}),\ \Eprint {http://arxiv.org/abs/2404.05644} {arXiv:2404.05644 [cond-mat.mtrl-sci]} \BibitemShut {NoStop}%
\bibitem [{\citenamefont {Chen}\ \emph {et~al.}(2024{\natexlab{b}})\citenamefont {Chen}, \citenamefont {Jin}, \citenamefont {Zhao},\ and\ \citenamefont {Li}}]{chen2024intricate}%
  \BibitemOpen
  \bibfield  {author} {\bibinfo {author} {\bibfnamefont {Shunda}\ \bibnamefont {Chen}}, \bibinfo {author} {\bibfnamefont {Xiaochen}\ \bibnamefont {Jin}}, \bibinfo {author} {\bibfnamefont {Wanyu}\ \bibnamefont {Zhao}}, \ and\ \bibinfo {author} {\bibfnamefont {Tianshu}\ \bibnamefont {Li}},\ }\bibfield  {title} {\enquote {\bibinfo {title} {Intricate short-range order in {GeSn} alloys revealed by atomistic simulations with highly accurate and efficient machine-learning potentials},}\ }\href {\doibase 10.1103/PhysRevMaterials.8.043805} {\bibfield  {journal} {\bibinfo  {journal} {Phys. Rev. Mater.}\ }\textbf {\bibinfo {volume} {8}},\ \bibinfo {pages} {043805} (\bibinfo {year} {2024}{\natexlab{b}})}\BibitemShut {NoStop}%
\bibitem [{\citenamefont {Song}\ \emph {et~al.}(2024)\citenamefont {Song}, \citenamefont {Liu}, \citenamefont {Chen}, \citenamefont {Fan}, \citenamefont {Su},\ and\ \citenamefont {Qian}}]{song2024solute}%
  \BibitemOpen
  \bibfield  {author} {\bibinfo {author} {\bibfnamefont {Keke}\ \bibnamefont {Song}}, \bibinfo {author} {\bibfnamefont {Jiahui}\ \bibnamefont {Liu}}, \bibinfo {author} {\bibfnamefont {Shunda}\ \bibnamefont {Chen}}, \bibinfo {author} {\bibfnamefont {Zheyong}\ \bibnamefont {Fan}}, \bibinfo {author} {\bibfnamefont {Yanjing}\ \bibnamefont {Su}}, \ and\ \bibinfo {author} {\bibfnamefont {Ping}\ \bibnamefont {Qian}},\ }\href {https://arxiv.org/abs/2404.13694} {\enquote {\bibinfo {title} {Solute segregation in polycrystalline aluminum from hybrid monte carlo and molecular dynamics simulations with a unified neuroevolution potential},}\ } (\bibinfo {year} {2024}),\ \Eprint {http://arxiv.org/abs/2404.13694} {arXiv:2404.13694 [cond-mat.mtrl-sci]} \BibitemShut {NoStop}%
\bibitem [{\citenamefont {Lindsay}(2016)}]{lindsay2016isotope}%
  \BibitemOpen
  \bibfield  {author} {\bibinfo {author} {\bibfnamefont {L}~\bibnamefont {Lindsay}},\ }\bibfield  {title} {\enquote {\bibinfo {title} {{Isotope scattering and phonon thermal conductivity in light atom compounds: LiH and LiF}},}\ }\href {\doibase https://doi.org/10.1103/PhysRevB.94.174304} {\bibfield  {journal} {\bibinfo  {journal} {Physical Review B}\ }\textbf {\bibinfo {volume} {94}},\ \bibinfo {pages} {174304} (\bibinfo {year} {2016})}\BibitemShut {NoStop}%
\bibitem [{\citenamefont {Zhou}\ \emph {et~al.}(2024)\citenamefont {Zhou}, \citenamefont {Liu},\ and\ \citenamefont {Song}}]{zhou2024isotope}%
  \BibitemOpen
  \bibfield  {author} {\bibinfo {author} {\bibfnamefont {Wenjiang}\ \bibnamefont {Zhou}}, \bibinfo {author} {\bibfnamefont {Te-Huan}\ \bibnamefont {Liu}}, \ and\ \bibinfo {author} {\bibfnamefont {Bai}\ \bibnamefont {Song}},\ }\bibfield  {title} {\enquote {\bibinfo {title} {{Isotope engineering of carrier mobility via Fr{\"o}hlich electron-phonon interaction}},}\ }\href {\doibase https://doi.org/10.1103/PhysRevB.109.L121201} {\bibfield  {journal} {\bibinfo  {journal} {Physical Review B}\ }\textbf {\bibinfo {volume} {109}},\ \bibinfo {pages} {L121201} (\bibinfo {year} {2024})}\BibitemShut {NoStop}%
\bibitem [{\citenamefont {Vidal}\ and\ \citenamefont {Vidal-Valat}(1986)}]{vidal1986accurate}%
  \BibitemOpen
  \bibfield  {author} {\bibinfo {author} {\bibfnamefont {Jean~Pierre}\ \bibnamefont {Vidal}}\ and\ \bibinfo {author} {\bibfnamefont {G}~\bibnamefont {Vidal-Valat}},\ }\bibfield  {title} {\enquote {\bibinfo {title} {{Accurate Debye--Waller factors of $^7$LiH and $^7$LiD by neutron diffraction at three temperatures}},}\ }\href {\doibase 10.1107/S0108768186098476} {\bibfield  {journal} {\bibinfo  {journal} {Acta Crystallographica Section B: Structural Science}\ }\textbf {\bibinfo {volume} {42}},\ \bibinfo {pages} {131--137} (\bibinfo {year} {1986})}\BibitemShut {NoStop}%
\bibitem [{\citenamefont {Anderson}\ \emph {et~al.}(1970)\citenamefont {Anderson}, \citenamefont {Nasise}, \citenamefont {Phillipson},\ and\ \citenamefont {Pretzel}}]{anderson1970isotopic}%
  \BibitemOpen
  \bibfield  {author} {\bibinfo {author} {\bibfnamefont {J~L}\ \bibnamefont {Anderson}}, \bibinfo {author} {\bibfnamefont {J}~\bibnamefont {Nasise}}, \bibinfo {author} {\bibfnamefont {K}~\bibnamefont {Phillipson}}, \ and\ \bibinfo {author} {\bibfnamefont {F~E}\ \bibnamefont {Pretzel}},\ }\bibfield  {title} {\enquote {\bibinfo {title} {Isotopic effects on the thermal expansion of lithium hydride},}\ }\href {\doibase https://doi.org/10.1016/0022-3697(70)90195-2} {\bibfield  {journal} {\bibinfo  {journal} {Journal of Physics and Chemistry of Solids}\ }\textbf {\bibinfo {volume} {31}},\ \bibinfo {pages} {613--618} (\bibinfo {year} {1970})}\BibitemShut {NoStop}%
\bibitem [{\citenamefont {Rowsell}\ \emph {et~al.}(2005)\citenamefont {Rowsell}, \citenamefont {Spencer}, \citenamefont {Eckert}, \citenamefont {Howard},\ and\ \citenamefont {Yaghi}}]{rowsell2005gas}%
  \BibitemOpen
  \bibfield  {author} {\bibinfo {author} {\bibfnamefont {Jesse~LC}\ \bibnamefont {Rowsell}}, \bibinfo {author} {\bibfnamefont {Elinor~C}\ \bibnamefont {Spencer}}, \bibinfo {author} {\bibfnamefont {Juergen}\ \bibnamefont {Eckert}}, \bibinfo {author} {\bibfnamefont {Judith~AK}\ \bibnamefont {Howard}}, \ and\ \bibinfo {author} {\bibfnamefont {Omar~M}\ \bibnamefont {Yaghi}},\ }\bibfield  {title} {\enquote {\bibinfo {title} {Gas adsorption sites in a large-pore metal-organic framework},}\ }\href {\doibase 10.1126/science.1113247} {\bibfield  {journal} {\bibinfo  {journal} {Science}\ }\textbf {\bibinfo {volume} {309}},\ \bibinfo {pages} {1350--1354} (\bibinfo {year} {2005})}\BibitemShut {NoStop}%
\bibitem [{\citenamefont {Zhou}\ \emph {et~al.}(2008)\citenamefont {Zhou}, \citenamefont {Wu}, \citenamefont {Yildirim}, \citenamefont {Simpson},\ and\ \citenamefont {Walker}}]{zhou2008origin}%
  \BibitemOpen
  \bibfield  {author} {\bibinfo {author} {\bibfnamefont {Wei}\ \bibnamefont {Zhou}}, \bibinfo {author} {\bibfnamefont {Hui}\ \bibnamefont {Wu}}, \bibinfo {author} {\bibfnamefont {Tanner}\ \bibnamefont {Yildirim}}, \bibinfo {author} {\bibfnamefont {Jeffrey~R}\ \bibnamefont {Simpson}}, \ and\ \bibinfo {author} {\bibfnamefont {AR~Hight}\ \bibnamefont {Walker}},\ }\bibfield  {title} {\enquote {\bibinfo {title} {Origin of the exceptional negative thermal expansion in metal-organic framework-5 \textnormal{Zn}$_\textnormal{4}$\textnormal{O}(1,4-benzenedicarboxylate)$_\textnormal{3}$},}\ }\href {\doibase 10.1103/PhysRevB.78.054114} {\bibfield  {journal} {\bibinfo  {journal} {Physical Review B}\ }\textbf {\bibinfo {volume} {78}},\ \bibinfo {pages} {054114} (\bibinfo {year} {2008})}\BibitemShut {NoStop}%
\bibitem [{\citenamefont {Lock}\ \emph {et~al.}(2010)\citenamefont {Lock}, \citenamefont {Wu}, \citenamefont {Christensen}, \citenamefont {Cameron}, \citenamefont {Peterson}, \citenamefont {Bridgeman}, \citenamefont {Kepert},\ and\ \citenamefont {Iversen}}]{lock2010elucidating}%
  \BibitemOpen
  \bibfield  {author} {\bibinfo {author} {\bibfnamefont {Nina}\ \bibnamefont {Lock}}, \bibinfo {author} {\bibfnamefont {Yue}\ \bibnamefont {Wu}}, \bibinfo {author} {\bibfnamefont {Mogens}\ \bibnamefont {Christensen}}, \bibinfo {author} {\bibfnamefont {Lisa~J}\ \bibnamefont {Cameron}}, \bibinfo {author} {\bibfnamefont {Vanessa~K}\ \bibnamefont {Peterson}}, \bibinfo {author} {\bibfnamefont {Adam~J}\ \bibnamefont {Bridgeman}}, \bibinfo {author} {\bibfnamefont {Cameron~J}\ \bibnamefont {Kepert}}, \ and\ \bibinfo {author} {\bibfnamefont {Bo~B}\ \bibnamefont {Iversen}},\ }\bibfield  {title} {\enquote {\bibinfo {title} {Elucidating negative thermal expansion in {MOF}-5},}\ }\href {\doibase 10.1021/jp103212z} {\bibfield  {journal} {\bibinfo  {journal} {The Journal of Physical Chemistry C}\ }\textbf {\bibinfo {volume} {114}},\ \bibinfo {pages} {16181--16186} (\bibinfo {year} {2010})}\BibitemShut {NoStop}%
\bibitem [{\citenamefont {Lock}\ \emph {et~al.}(2013)\citenamefont {Lock}, \citenamefont {Christensen}, \citenamefont {Wu}, \citenamefont {Peterson}, \citenamefont {Thomsen}, \citenamefont {Piltz}, \citenamefont {Ramirez-Cuesta}, \citenamefont {McIntyre}, \citenamefont {Nor{\'e}n}, \citenamefont {Kutteh} \emph {et~al.}}]{lock2013scrutinizing}%
  \BibitemOpen
  \bibfield  {author} {\bibinfo {author} {\bibfnamefont {Nina}\ \bibnamefont {Lock}}, \bibinfo {author} {\bibfnamefont {Mogens}\ \bibnamefont {Christensen}}, \bibinfo {author} {\bibfnamefont {Yue}\ \bibnamefont {Wu}}, \bibinfo {author} {\bibfnamefont {Vanessa~K}\ \bibnamefont {Peterson}}, \bibinfo {author} {\bibfnamefont {Maja~K}\ \bibnamefont {Thomsen}}, \bibinfo {author} {\bibfnamefont {Ross~O}\ \bibnamefont {Piltz}}, \bibinfo {author} {\bibfnamefont {Anibal~J}\ \bibnamefont {Ramirez-Cuesta}}, \bibinfo {author} {\bibfnamefont {Garry~J}\ \bibnamefont {McIntyre}}, \bibinfo {author} {\bibfnamefont {Katarina}\ \bibnamefont {Nor{\'e}n}}, \bibinfo {author} {\bibfnamefont {Ramzi}\ \bibnamefont {Kutteh}},  \emph {et~al.},\ }\bibfield  {title} {\enquote {\bibinfo {title} {Scrutinizing negative thermal expansion in {MOF}-5 by scattering techniques and ab initio calculations},}\ }\href {\doibase 10.1039/C2DT31491F} {\bibfield  {journal} {\bibinfo  {journal} {Dalton transactions}\ }\textbf {\bibinfo {volume} {42}},\
  \bibinfo {pages} {1996--2007} (\bibinfo {year} {2013})}\BibitemShut {NoStop}%
\bibitem [{\citenamefont {Wu}\ \emph {et~al.}(2008)\citenamefont {Wu}, \citenamefont {Kobayashi}, \citenamefont {Halder}, \citenamefont {Peterson}, \citenamefont {Chapman}, \citenamefont {Lock}, \citenamefont {Southon},\ and\ \citenamefont {Kepert}}]{wu2008negative}%
  \BibitemOpen
  \bibfield  {author} {\bibinfo {author} {\bibfnamefont {Yue}\ \bibnamefont {Wu}}, \bibinfo {author} {\bibfnamefont {Atsushi}\ \bibnamefont {Kobayashi}}, \bibinfo {author} {\bibfnamefont {Gregory~J}\ \bibnamefont {Halder}}, \bibinfo {author} {\bibfnamefont {Vanessa~K}\ \bibnamefont {Peterson}}, \bibinfo {author} {\bibfnamefont {Karena~W}\ \bibnamefont {Chapman}}, \bibinfo {author} {\bibfnamefont {Nina}\ \bibnamefont {Lock}}, \bibinfo {author} {\bibfnamefont {Peter~D}\ \bibnamefont {Southon}}, \ and\ \bibinfo {author} {\bibfnamefont {Cameron~J}\ \bibnamefont {Kepert}},\ }\bibfield  {title} {\enquote {\bibinfo {title} {Negative thermal expansion in the metal--organic framework material \textnormal{Cu}$_\textnormal{3}$(1,3,5‐benzenetricarboxylate)$_\textnormal{2}$},}\ }\href {\doibase 10.1002/anie.200803925} {\bibfield  {journal} {\bibinfo  {journal} {Angewandte Chemie International Edition}\ }\textbf {\bibinfo {volume} {47}},\ \bibinfo {pages} {8929--8932} (\bibinfo {year} {2008})}\BibitemShut {NoStop}%
\bibitem [{\citenamefont {Peterson}\ \emph {et~al.}(2010)\citenamefont {Peterson}, \citenamefont {Kearley}, \citenamefont {Wu}, \citenamefont {Ramirez-Cuesta}, \citenamefont {Kemner},\ and\ \citenamefont {Kepert}}]{peterson2010local}%
  \BibitemOpen
  \bibfield  {author} {\bibinfo {author} {\bibfnamefont {Vanessa~K}\ \bibnamefont {Peterson}}, \bibinfo {author} {\bibfnamefont {Gordon~J}\ \bibnamefont {Kearley}}, \bibinfo {author} {\bibfnamefont {Yue}\ \bibnamefont {Wu}}, \bibinfo {author} {\bibfnamefont {Anibal~Javier}\ \bibnamefont {Ramirez-Cuesta}}, \bibinfo {author} {\bibfnamefont {Ewout}\ \bibnamefont {Kemner}}, \ and\ \bibinfo {author} {\bibfnamefont {Cameron~J}\ \bibnamefont {Kepert}},\ }\bibfield  {title} {\enquote {\bibinfo {title} {Local vibrational mechanism for negative thermal expansion: A combined neutron scattering and first-principles study},}\ }\href {\doibase 10.1002/anie.200903366} {\bibfield  {journal} {\bibinfo  {journal} {Angewandte Chemie}\ }\textbf {\bibinfo {volume} {122}},\ \bibinfo {pages} {595--598} (\bibinfo {year} {2010})}\BibitemShut {NoStop}%
\bibitem [{\citenamefont {Schneider}\ \emph {et~al.}(2019)\citenamefont {Schneider}, \citenamefont {Bodesheim}, \citenamefont {Ehrenreich}, \citenamefont {Crocell{\`a}}, \citenamefont {Mink}, \citenamefont {Fischer}, \citenamefont {Butler},\ and\ \citenamefont {Kieslich}}]{schneider2019tuning}%
  \BibitemOpen
  \bibfield  {author} {\bibinfo {author} {\bibfnamefont {Christian}\ \bibnamefont {Schneider}}, \bibinfo {author} {\bibfnamefont {David}\ \bibnamefont {Bodesheim}}, \bibinfo {author} {\bibfnamefont {Michael~G}\ \bibnamefont {Ehrenreich}}, \bibinfo {author} {\bibfnamefont {Valentina}\ \bibnamefont {Crocell{\`a}}}, \bibinfo {author} {\bibfnamefont {J{\'a}nos}\ \bibnamefont {Mink}}, \bibinfo {author} {\bibfnamefont {Roland~A}\ \bibnamefont {Fischer}}, \bibinfo {author} {\bibfnamefont {Keith~T}\ \bibnamefont {Butler}}, \ and\ \bibinfo {author} {\bibfnamefont {Gregor}\ \bibnamefont {Kieslich}},\ }\bibfield  {title} {\enquote {\bibinfo {title} {Tuning the negative thermal expansion behavior of the metal--organic framework \textnormal{Cu}$_\textnormal{3}$\textnormal{BTC}$_\textnormal{2}$ by retrofitting},}\ }\href {\doibase 10.1021/jacs.9b04755} {\bibfield  {journal} {\bibinfo  {journal} {Journal of the American Chemical Society}\ }\textbf {\bibinfo {volume} {141}},\ \bibinfo {pages} {10504--10509} (\bibinfo {year}
  {2019})}\BibitemShut {NoStop}%
\bibitem [{\citenamefont {Sapnik}\ \emph {et~al.}(2018)\citenamefont {Sapnik}, \citenamefont {Geddes}, \citenamefont {Reynolds}, \citenamefont {Yeung},\ and\ \citenamefont {Goodwin}}]{sapnik2018compositional}%
  \BibitemOpen
  \bibfield  {author} {\bibinfo {author} {\bibfnamefont {Adam~F}\ \bibnamefont {Sapnik}}, \bibinfo {author} {\bibfnamefont {Harry~S}\ \bibnamefont {Geddes}}, \bibinfo {author} {\bibfnamefont {Emily~M}\ \bibnamefont {Reynolds}}, \bibinfo {author} {\bibfnamefont {Hamish H-M}\ \bibnamefont {Yeung}}, \ and\ \bibinfo {author} {\bibfnamefont {Andrew~L}\ \bibnamefont {Goodwin}},\ }\bibfield  {title} {\enquote {\bibinfo {title} {Compositional inhomogeneity and tuneable thermal expansion in mixed-metal {ZIF}-8 analogues},}\ }\href {\doibase 10.1039/C8CC04172E} {\bibfield  {journal} {\bibinfo  {journal} {Chemical communications}\ }\textbf {\bibinfo {volume} {54}},\ \bibinfo {pages} {9651--9654} (\bibinfo {year} {2018})}\BibitemShut {NoStop}%
\bibitem [{\citenamefont {Burtch}\ \emph {et~al.}(2019)\citenamefont {Burtch}, \citenamefont {Baxter}, \citenamefont {Heinen}, \citenamefont {Bird}, \citenamefont {Schneemann}, \citenamefont {Dubbeldam},\ and\ \citenamefont {Wilkinson}}]{burtch2019negative}%
  \BibitemOpen
  \bibfield  {author} {\bibinfo {author} {\bibfnamefont {Nicholas~C}\ \bibnamefont {Burtch}}, \bibinfo {author} {\bibfnamefont {Samuel~J}\ \bibnamefont {Baxter}}, \bibinfo {author} {\bibfnamefont {Jurn}\ \bibnamefont {Heinen}}, \bibinfo {author} {\bibfnamefont {Ashley}\ \bibnamefont {Bird}}, \bibinfo {author} {\bibfnamefont {Andreas}\ \bibnamefont {Schneemann}}, \bibinfo {author} {\bibfnamefont {David}\ \bibnamefont {Dubbeldam}}, \ and\ \bibinfo {author} {\bibfnamefont {Angus~P}\ \bibnamefont {Wilkinson}},\ }\bibfield  {title} {\enquote {\bibinfo {title} {Negative thermal expansion design strategies in a diverse series of metal--organic frameworks},}\ }\href {\doibase 10.1002/adfm.201904669} {\bibfield  {journal} {\bibinfo  {journal} {Advanced Functional Materials}\ }\textbf {\bibinfo {volume} {29}},\ \bibinfo {pages} {1904669} (\bibinfo {year} {2019})}\BibitemShut {NoStop}%
\bibitem [{\citenamefont {Lamaire}\ \emph {et~al.}(2019)\citenamefont {Lamaire}, \citenamefont {Wieme}, \citenamefont {Rogge}, \citenamefont {Waroquier},\ and\ \citenamefont {Van~Speybroeck}}]{lamaire2019importance}%
  \BibitemOpen
  \bibfield  {author} {\bibinfo {author} {\bibfnamefont {Aran}\ \bibnamefont {Lamaire}}, \bibinfo {author} {\bibfnamefont {Jelle}\ \bibnamefont {Wieme}}, \bibinfo {author} {\bibfnamefont {Sven~MJ}\ \bibnamefont {Rogge}}, \bibinfo {author} {\bibfnamefont {Michel}\ \bibnamefont {Waroquier}}, \ and\ \bibinfo {author} {\bibfnamefont {Veronique}\ \bibnamefont {Van~Speybroeck}},\ }\bibfield  {title} {\enquote {\bibinfo {title} {On the importance of anharmonicities and nuclear quantum effects in modelling the structural properties and thermal expansion of {MOF}-5},}\ }\href {\doibase 10.1063/1.5085649} {\bibfield  {journal} {\bibinfo  {journal} {The Journal of Chemical Physics}\ }\textbf {\bibinfo {volume} {150}},\ \bibinfo {pages} {094503} (\bibinfo {year} {2019})}\BibitemShut {NoStop}%
\bibitem [{\citenamefont {Ceriotti}\ \emph {et~al.}(2010)\citenamefont {Ceriotti}, \citenamefont {Parrinello}, \citenamefont {Markland},\ and\ \citenamefont {Manolopoulos}}]{ceriotti2010jcp}%
  \BibitemOpen
  \bibfield  {author} {\bibinfo {author} {\bibfnamefont {Michele}\ \bibnamefont {Ceriotti}}, \bibinfo {author} {\bibfnamefont {Michele}\ \bibnamefont {Parrinello}}, \bibinfo {author} {\bibfnamefont {Thomas~E.}\ \bibnamefont {Markland}}, \ and\ \bibinfo {author} {\bibfnamefont {David~E.}\ \bibnamefont {Manolopoulos}},\ }\bibfield  {title} {\enquote {\bibinfo {title} {{Efficient stochastic thermostatting of path integral molecular dynamics}},}\ }\href {\doibase 10.1063/1.3489925} {\bibfield  {journal} {\bibinfo  {journal} {The Journal of Chemical Physics}\ }\textbf {\bibinfo {volume} {133}},\ \bibinfo {pages} {124104} (\bibinfo {year} {2010})}\BibitemShut {NoStop}%
\bibitem [{\citenamefont {Korol}\ \emph {et~al.}(2019)\citenamefont {Korol}, \citenamefont {Bou-Rabee},\ and\ \citenamefont {Miller}}]{Korol2019jcp}%
  \BibitemOpen
  \bibfield  {author} {\bibinfo {author} {\bibfnamefont {Roman}\ \bibnamefont {Korol}}, \bibinfo {author} {\bibfnamefont {Nawaf}\ \bibnamefont {Bou-Rabee}}, \ and\ \bibinfo {author} {\bibfnamefont {III}\ \bibnamefont {Miller}, \bibfnamefont {Thomas~F.}},\ }\bibfield  {title} {\enquote {\bibinfo {title} {{Cayley modification for strongly stable path-integral and ring-polymer molecular dynamics}},}\ }\href {\doibase 10.1063/1.5120282} {\bibfield  {journal} {\bibinfo  {journal} {The Journal of Chemical Physics}\ }\textbf {\bibinfo {volume} {151}},\ \bibinfo {pages} {124103} (\bibinfo {year} {2019})}\BibitemShut {NoStop}%
\bibitem [{\citenamefont {Tuckerman}(2023)}]{tuckerman_book}%
  \BibitemOpen
  \bibfield  {author} {\bibinfo {author} {\bibfnamefont {Mark~E}\ \bibnamefont {Tuckerman}},\ }\href@noop {} {\emph {\bibinfo {title} {Statistical mechanics: theory and molecular simulation}}}\ (\bibinfo  {publisher} {Oxford university press},\ \bibinfo {address} {Oxford},\ \bibinfo {year} {2023})\BibitemShut {NoStop}%
\bibitem [{\citenamefont {Craig}\ and\ \citenamefont {Manolopoulos}(2004)}]{Craig2004JCP}%
  \BibitemOpen
  \bibfield  {author} {\bibinfo {author} {\bibfnamefont {Ian~R.}\ \bibnamefont {Craig}}\ and\ \bibinfo {author} {\bibfnamefont {David~E.}\ \bibnamefont {Manolopoulos}},\ }\bibfield  {title} {\enquote {\bibinfo {title} {{Quantum statistics and classical mechanics: Real time correlation functions from ring polymer molecular dynamics}},}\ }\href {\doibase 10.1063/1.1777575} {\bibfield  {journal} {\bibinfo  {journal} {The Journal of Chemical Physics}\ }\textbf {\bibinfo {volume} {121}},\ \bibinfo {pages} {3368--3373} (\bibinfo {year} {2004})}\BibitemShut {NoStop}%
\bibitem [{\citenamefont {Habershon}\ \emph {et~al.}(2013)\citenamefont {Habershon}, \citenamefont {Manolopoulos}, \citenamefont {Markland},\ and\ \citenamefont {Miller}}]{Habershon2014ARPC}%
  \BibitemOpen
  \bibfield  {author} {\bibinfo {author} {\bibfnamefont {Scott}\ \bibnamefont {Habershon}}, \bibinfo {author} {\bibfnamefont {David~E.}\ \bibnamefont {Manolopoulos}}, \bibinfo {author} {\bibfnamefont {Thomas~E.}\ \bibnamefont {Markland}}, \ and\ \bibinfo {author} {\bibfnamefont {Thomas~F.}\ \bibnamefont {Miller}},\ }\bibfield  {title} {\enquote {\bibinfo {title} {{Ring-polymer molecular dynamics: Quantum effects in chemical dynamics from classical trajectories in an extended phase space}},}\ }\href {\doibase https://doi.org/10.1146/annurev-physchem-040412-110122} {\bibfield  {journal} {\bibinfo  {journal} {Annual Review of Physical Chemistry}\ }\textbf {\bibinfo {volume} {64}},\ \bibinfo {pages} {387--413} (\bibinfo {year} {2013})}\BibitemShut {NoStop}%
\bibitem [{\citenamefont {Rossi}\ \emph {et~al.}(2014)\citenamefont {Rossi}, \citenamefont {Ceriotti},\ and\ \citenamefont {Manolopoulos}}]{rossi_how_2014}%
  \BibitemOpen
  \bibfield  {author} {\bibinfo {author} {\bibfnamefont {Mariana}\ \bibnamefont {Rossi}}, \bibinfo {author} {\bibfnamefont {Michele}\ \bibnamefont {Ceriotti}}, \ and\ \bibinfo {author} {\bibfnamefont {David~E.}\ \bibnamefont {Manolopoulos}},\ }\bibfield  {title} {\enquote {\bibinfo {title} {How to remove the spurious resonances from ring polymer molecular dynamics},}\ }\href {\doibase 10.1063/1.4883861} {\bibfield  {journal} {\bibinfo  {journal} {The Journal of Chemical Physics}\ }\textbf {\bibinfo {volume} {140}},\ \bibinfo {pages} {234116} (\bibinfo {year} {2014})}\BibitemShut {NoStop}%
\bibitem [{\citenamefont {Berendsen}\ \emph {et~al.}(1984)\citenamefont {Berendsen}, \citenamefont {Postma}, \citenamefont {Van~Gunsteren}, \citenamefont {DiNola},\ and\ \citenamefont {Haak}}]{berendsen1984molecular}%
  \BibitemOpen
  \bibfield  {author} {\bibinfo {author} {\bibfnamefont {Herman J~C}\ \bibnamefont {Berendsen}}, \bibinfo {author} {\bibfnamefont {J~P M~van}\ \bibnamefont {Postma}}, \bibinfo {author} {\bibfnamefont {Wilfred~F}\ \bibnamefont {Van~Gunsteren}}, \bibinfo {author} {\bibfnamefont {A~R H~J}\ \bibnamefont {DiNola}}, \ and\ \bibinfo {author} {\bibfnamefont {Jan~R}\ \bibnamefont {Haak}},\ }\bibfield  {title} {\enquote {\bibinfo {title} {Molecular dynamics with coupling to an external bath},}\ }\href {\doibase 10.1063/1.448118} {\bibfield  {journal} {\bibinfo  {journal} {The Journal of Chemical Physics}\ }\textbf {\bibinfo {volume} {81}},\ \bibinfo {pages} {3684--3690} (\bibinfo {year} {1984})}\BibitemShut {NoStop}%
\bibitem [{\citenamefont {Fransson}\ \emph {et~al.}(2021)\citenamefont {Fransson}, \citenamefont {Slabanja}, \citenamefont {Erhart},\ and\ \citenamefont {Wahnström}}]{fransson_dynasortool_2021}%
  \BibitemOpen
  \bibfield  {author} {\bibinfo {author} {\bibfnamefont {Erik}\ \bibnamefont {Fransson}}, \bibinfo {author} {\bibfnamefont {Mattias}\ \bibnamefont {Slabanja}}, \bibinfo {author} {\bibfnamefont {Paul}\ \bibnamefont {Erhart}}, \ and\ \bibinfo {author} {\bibfnamefont {Göran}\ \bibnamefont {Wahnström}},\ }\bibfield  {title} {\enquote {\bibinfo {title} {{DYNASOR}—{A} {tool} for {extracting} {dynamical} {structure} {factors} and {current} {correlation} {functions} from {molecular} {dynamics} {simulations}},}\ }\href {\doibase 10.1002/adts.202000240} {\bibfield  {journal} {\bibinfo  {journal} {Advanced Theory and Simulations}\ }\textbf {\bibinfo {volume} {4}},\ \bibinfo {pages} {2000240} (\bibinfo {year} {2021})}\BibitemShut {NoStop}%
\bibitem [{\citenamefont {Thomas}\ \emph {et~al.}(2010)\citenamefont {Thomas}, \citenamefont {Turney}, \citenamefont {Iutzi}, \citenamefont {Amon},\ and\ \citenamefont {McGaughey}}]{thomas_predicting_2010}%
  \BibitemOpen
  \bibfield  {author} {\bibinfo {author} {\bibfnamefont {John~A.}\ \bibnamefont {Thomas}}, \bibinfo {author} {\bibfnamefont {Joseph~E.}\ \bibnamefont {Turney}}, \bibinfo {author} {\bibfnamefont {Ryan~M.}\ \bibnamefont {Iutzi}}, \bibinfo {author} {\bibfnamefont {Cristina~H.}\ \bibnamefont {Amon}}, \ and\ \bibinfo {author} {\bibfnamefont {Alan J.~H.}\ \bibnamefont {McGaughey}},\ }\bibfield  {title} {\enquote {\bibinfo {title} {Predicting phonon dispersion relations and lifetimes from the spectral energy density},}\ }\href {\doibase 10.1103/PhysRevB.81.081411} {\bibfield  {journal} {\bibinfo  {journal} {Physical Review B}\ }\textbf {\bibinfo {volume} {81}},\ \bibinfo {pages} {081411} (\bibinfo {year} {2010})}\BibitemShut {NoStop}%
\bibitem [{\citenamefont {Schaul}\ \emph {et~al.}(2011)\citenamefont {Schaul}, \citenamefont {Glasmachers},\ and\ \citenamefont {Schmidhuber}}]{Schaul2011High}%
  \BibitemOpen
  \bibfield  {author} {\bibinfo {author} {\bibfnamefont {Tom}\ \bibnamefont {Schaul}}, \bibinfo {author} {\bibfnamefont {Tobias}\ \bibnamefont {Glasmachers}}, \ and\ \bibinfo {author} {\bibfnamefont {J\"{u}rgen}\ \bibnamefont {Schmidhuber}},\ }\bibfield  {title} {\enquote {\bibinfo {title} {High dimensions and heavy tails for natural evolution strategies},}\ }in\ \href {\doibase 10.1145/2001576.2001692} {\emph {\bibinfo {booktitle} {Proceedings of the 13th Annual Conference on Genetic and Evolutionary Computation}}},\ \bibinfo {series and number} {GECCO '11}\ (\bibinfo  {publisher} {Association for Computing Machinery},\ \bibinfo {address} {New York, NY, USA},\ \bibinfo {year} {2011})\ p.\ \bibinfo {pages} {845–852}\BibitemShut {NoStop}%
\bibitem [{\citenamefont {Behler}\ and\ \citenamefont {Parrinello}(2007)}]{Behler2007prl}%
  \BibitemOpen
  \bibfield  {author} {\bibinfo {author} {\bibfnamefont {J\"org}\ \bibnamefont {Behler}}\ and\ \bibinfo {author} {\bibfnamefont {Michele}\ \bibnamefont {Parrinello}},\ }\bibfield  {title} {\enquote {\bibinfo {title} {Generalized neural-network representation of high-dimensional potential-energy surfaces},}\ }\href {\doibase 10.1103/PhysRevLett.98.146401} {\bibfield  {journal} {\bibinfo  {journal} {Physical Review Letters}\ }\textbf {\bibinfo {volume} {98}},\ \bibinfo {pages} {146401} (\bibinfo {year} {2007})}\BibitemShut {NoStop}%
\bibitem [{\citenamefont {Bl\"ochl}(1994)}]{PAW1994}%
  \BibitemOpen
  \bibfield  {author} {\bibinfo {author} {\bibfnamefont {P.~E.}\ \bibnamefont {Bl\"ochl}},\ }\bibfield  {title} {\enquote {\bibinfo {title} {Projector augmented-wave method},}\ }\href {\doibase 10.1103/PhysRevB.50.17953} {\bibfield  {journal} {\bibinfo  {journal} {Physical Review B}\ }\textbf {\bibinfo {volume} {50}},\ \bibinfo {pages} {17953--17979} (\bibinfo {year} {1994})}\BibitemShut {NoStop}%
\bibitem [{\citenamefont {Kresse}\ and\ \citenamefont {Furthmüller}(1996)}]{planewavebasis1996}%
  \BibitemOpen
  \bibfield  {author} {\bibinfo {author} {\bibfnamefont {G.}~\bibnamefont {Kresse}}\ and\ \bibinfo {author} {\bibfnamefont {J.}~\bibnamefont {Furthmüller}},\ }\bibfield  {title} {\enquote {\bibinfo {title} {Efficiency of ab-initio total energy calculations for metals and semiconductors using a plane-wave basis set},}\ }\href {\doibase https://doi.org/10.1016/0927-0256(96)00008-0} {\bibfield  {journal} {\bibinfo  {journal} {Computational Materials Science}\ }\textbf {\bibinfo {volume} {6}},\ \bibinfo {pages} {15--50} (\bibinfo {year} {1996})}\BibitemShut {NoStop}%
\bibitem [{\citenamefont {Perdew}\ and\ \citenamefont {Zunger}(1981)}]{PerdewZungerfunctional1981}%
  \BibitemOpen
  \bibfield  {author} {\bibinfo {author} {\bibfnamefont {J.~P.}\ \bibnamefont {Perdew}}\ and\ \bibinfo {author} {\bibfnamefont {Alex}\ \bibnamefont {Zunger}},\ }\bibfield  {title} {\enquote {\bibinfo {title} {Self-interaction correction to density-functional approximations for many-electron systems},}\ }\href {\doibase 10.1103/PhysRevB.23.5048} {\bibfield  {journal} {\bibinfo  {journal} {Physical Review B}\ }\textbf {\bibinfo {volume} {23}},\ \bibinfo {pages} {5048--5079} (\bibinfo {year} {1981})}\BibitemShut {NoStop}%
\bibitem [{\citenamefont {Walker}\ \emph {et~al.}(2010)\citenamefont {Walker}, \citenamefont {Civalleri}, \citenamefont {Slater}, \citenamefont {Mellot-Draznieks}, \citenamefont {Cor{\`a}}, \citenamefont {Zicovich-Wilson}, \citenamefont {Rom{\'a}n-P{\'e}rez}, \citenamefont {Soler},\ and\ \citenamefont {Gale}}]{walker2010flexibility}%
  \BibitemOpen
  \bibfield  {author} {\bibinfo {author} {\bibfnamefont {Andrew~M}\ \bibnamefont {Walker}}, \bibinfo {author} {\bibfnamefont {Bartolomeo}\ \bibnamefont {Civalleri}}, \bibinfo {author} {\bibfnamefont {Ben}\ \bibnamefont {Slater}}, \bibinfo {author} {\bibfnamefont {Caroline}\ \bibnamefont {Mellot-Draznieks}}, \bibinfo {author} {\bibfnamefont {Furio}\ \bibnamefont {Cor{\`a}}}, \bibinfo {author} {\bibfnamefont {Claudio~M}\ \bibnamefont {Zicovich-Wilson}}, \bibinfo {author} {\bibfnamefont {Guillermo}\ \bibnamefont {Rom{\'a}n-P{\'e}rez}}, \bibinfo {author} {\bibfnamefont {Jos{\'e}~M}\ \bibnamefont {Soler}}, \ and\ \bibinfo {author} {\bibfnamefont {Julian~D}\ \bibnamefont {Gale}},\ }\bibfield  {title} {\enquote {\bibinfo {title} {{Flexibility in a metal--organic framework material controlled by weak dispersion forces: the bistability of MIL-53 (Al)}},}\ }\href {\doibase 10.1002/anie.201002413} {\bibfield  {journal} {\bibinfo  {journal} {Angewandte Chemie International Edition}\ }\textbf {\bibinfo {volume} {49}},\
  \bibinfo {pages} {7501--7503} (\bibinfo {year} {2010})}\BibitemShut {NoStop}%
\bibitem [{\citenamefont {Wieme}\ \emph {et~al.}(2018)\citenamefont {Wieme}, \citenamefont {Lejaeghere}, \citenamefont {Kresse},\ and\ \citenamefont {Van~Speybroeck}}]{wieme2018tuning}%
  \BibitemOpen
  \bibfield  {author} {\bibinfo {author} {\bibfnamefont {Jelle}\ \bibnamefont {Wieme}}, \bibinfo {author} {\bibfnamefont {Kurt}\ \bibnamefont {Lejaeghere}}, \bibinfo {author} {\bibfnamefont {Georg}\ \bibnamefont {Kresse}}, \ and\ \bibinfo {author} {\bibfnamefont {Veronique}\ \bibnamefont {Van~Speybroeck}},\ }\bibfield  {title} {\enquote {\bibinfo {title} {{Tuning the balance between dispersion and entropy to design temperature-responsive flexible metal-organic frameworks}},}\ }\href {\doibase 10.1038/s41467-018-07298-4} {\bibfield  {journal} {\bibinfo  {journal} {Nature Communications}\ }\textbf {\bibinfo {volume} {9}},\ \bibinfo {pages} {4899} (\bibinfo {year} {2018})}\BibitemShut {NoStop}%
\bibitem [{\citenamefont {Ying}\ and\ \citenamefont {Fan}(2023)}]{ying2023combining}%
  \BibitemOpen
  \bibfield  {author} {\bibinfo {author} {\bibfnamefont {Penghua}\ \bibnamefont {Ying}}\ and\ \bibinfo {author} {\bibfnamefont {Zheyong}\ \bibnamefont {Fan}},\ }\bibfield  {title} {\enquote {\bibinfo {title} {{Combining the D3 dispersion correction with the neuroevolution machine-learned potential}},}\ }\href {\doibase 10.1088/1361-648X/ad1278} {\bibfield  {journal} {\bibinfo  {journal} {Journal of Physics: Condensed Matter}\ }\textbf {\bibinfo {volume} {36}},\ \bibinfo {pages} {125901} (\bibinfo {year} {2023})}\BibitemShut {NoStop}%
\bibitem [{\citenamefont {Grimme}\ \emph {et~al.}(2011)\citenamefont {Grimme}, \citenamefont {Ehrlich},\ and\ \citenamefont {Goerigk}}]{grimme2011effect}%
  \BibitemOpen
  \bibfield  {author} {\bibinfo {author} {\bibfnamefont {Stefan}\ \bibnamefont {Grimme}}, \bibinfo {author} {\bibfnamefont {Stephan}\ \bibnamefont {Ehrlich}}, \ and\ \bibinfo {author} {\bibfnamefont {Lars}\ \bibnamefont {Goerigk}},\ }\bibfield  {title} {\enquote {\bibinfo {title} {Effect of the damping function in dispersion corrected density functional theory},}\ }\href {\doibase 10.1002/jcc.21759} {\bibfield  {journal} {\bibinfo  {journal} {Journal of computational chemistry}\ }\textbf {\bibinfo {volume} {32}},\ \bibinfo {pages} {1456--1465} (\bibinfo {year} {2011})}\BibitemShut {NoStop}%
\bibitem [{\citenamefont {Xu}\ \emph {et~al.}(2024)\citenamefont {Xu}, \citenamefont {Rosander}, \citenamefont {Sch{\"a}fer}, \citenamefont {Lindgren}, \citenamefont {{\"O}sterbacka}, \citenamefont {Fang}, \citenamefont {Chen}, \citenamefont {He}, \citenamefont {Fan},\ and\ \citenamefont {Erhart}}]{XuRosSch24}%
  \BibitemOpen
  \bibfield  {author} {\bibinfo {author} {\bibfnamefont {Nan}\ \bibnamefont {Xu}}, \bibinfo {author} {\bibfnamefont {Petter}\ \bibnamefont {Rosander}}, \bibinfo {author} {\bibfnamefont {Christian}\ \bibnamefont {Sch{\"a}fer}}, \bibinfo {author} {\bibfnamefont {Eric}\ \bibnamefont {Lindgren}}, \bibinfo {author} {\bibfnamefont {Nicklas}\ \bibnamefont {{\"O}sterbacka}}, \bibinfo {author} {\bibfnamefont {Mandi}\ \bibnamefont {Fang}}, \bibinfo {author} {\bibfnamefont {Wei}\ \bibnamefont {Chen}}, \bibinfo {author} {\bibfnamefont {Yi}~\bibnamefont {He}}, \bibinfo {author} {\bibfnamefont {Zheyong}\ \bibnamefont {Fan}}, \ and\ \bibinfo {author} {\bibfnamefont {Paul}\ \bibnamefont {Erhart}},\ }\bibfield  {title} {\enquote {\bibinfo {title} {{Tensorial Properties via the Neuroevolution Potential Framework: Fast Simulation of Infrared and Raman Spectra}},}\ }\href {\doibase 10.1021/acs.jctc.3c01343} {\bibfield  {journal} {\bibinfo  {journal} {Journal of Chemical Theory and Computation}\ }\textbf {\bibinfo {volume}
  {20}},\ \bibinfo {pages} {3273--3284} (\bibinfo {year} {2024})}\BibitemShut {NoStop}%
\bibitem [{\citenamefont {Zhang}\ \emph {et~al.}(2021)\citenamefont {Zhang}, \citenamefont {Wang}, \citenamefont {Car},\ and\ \citenamefont {E}}]{zhang2018prl}%
  \BibitemOpen
  \bibfield  {author} {\bibinfo {author} {\bibfnamefont {Linfeng}\ \bibnamefont {Zhang}}, \bibinfo {author} {\bibfnamefont {Han}\ \bibnamefont {Wang}}, \bibinfo {author} {\bibfnamefont {Roberto}\ \bibnamefont {Car}}, \ and\ \bibinfo {author} {\bibfnamefont {Weinan}\ \bibnamefont {E}},\ }\bibfield  {title} {\enquote {\bibinfo {title} {Phase diagram of a deep potential water model},}\ }\href {\doibase 10.1103/PhysRevLett.126.236001} {\bibfield  {journal} {\bibinfo  {journal} {Phys. Rev. Lett.}\ }\textbf {\bibinfo {volume} {126}},\ \bibinfo {pages} {236001} (\bibinfo {year} {2021})}\BibitemShut {NoStop}%
\bibitem [{\citenamefont {Hui}\ and\ \citenamefont {Chai}(2016)}]{HuiCha16}%
  \BibitemOpen
  \bibfield  {author} {\bibinfo {author} {\bibfnamefont {Kerwin}\ \bibnamefont {Hui}}\ and\ \bibinfo {author} {\bibfnamefont {Jeng-Da}\ \bibnamefont {Chai}},\ }\bibfield  {title} {\enquote {\bibinfo {title} {{{SCAN-based}} hybrid and double-hybrid density functionals from models without fitted parameters},}\ }\href {\doibase 10.1063/1.4940734} {\bibfield  {journal} {\bibinfo  {journal} {The Journal of Chemical Physics}\ }\textbf {\bibinfo {volume} {144}},\ \bibinfo {pages} {044114} (\bibinfo {year} {2016})}\BibitemShut {NoStop}%
\bibitem [{\citenamefont {Dion}\ \emph {et~al.}(2004)\citenamefont {Dion}, \citenamefont {Rydberg}, \citenamefont {Schr\"oder}, \citenamefont {Langreth},\ and\ \citenamefont {Lundqvist}}]{DioRydSch04}%
  \BibitemOpen
  \bibfield  {author} {\bibinfo {author} {\bibfnamefont {M.}~\bibnamefont {Dion}}, \bibinfo {author} {\bibfnamefont {H.}~\bibnamefont {Rydberg}}, \bibinfo {author} {\bibfnamefont {E.}~\bibnamefont {Schr\"oder}}, \bibinfo {author} {\bibfnamefont {D.~C.}\ \bibnamefont {Langreth}}, \ and\ \bibinfo {author} {\bibfnamefont {B.~I.}\ \bibnamefont {Lundqvist}},\ }\bibfield  {title} {\enquote {\bibinfo {title} {Van der {Waals} density functional for general geometries},}\ }\href {\doibase 10.1103/PhysRevLett.92.246401} {\bibfield  {journal} {\bibinfo  {journal} {Physical Review Letters}\ }\textbf {\bibinfo {volume} {92}},\ \bibinfo {pages} {246401} (\bibinfo {year} {2004})}\BibitemShut {NoStop}%
\bibitem [{\citenamefont {Berland}\ and\ \citenamefont {Hyldgaard}(2014)}]{BerHyl2014}%
  \BibitemOpen
  \bibfield  {author} {\bibinfo {author} {\bibfnamefont {Kristian}\ \bibnamefont {Berland}}\ and\ \bibinfo {author} {\bibfnamefont {Per}\ \bibnamefont {Hyldgaard}},\ }\bibfield  {title} {\enquote {\bibinfo {title} {Exchange functional that tests the robustness of the plasmon description of the {van der Waals} density functional},}\ }\href {\doibase 10.1103/PhysRevB.89.035412} {\bibfield  {journal} {\bibinfo  {journal} {Physical Review B}\ }\textbf {\bibinfo {volume} {89}},\ \bibinfo {pages} {035412} (\bibinfo {year} {2014})}\BibitemShut {NoStop}%
\bibitem [{\citenamefont {Li}\ \emph {et~al.}(1999)\citenamefont {Li}, \citenamefont {Eddaoudi}, \citenamefont {O'Keeffe},\ and\ \citenamefont {Yaghi}}]{li1999design}%
  \BibitemOpen
  \bibfield  {author} {\bibinfo {author} {\bibfnamefont {Hailian}\ \bibnamefont {Li}}, \bibinfo {author} {\bibfnamefont {Mohamed}\ \bibnamefont {Eddaoudi}}, \bibinfo {author} {\bibfnamefont {Michael}\ \bibnamefont {O'Keeffe}}, \ and\ \bibinfo {author} {\bibfnamefont {Omar~M}\ \bibnamefont {Yaghi}},\ }\bibfield  {title} {\enquote {\bibinfo {title} {{Design and synthesis of an exceptionally stable and highly porous metal-organic framework}},}\ }\href {\doibase 10.1038/46248} {\bibfield  {journal} {\bibinfo  {journal} {Nature}\ }\textbf {\bibinfo {volume} {402}},\ \bibinfo {pages} {276--279} (\bibinfo {year} {1999})}\BibitemShut {NoStop}%
\bibitem [{\citenamefont {Chui}\ \emph {et~al.}(1999)\citenamefont {Chui}, \citenamefont {Lo}, \citenamefont {Charmant}, \citenamefont {Orpen},\ and\ \citenamefont {Williams}}]{chui1999chemically}%
  \BibitemOpen
  \bibfield  {author} {\bibinfo {author} {\bibfnamefont {Stephen S-Y}\ \bibnamefont {Chui}}, \bibinfo {author} {\bibfnamefont {Samuel M-F}\ \bibnamefont {Lo}}, \bibinfo {author} {\bibfnamefont {Jonathan~PH}\ \bibnamefont {Charmant}}, \bibinfo {author} {\bibfnamefont {A~Guy}\ \bibnamefont {Orpen}}, \ and\ \bibinfo {author} {\bibfnamefont {Ian~D}\ \bibnamefont {Williams}},\ }\bibfield  {title} {\enquote {\bibinfo {title} {{A chemically functionalizable nanoporous material [\textnormal{Cu}$_\textnormal{3}$(TMA)$_\textnormal{2}$(H$_\textnormal{2}$O)$_\textnormal{3}$]$_\textnormal{n}$}},}\ }\href {\doibase 10.1126/science.283.5405.1148} {\bibfield  {journal} {\bibinfo  {journal} {Science}\ }\textbf {\bibinfo {volume} {283}},\ \bibinfo {pages} {1148--1150} (\bibinfo {year} {1999})}\BibitemShut {NoStop}%
\bibitem [{\citenamefont {Huang}\ \emph {et~al.}(2006)\citenamefont {Huang}, \citenamefont {Lin}, \citenamefont {Zhang},\ and\ \citenamefont {Chen}}]{huang2006ligand}%
  \BibitemOpen
  \bibfield  {author} {\bibinfo {author} {\bibfnamefont {Xiao-Chun}\ \bibnamefont {Huang}}, \bibinfo {author} {\bibfnamefont {Yan-Yong}\ \bibnamefont {Lin}}, \bibinfo {author} {\bibfnamefont {Jie-Peng}\ \bibnamefont {Zhang}}, \ and\ \bibinfo {author} {\bibfnamefont {Xiao-Ming}\ \bibnamefont {Chen}},\ }\bibfield  {title} {\enquote {\bibinfo {title} {{Ligand-directed strategy for zeolite-type metal--organic frameworks: Zinc (II) imidazolates with unusual zeolitic topologies}},}\ }\href {\doibase 10.1002/anie.200503778} {\bibfield  {journal} {\bibinfo  {journal} {Angewandte Chemie International Edition}\ }\textbf {\bibinfo {volume} {45}},\ \bibinfo {pages} {1557--1559} (\bibinfo {year} {2006})}\BibitemShut {NoStop}%
\bibitem [{\citenamefont {Br{\"u}ckner}\ \emph {et~al.}(1966)\citenamefont {Br{\"u}ckner}, \citenamefont {Kleinst{\"u}ck},\ and\ \citenamefont {Schulze}}]{bruckner1966untersuchungen}%
  \BibitemOpen
  \bibfield  {author} {\bibinfo {author} {\bibfnamefont {W}~\bibnamefont {Br{\"u}ckner}}, \bibinfo {author} {\bibfnamefont {K}~\bibnamefont {Kleinst{\"u}ck}}, \ and\ \bibinfo {author} {\bibfnamefont {GER}\ \bibnamefont {Schulze}},\ }\bibfield  {title} {\enquote {\bibinfo {title} {{Untersuchungen von Gittereigenschaften im System LiH-LiD}},}\ }\href {\doibase https://doi.org/10.1002/pssb.19660140205} {\bibfield  {journal} {\bibinfo  {journal} {Physica Status Solidi (b)}\ }\textbf {\bibinfo {volume} {14}},\ \bibinfo {pages} {297--302} (\bibinfo {year} {1966})}\BibitemShut {NoStop}%
\bibitem [{\citenamefont {Ravindra}\ \emph {et~al.}(2024)\citenamefont {Ravindra}, \citenamefont {Advincula}, \citenamefont {Schran}, \citenamefont {Michaelides},\ and\ \citenamefont {Kapil}}]{RavAdvSch24}%
  \BibitemOpen
  \bibfield  {author} {\bibinfo {author} {\bibfnamefont {Pavan}\ \bibnamefont {Ravindra}}, \bibinfo {author} {\bibfnamefont {Xavier~R.}\ \bibnamefont {Advincula}}, \bibinfo {author} {\bibfnamefont {Christoph}\ \bibnamefont {Schran}}, \bibinfo {author} {\bibfnamefont {Angelos}\ \bibnamefont {Michaelides}}, \ and\ \bibinfo {author} {\bibfnamefont {Venkat}\ \bibnamefont {Kapil}},\ }\href {https://arxiv.org/abs/2312.01340} {\enquote {\bibinfo {title} {Quasi-one-dimensional hydrogen bonding in nanoconfined ice},}\ } (\bibinfo {year} {2024}),\ \Eprint {http://arxiv.org/abs/2312.01340} {arXiv:2312.01340 [cond-mat.stat-mech]} \BibitemShut {NoStop}%
\bibitem [{\citenamefont {Han}\ \emph {et~al.}(2018)\citenamefont {Han}, \citenamefont {Zhang}, \citenamefont {Car},\ and\ \citenamefont {E}}]{HanZhaCar18}%
  \BibitemOpen
  \bibfield  {author} {\bibinfo {author} {\bibfnamefont {Jiequn}\ \bibnamefont {Han}}, \bibinfo {author} {\bibfnamefont {Linfeng}\ \bibnamefont {Zhang}}, \bibinfo {author} {\bibfnamefont {Roberto}\ \bibnamefont {Car}}, \ and\ \bibinfo {author} {\bibfnamefont {Weinan}\ \bibnamefont {E}},\ }\bibfield  {title} {\enquote {\bibinfo {title} {Deep {{Potential}}: {{A General Representation}} of a {{Many-Body Potential Energy Surface}}},}\ }\href {\doibase 10.4208/cicp.oa-2017-0213} {\bibfield  {journal} {\bibinfo  {journal} {Communications in Computational Physics}\ }\textbf {\bibinfo {volume} {23}},\ \bibinfo {pages} {629--639} (\bibinfo {year} {2018})}\BibitemShut {NoStop}%
\bibitem [{\citenamefont {Batatia}\ \emph {et~al.}(2024)\citenamefont {Batatia}, \citenamefont {Benner}, \citenamefont {Chiang}, \citenamefont {Elena}, \citenamefont {Kov{\'a}cs}, \citenamefont {Riebesell}, \citenamefont {Advincula}, \citenamefont {Asta}, \citenamefont {Avaylon}, \citenamefont {Baldwin} \emph {et~al.}}]{BatBenChi24}%
  \BibitemOpen
  \bibfield  {author} {\bibinfo {author} {\bibfnamefont {Ilyes}\ \bibnamefont {Batatia}}, \bibinfo {author} {\bibfnamefont {Philipp}\ \bibnamefont {Benner}}, \bibinfo {author} {\bibfnamefont {Yuan}\ \bibnamefont {Chiang}}, \bibinfo {author} {\bibfnamefont {Alin~M.}\ \bibnamefont {Elena}}, \bibinfo {author} {\bibfnamefont {D{\'a}vid~P.}\ \bibnamefont {Kov{\'a}cs}}, \bibinfo {author} {\bibfnamefont {Janosh}\ \bibnamefont {Riebesell}}, \bibinfo {author} {\bibfnamefont {Xavier~R.}\ \bibnamefont {Advincula}}, \bibinfo {author} {\bibfnamefont {Mark}\ \bibnamefont {Asta}}, \bibinfo {author} {\bibfnamefont {Matthew}\ \bibnamefont {Avaylon}}, \bibinfo {author} {\bibfnamefont {William~J.}\ \bibnamefont {Baldwin}},  \emph {et~al.},\ }\href {https://arxiv.org/abs/2401.00096} {\enquote {\bibinfo {title} {A foundation model for atomistic materials chemistry},}\ } (\bibinfo {year} {2024}),\ \Eprint {http://arxiv.org/abs/2401.00096} {arXiv:2401.00096 [physics.chem-ph]} \BibitemShut {NoStop}%
\bibitem [{\citenamefont {Chen}\ \emph {et~al.}(2016)\citenamefont {Chen}, \citenamefont {Ambrosio}, \citenamefont {Miceli},\ and\ \citenamefont {Pasquarello}}]{CheAmbMic16}%
  \BibitemOpen
  \bibfield  {author} {\bibinfo {author} {\bibfnamefont {Wei}\ \bibnamefont {Chen}}, \bibinfo {author} {\bibfnamefont {Francesco}\ \bibnamefont {Ambrosio}}, \bibinfo {author} {\bibfnamefont {Giacomo}\ \bibnamefont {Miceli}}, \ and\ \bibinfo {author} {\bibfnamefont {Alfredo}\ \bibnamefont {Pasquarello}},\ }\bibfield  {title} {\enquote {\bibinfo {title} {{\emph{Ab Initio}} {{Electronic Structure}} of {{Liquid Water}}},}\ }\href {\doibase 10.1103/PhysRevLett.117.186401} {\bibfield  {journal} {\bibinfo  {journal} {Physical Review Letters}\ }\textbf {\bibinfo {volume} {117}},\ \bibinfo {pages} {186401} (\bibinfo {year} {2016})}\BibitemShut {NoStop}%
\bibitem [{\citenamefont {Morrone}\ and\ \citenamefont {Car}(2008)}]{MorCar08}%
  \BibitemOpen
  \bibfield  {author} {\bibinfo {author} {\bibfnamefont {Joseph~A.}\ \bibnamefont {Morrone}}\ and\ \bibinfo {author} {\bibfnamefont {Roberto}\ \bibnamefont {Car}},\ }\bibfield  {title} {\enquote {\bibinfo {title} {Nuclear {{Quantum Effects}} in {{Water}}},}\ }\href {\doibase 10.1103/PhysRevLett.101.017801} {\bibfield  {journal} {\bibinfo  {journal} {Physical Review Letters}\ }\textbf {\bibinfo {volume} {101}},\ \bibinfo {pages} {017801} (\bibinfo {year} {2008})}\BibitemShut {NoStop}%
\bibitem [{\citenamefont {Ceriotti}\ \emph {et~al.}(2016)\citenamefont {Ceriotti}, \citenamefont {Fang}, \citenamefont {Kusalik}, \citenamefont {McKenzie}, \citenamefont {Michaelides}, \citenamefont {Morales},\ and\ \citenamefont {Markland}}]{CerFanKus16}%
  \BibitemOpen
  \bibfield  {author} {\bibinfo {author} {\bibfnamefont {Michele}\ \bibnamefont {Ceriotti}}, \bibinfo {author} {\bibfnamefont {Wei}\ \bibnamefont {Fang}}, \bibinfo {author} {\bibfnamefont {Peter~G.}\ \bibnamefont {Kusalik}}, \bibinfo {author} {\bibfnamefont {Ross~H.}\ \bibnamefont {McKenzie}}, \bibinfo {author} {\bibfnamefont {Angelos}\ \bibnamefont {Michaelides}}, \bibinfo {author} {\bibfnamefont {Miguel~A.}\ \bibnamefont {Morales}}, \ and\ \bibinfo {author} {\bibfnamefont {Thomas~E.}\ \bibnamefont {Markland}},\ }\bibfield  {title} {\enquote {\bibinfo {title} {Nuclear {{Quantum Effects}} in {{Water}} and {{Aqueous Systems}}: {{Experiment}}, {{Theory}}, and {{Current Challenges}}},}\ }\href {\doibase 10.1021/acs.chemrev.5b00674} {\bibfield  {journal} {\bibinfo  {journal} {Chemical Reviews}\ }\textbf {\bibinfo {volume} {116}},\ \bibinfo {pages} {7529--7550} (\bibinfo {year} {2016})}\BibitemShut {NoStop}%
\bibitem [{\citenamefont {Togo}\ \emph {et~al.}(2023)\citenamefont {Togo}, \citenamefont {Chaput}, \citenamefont {Tadano},\ and\ \citenamefont {Tanaka}}]{phonopy-phono3py-JPCM}%
  \BibitemOpen
  \bibfield  {author} {\bibinfo {author} {\bibfnamefont {Atsushi}\ \bibnamefont {Togo}}, \bibinfo {author} {\bibfnamefont {Laurent}\ \bibnamefont {Chaput}}, \bibinfo {author} {\bibfnamefont {Terumasa}\ \bibnamefont {Tadano}}, \ and\ \bibinfo {author} {\bibfnamefont {Isao}\ \bibnamefont {Tanaka}},\ }\bibfield  {title} {\enquote {\bibinfo {title} {Implementation strategies in phonopy and phono3py},}\ }\href {\doibase 10.1088/1361-648X/acd831} {\bibfield  {journal} {\bibinfo  {journal} {Journal of Physics: Condensed Matter}\ }\textbf {\bibinfo {volume} {35}},\ \bibinfo {pages} {353001} (\bibinfo {year} {2023})}\BibitemShut {NoStop}%
\bibitem [{\citenamefont {Togo}(2023)}]{phonopy-phono3py-JPSJ}%
  \BibitemOpen
  \bibfield  {author} {\bibinfo {author} {\bibfnamefont {Atsushi}\ \bibnamefont {Togo}},\ }\bibfield  {title} {\enquote {\bibinfo {title} {First-principles phonon calculations with phonopy and phono3py},}\ }\href {\doibase 10.7566/JPSJ.92.012001} {\bibfield  {journal} {\bibinfo  {journal} {Journal of the Physical Society of Japan}\ }\textbf {\bibinfo {volume} {92}},\ \bibinfo {pages} {012001} (\bibinfo {year} {2023})}\BibitemShut {NoStop}%
\bibitem [{\citenamefont {Eriksson}\ \emph {et~al.}(2019)\citenamefont {Eriksson}, \citenamefont {Fransson},\ and\ \citenamefont {Erhart}}]{eriksson_hiphive_2019}%
  \BibitemOpen
  \bibfield  {author} {\bibinfo {author} {\bibfnamefont {Fredrik}\ \bibnamefont {Eriksson}}, \bibinfo {author} {\bibfnamefont {Erik}\ \bibnamefont {Fransson}}, \ and\ \bibinfo {author} {\bibfnamefont {Paul}\ \bibnamefont {Erhart}},\ }\bibfield  {title} {\enquote {\bibinfo {title} {The {Hiphive} {package} for the {extraction} of {high}-{order} {force} {constants} by {machine} {learning}},}\ }\href {\doibase 10.1002/adts.201800184} {\bibfield  {journal} {\bibinfo  {journal} {Advanced Theory and Simulations}\ }\textbf {\bibinfo {volume} {2}},\ \bibinfo {pages} {1800184} (\bibinfo {year} {2019})}\BibitemShut {NoStop}%
\bibitem [{\citenamefont {Fransson}\ \emph {et~al.}(2023{\natexlab{c}})\citenamefont {Fransson}, \citenamefont {Rosander}, \citenamefont {Eriksson}, \citenamefont {Rahm}, \citenamefont {Tadano},\ and\ \citenamefont {Erhart}}]{FraRosEri23}%
  \BibitemOpen
  \bibfield  {author} {\bibinfo {author} {\bibfnamefont {Erik}\ \bibnamefont {Fransson}}, \bibinfo {author} {\bibfnamefont {Petter}\ \bibnamefont {Rosander}}, \bibinfo {author} {\bibfnamefont {Fredrik}\ \bibnamefont {Eriksson}}, \bibinfo {author} {\bibfnamefont {J.~Magnus}\ \bibnamefont {Rahm}}, \bibinfo {author} {\bibfnamefont {Terumasa}\ \bibnamefont {Tadano}}, \ and\ \bibinfo {author} {\bibfnamefont {Paul}\ \bibnamefont {Erhart}},\ }\bibfield  {title} {\enquote {\bibinfo {title} {Limits of the phonon quasi-particle picture at the cubic-to-tetragonal phase transition in halide perovskites},}\ }\href {\doibase 10.1038/s42005-023-01297-8} {\bibfield  {journal} {\bibinfo  {journal} {Communications Physics}\ }\textbf {\bibinfo {volume} {6}},\ \bibinfo {pages} {173} (\bibinfo {year} {2023}{\natexlab{c}})}\BibitemShut {NoStop}%
\bibitem [{\citenamefont {Barbalinardo}\ \emph {et~al.}(2020)\citenamefont {Barbalinardo}, \citenamefont {Chen}, \citenamefont {Lundgren},\ and\ \citenamefont {Donadio}}]{kaldo}%
  \BibitemOpen
  \bibfield  {author} {\bibinfo {author} {\bibfnamefont {Giuseppe}\ \bibnamefont {Barbalinardo}}, \bibinfo {author} {\bibfnamefont {Zekun}\ \bibnamefont {Chen}}, \bibinfo {author} {\bibfnamefont {Nicholas~W}\ \bibnamefont {Lundgren}}, \ and\ \bibinfo {author} {\bibfnamefont {Davide}\ \bibnamefont {Donadio}},\ }\bibfield  {title} {\enquote {\bibinfo {title} {{Efficient anharmonic lattice dynamics calculations of thermal transport in crystalline and disordered solids}},}\ }\href {\doibase 10.1063/5.0020443} {\bibfield  {journal} {\bibinfo  {journal} {Journal of Applied Physics}\ }\textbf {\bibinfo {volume} {128}},\ \bibinfo {pages} {135104--12} (\bibinfo {year} {2020})}\BibitemShut {NoStop}%
\bibitem [{\citenamefont {Wang}\ \emph {et~al.}(2024)\citenamefont {Wang}, \citenamefont {Tian},\ and\ \citenamefont {Zhou}}]{wangzhou2024jctc}%
  \BibitemOpen
  \bibfield  {author} {\bibinfo {author} {\bibfnamefont {Chenyu}\ \bibnamefont {Wang}}, \bibinfo {author} {\bibfnamefont {Wei}\ \bibnamefont {Tian}}, \ and\ \bibinfo {author} {\bibfnamefont {Ke}~\bibnamefont {Zhou}},\ }\bibfield  {title} {\enquote {\bibinfo {title} {Ab initio simulation of liquid water without artificial high temperature},}\ }\href {\doibase 10.1021/acs.jctc.4c00650} {\bibfield  {journal} {\bibinfo  {journal} {Journal of Chemical Theory and Computation}\ } (\bibinfo {year} {2024}),\ 10.1021/acs.jctc.4c00650}\BibitemShut {NoStop}%
\bibitem [{\citenamefont {Folkner}\ \emph {et~al.}(2024)\citenamefont {Folkner}, \citenamefont {Chen}, \citenamefont {Barbalinardo}, \citenamefont {Knoop},\ and\ \citenamefont {Donadio}}]{folkner2024elastic}%
  \BibitemOpen
  \bibfield  {author} {\bibinfo {author} {\bibfnamefont {Dylan~A.}\ \bibnamefont {Folkner}}, \bibinfo {author} {\bibfnamefont {Zekun}\ \bibnamefont {Chen}}, \bibinfo {author} {\bibfnamefont {Giuseppe}\ \bibnamefont {Barbalinardo}}, \bibinfo {author} {\bibfnamefont {Florian}\ \bibnamefont {Knoop}}, \ and\ \bibinfo {author} {\bibfnamefont {Davide}\ \bibnamefont {Donadio}},\ }\href {https://arxiv.org/abs/2409.09551} {\enquote {\bibinfo {title} {Elastic moduli and thermal conductivity of quantum materials at finite temperature},}\ } (\bibinfo {year} {2024}),\ \Eprint {http://arxiv.org/abs/2409.09551} {arXiv:2409.09551 [cond-mat.mtrl-sci]} \BibitemShut {NoStop}%
\bibitem [{\citenamefont {Zeng}\ \emph {et~al.}(2024)\citenamefont {Zeng}, \citenamefont {Fan}, \citenamefont {Chen}, \citenamefont {Liang}, \citenamefont {Chen}, \citenamefont {Thornton},\ and\ \citenamefont {Cheng}}]{zeng2024lattice}%
  \BibitemOpen
  \bibfield  {author} {\bibinfo {author} {\bibfnamefont {Zezhu}\ \bibnamefont {Zeng}}, \bibinfo {author} {\bibfnamefont {Zheyong}\ \bibnamefont {Fan}}, \bibinfo {author} {\bibfnamefont {Chen}\ \bibnamefont {Chen}}, \bibinfo {author} {\bibfnamefont {Ting}\ \bibnamefont {Liang}}, \bibinfo {author} {\bibfnamefont {Yue}\ \bibnamefont {Chen}}, \bibinfo {author} {\bibfnamefont {Geoff}\ \bibnamefont {Thornton}}, \ and\ \bibinfo {author} {\bibfnamefont {Bingqing}\ \bibnamefont {Cheng}},\ }\href {https://arxiv.org/abs/2407.18510} {\enquote {\bibinfo {title} {Lattice distortion leads to glassy thermal transport in crystalline {Cs$_3$Bi$_2$I$_6$Cl$_3$}},}\ } (\bibinfo {year} {2024}),\ \Eprint {http://arxiv.org/abs/2407.18510} {arXiv:2407.18510 [cond-mat.mtrl-sci]} \BibitemShut {NoStop}%
\end{thebibliography}%



\clearpage
\newpage

\newcommand{\stitlecaptionlabel}[4]{
    \phantomsection
    \addcontentsline{toc}{subsection}{\ref{#4}. #1}
    \caption[{#1}]{\textbf{#2.} #3}
    \label{#4}
}
\newcommand{\titlecaptionlabel}[3]{
    \stitlecaptionlabel{#1}{#1}{#2}{#3}
}

\newcommand{\manuallabel}[2]{\def\@currentlabel{#2}\label{#1}}
\makeatother

\newcounter{note}
\newcommand{\notetitlelabel}[2]{
    \phantomsection
    \refstepcounter{note}
    \addcontentsline{toc}{subsection}{\ref{#1}. #2}
    \manuallabel{#1}{\thenote}
    \subsection*{Supplemental Note \thenote: #2}
}
\newcommand*{\noteautorefname}{Supplemental Note}

\setcounter{secnumdepth}{0}
\renewcommand{\refname}{}
\renewcommand{\figurename}{Figure}
\renewcommand{\tablename}{Table}
\renewcommand\thefigure{S\arabic{figure}}
\renewcommand\thetable{S\arabic{table}}
\renewcommand\thenote{S\arabic{note}}
\renewcommand\theequation{S\arabic{equation}}
\renewcommand{\topfraction}{.9}
\renewcommand{\bottomfraction}{.9}
\renewcommand{\floatpagefraction}{.9}
\newcommand{\listreferencename}{Supplemental References}
\def\sectionautorefname{Sect.}
\def\figureautorefname{Fig.}
\def\tableautorefname{Table}
\def\equationautorefname{Eq.}

\renewcommand{\bibsection}{\section*{\listreferencename}}
\renewcommand{\vec}[1]{\ensuremath\boldsymbol{#1}}
\newcommand{\dd}{\mathrm{d}}
\providecommand{\todo}[1]{\textbf{\color{red}#1}}

\setcounter{figure}{0}
\setcounter{table}{0}

\onecolumngrid
\begin{center}
{\bf Supplemental Material}
\end{center}

\begin{table*}[h!]
\caption{Volumetric thermal expansion coefficients of three MOFs obtained from previous experimental measurements. 
}
\begin{center}
\begin{tabular}{ l  l  l  l  l  l}
\hline
\hline
Material &$\alpha_V$ (10$^{-6}$ K$^{-1}$) & Temperature range (K) & Approach &  Reference \\
\hline
MOF-5 & -45.9 & 30–293 & single-crystal X-ray diffraction & \cite{rowsell2005gas} \\

 & -42.0 & 4–600 & neutron powder diffraction &  \cite{zhou2008origin} \\

 & -39.3 & 80–500 & powder X-ray diffraction &  \cite{lock2010elucidating} \\

 & -43.8 & 100–425 & powder X-ray diffraction & \cite{lock2013scrutinizing} \\

 & -36.0 & 20–400 & powder neutron diffraction  & \cite{lock2013scrutinizing} \\
\hline
HKUST-1 & -12.3 & 80–500 & powder X-ray diffraction &  \cite{wu2008negative} \\

 & -14.7 & 100-300 & neutron powder diffraction  & \cite{peterson2010local}\\

 & -15.3 & 100-300 & powder X-ray diffraction &  \cite{schneider2019tuning} \\
\hline
ZIF-8 & 35.7 & 100-300 & powder X-ray diffraction &  \cite{sapnik2018compositional}\\
 & 19.6 & 280-380 & powder X-ray diffraction &  \cite{burtch2019negative} \\
\hline
\hline
\end{tabular}
\end{center}
\label{table:experiment}
\end{table*}


\vfill

\begin{figure*}[htbp]
\begin{center}
\includegraphics[width=\columnwidth]{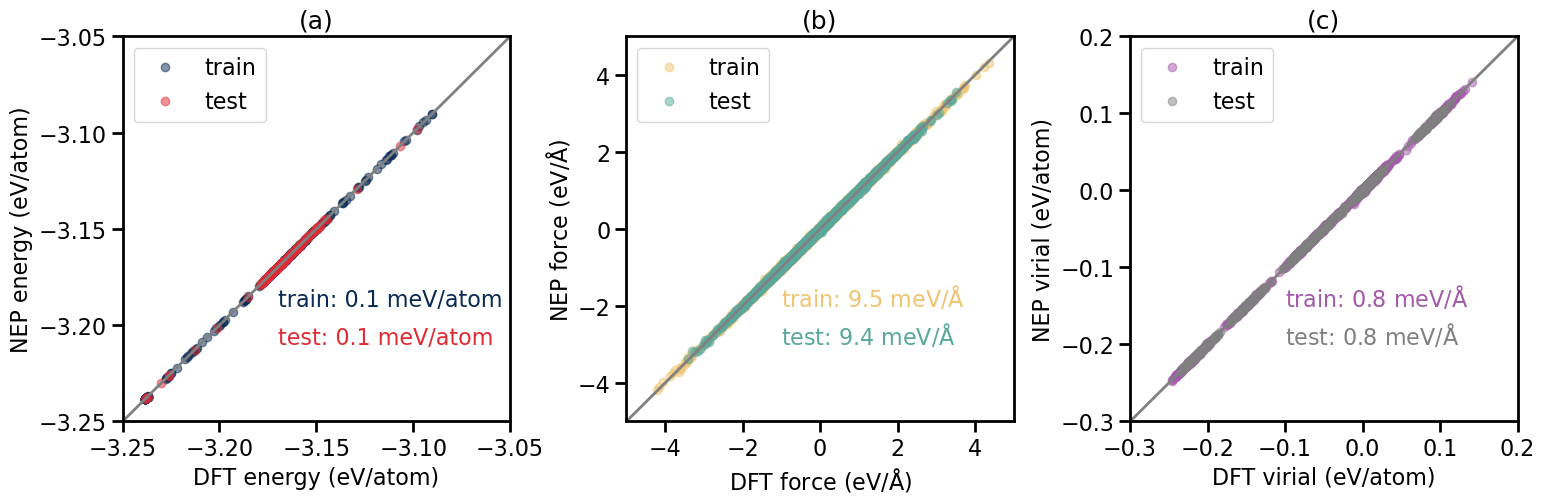}
\caption{The partial plots of (a) total energy, (b) atomic forces, (c) virial for LiH. The insets show the RMSE of the training and testing datasets.}
\end{center}
\end{figure*}

\vfill

\begin{figure}[htbp]
\begin{center}
\includegraphics[width=0.8\columnwidth]{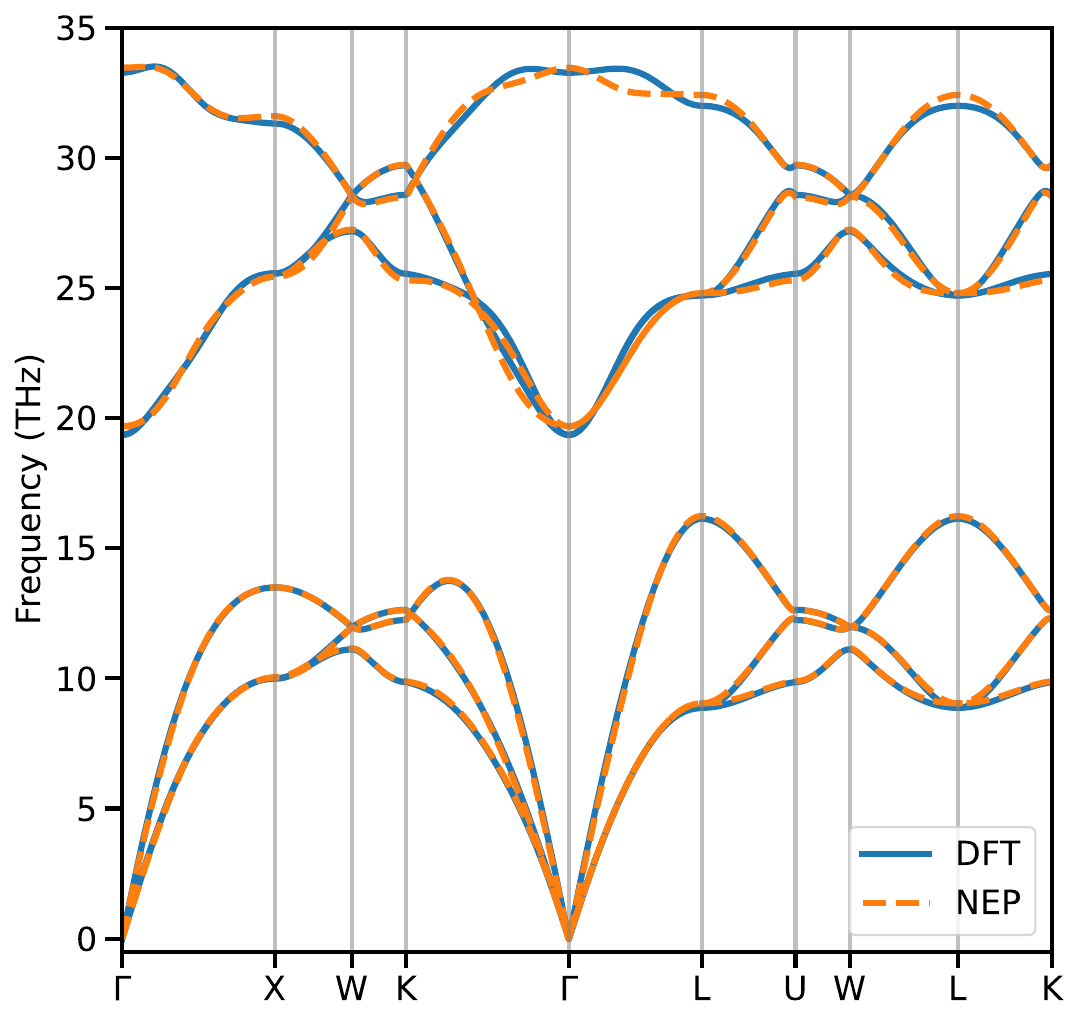}
\caption{Phonon dispersion relations for LiH. The blue and orange lines are calculated using DFT and NEP based finite-difference (FD) method.}
\end{center}
\end{figure}

\begin{figure}[htbp]
\begin{center}
\includegraphics[width=\columnwidth]{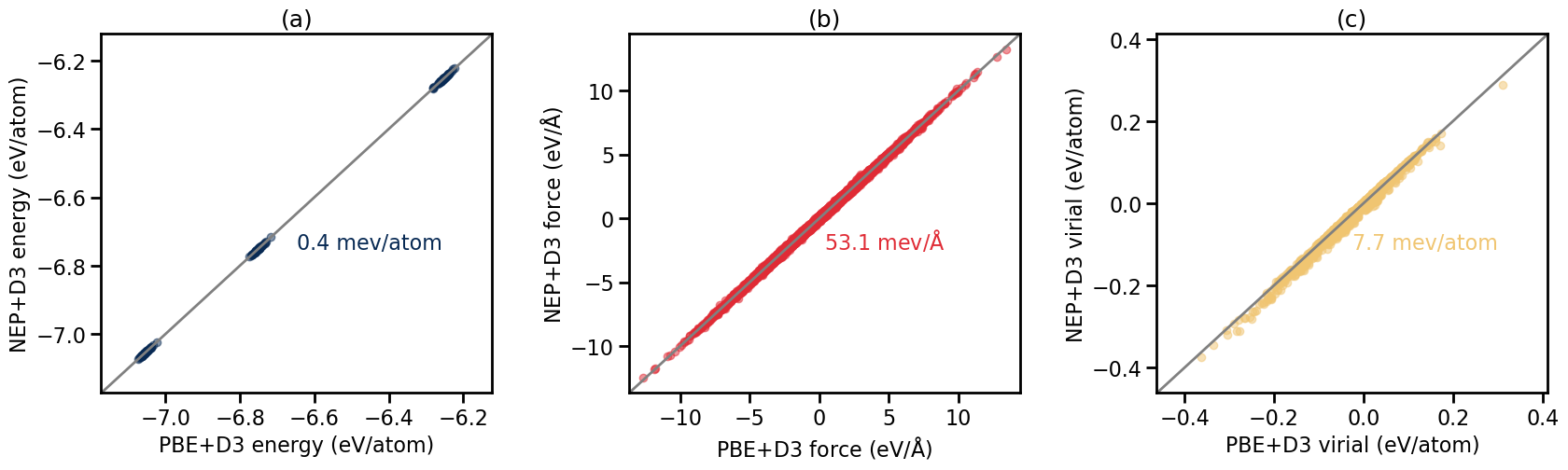}
\caption{The partial plots of (a) total energy, (b) atomic forces, (c) virial as predicted by the NEP-D3 approach, are compared against PBE-D3 calculations for snapshots extracted from PIMD simulations of three MOFs. }
\end{center}
\end{figure}

\begin{figure}[htbp]
\begin{center}
\includegraphics[width=0.6\columnwidth]{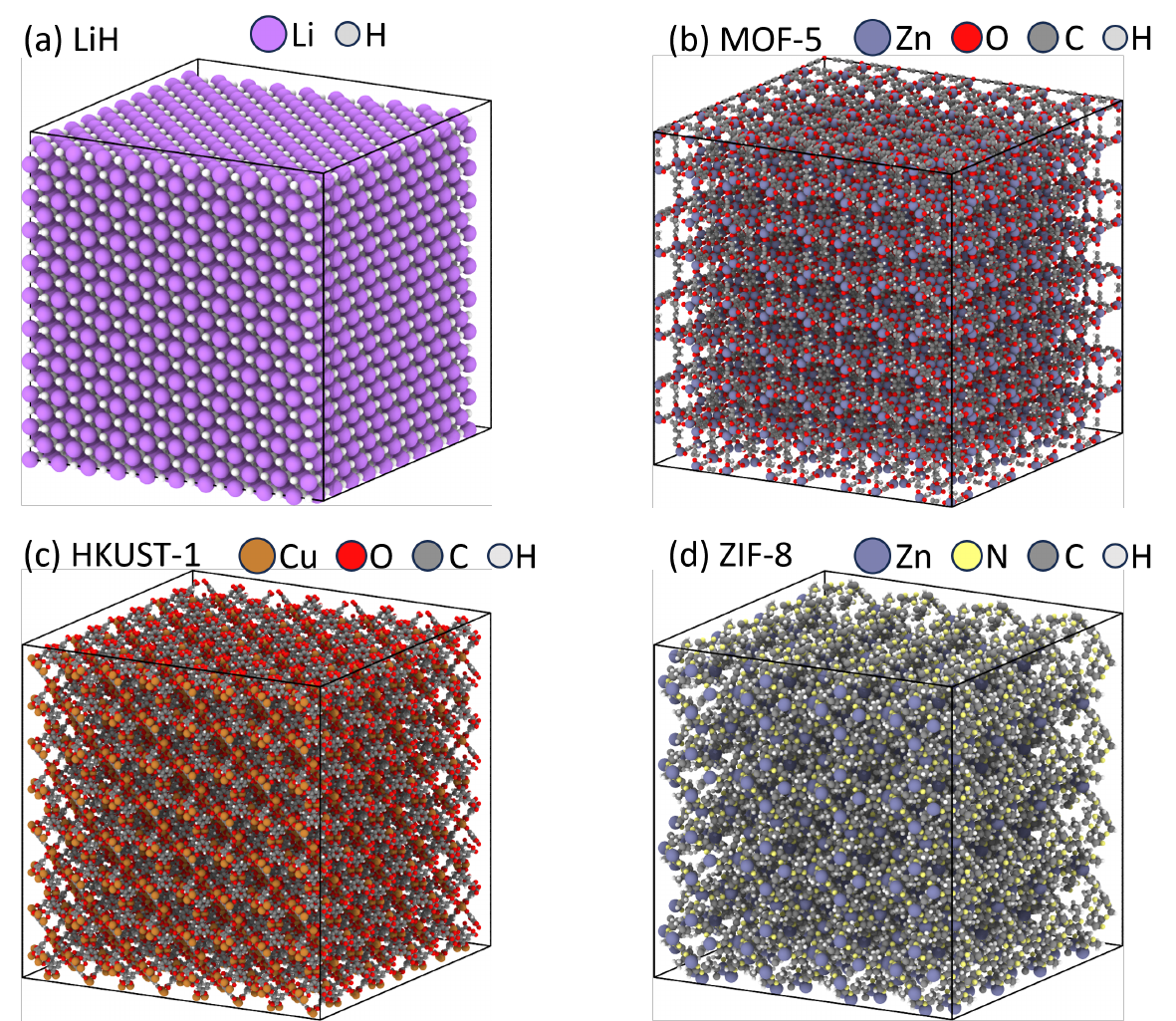}
\caption{The atomic supercells investigated in this work: (a) \numproduct{10x10x10} cubic supercell of LiH (8000 atoms), \numproduct{4x4x4} supercells of (b) MOF-5 (27136 atoms), (c) HKUST-1 (39936 atoms), and (d) ZIF-8 (17664 atoms).}
\end{center}
\end{figure}

\begin{figure}[htb]
\begin{center}
\includegraphics[width=\columnwidth]{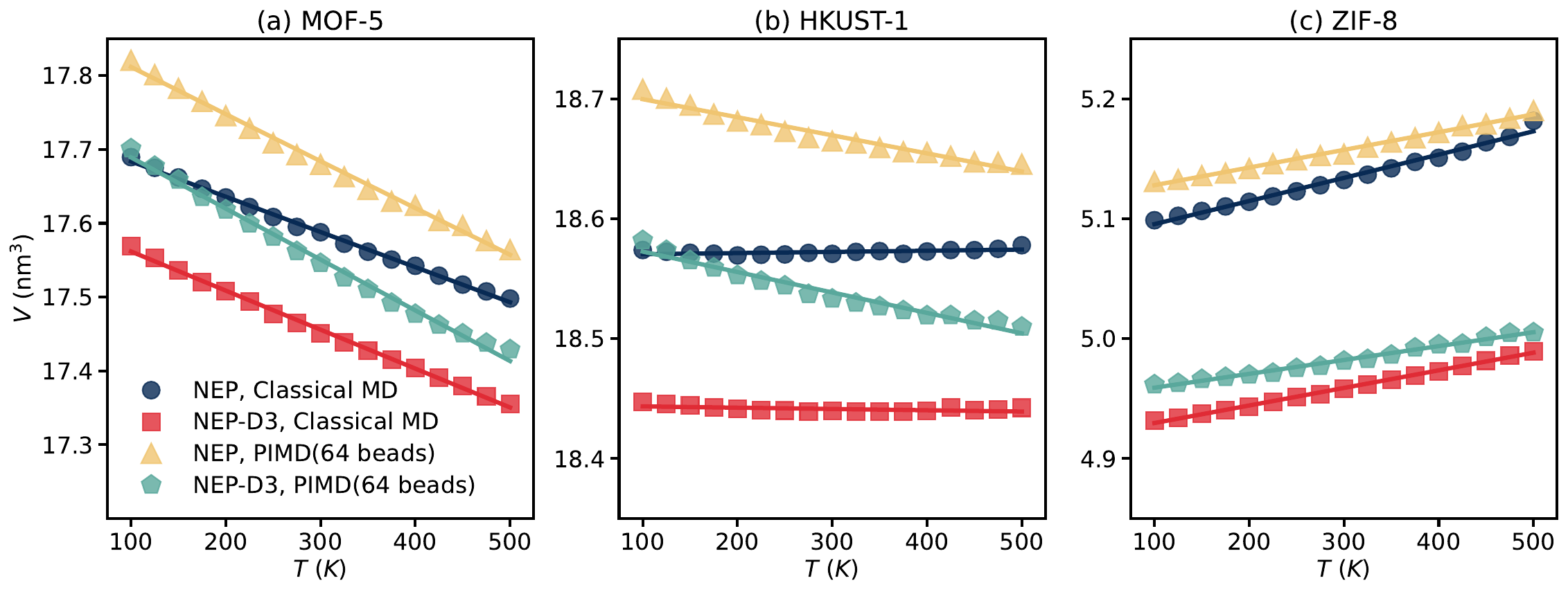}
\caption{\label{fig:temp}The volume of (a) MOF-5, (b) HKUST-1, and (c) ZIF-8 as a function of temperature, obtained from classical MD simulations driven by NEP (blue circles) or NEP-D3 (red squares), along with PIMD simulations using 64 beads, driven by NEP (orange triangles) or NEP-D3 (green pentagons). For each $V(T)$ result, a solid line is fitted using  Eq. \eqref{equation:tec} of main text.}
\end{center}
\end{figure}

\begin{figure}[htb]
    \centering
    \includegraphics[width=\columnwidth]{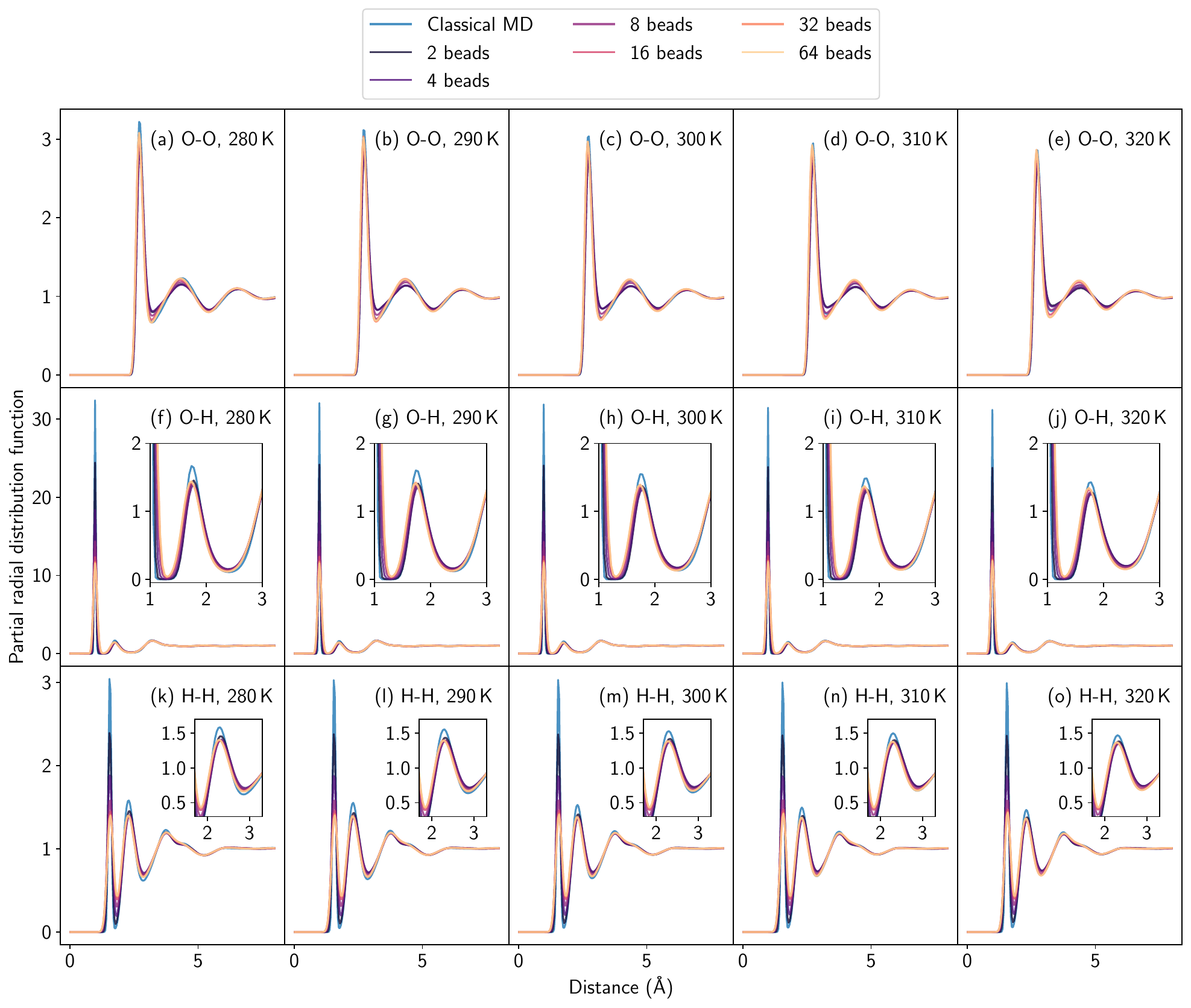}
    \caption{Partial oxygen--oxygen (a--e), oxygen-hydrogen (f--j) and hydrogen--hydrogen (k--o) radial distribution functions from classical MD and PIMD simulations with different numbers of beads, each for five different temperatures: \SI{280}{\kelvin} (a,f,k), \SI{290}{\kelvin} (b,g,l), \SI{300}{\kelvin} (c,h,m), \SI{310}{\kelvin} (d,i,n), and \SI{320}{\kelvin} (e,j,o). The insets show a zoomed-in view of the second peaks of the O--H and H--H RDFs at each temperature.}
\end{figure}

\begin{figure}[htb]
    \centering
    \includegraphics[width=0.5\columnwidth]{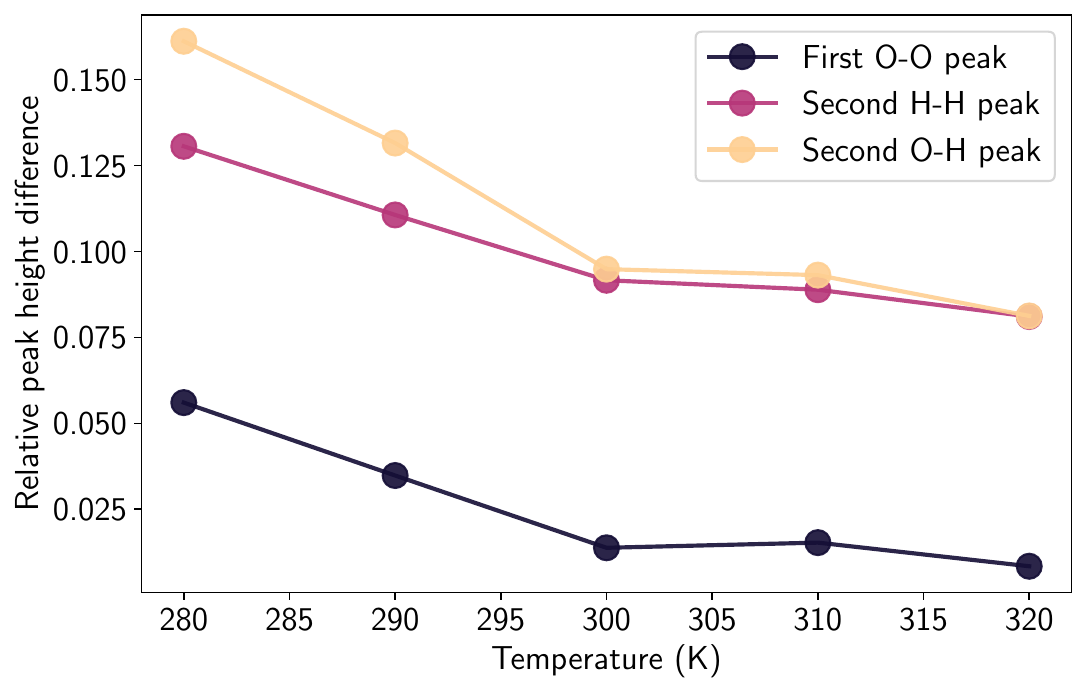}
    \caption{Relative peak height difference between classical MD and 64-bead PIMD for the first nearest-neighbor peak in the oxygen--oxygen (dark color), second peak of the hydrogen--hydrogen (intermediate color) and second peak of the oxygen--hydrogen (light color) RDFs. Note the peak height difference increasing when the temperature decreases.}
\end{figure}
\clearpage

\begin{figure}[!htb]
    \centering
    \includegraphics[width=0.7\columnwidth]{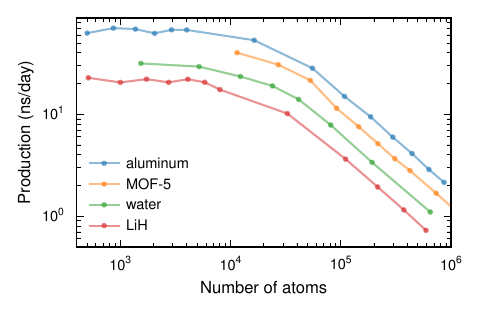}
    \caption{Computational efficiency on a single A100 GPU of the NEP models for aluminum, MOF-5, water, and LiH. The efficiency is measured in terms of simulated ns per day of wall-clock time, assuming a classical simulation (equivalent to one bead) and using a time step of \qty{2}{\femto\second} for aluminum and \qty{0.5}{\femto\second} for MOF-5, water and LiH.} 
\end{figure}

\clearpage
\phantomsection
\addcontentsline{toc}{section}{\listreferencename}

\end{document}